\documentstyle[12pt]{article}
\newcommand{\A}{{\cal A}}

\newcommand{\al}{\alpha}

\newcommand{\be}{\beta}

\newcommand{\bwe}{\mbox{$\bigwedge$}}
\newcommand{\C}{\;\mbox{{\sf I}}\!\!\!C}
\newcommand{\Ca}{{\cal C}}

\newcommand{\cent}{\centerline}

\newcommand{\de}{\delta}

\newcommand{\dis}{\displaystyle}

\newcommand{\divi}{{\rm div}}

\newcommand{\Fe}{{\cal F}}
\newcommand{\for}{\forall}
\newcommand{\Ga}{\Gamma}

\newcommand{\ga}{\gamma}
\newcommand{\grad}{{\rm grad}}

\newcommand{\h}{\hspace*{7mm}}
\newcommand{\hb}{\hbar}

\newcommand{\hs}{\hspace*}

\newcommand{\ia}{{\bf i}}
\newcommand{\infi}{\infty}

\newcommand{\La}{\Lambda}
\newcommand{\Le}{{\cal L}}
\newcommand{\Lfr}{\Leftrightarrow}

\newcommand{\la}{\lambda}
\newcommand{\lan}{\langle}
\newcommand{\ld}{\ldots}
\newcommand{\leri}{\leftrightarrow}

\newcommand{\Me}{{\cal M}}

\newcommand{\n}{{\noindent}}
\newcommand{\nab}{\nabla}
\newcommand{\nb}{\nonumber}

\newcommand{\Om}{\Omega}
\newcommand{\om}{\omega}

\newcommand{\ot}{\otimes}
\newcommand{\ov}{\overline}

\newcommand{\pa}{\partial}

\newcommand{\R}{I\!\!R}
\newcommand{\Rig}{\Rightarrow}
\newcommand{\ro}{\mbox{\boldmath $\rho$}}
\newcommand{\ran}{\rangle}

\newcommand{\rig}{\rightarrow}
\newcommand{\rot}{{\rm rot}}

\newcommand{\Sig}{\Sigma}

\newcommand{\sig}{\sigma}

\newcommand{\sub}{\subset}
\newcommand{\sups}{\supseteq}
\newcommand{\subq}{\subseteq}

\newcommand{\T}{{\bf T}}

\newcommand{\te}{\theta}
\newcommand{\teo}{\mbox{\boldmath $\theta$}}

\newcommand{\und}{\underline}

\newcommand{\upa}{\uparrow}

\newcommand{\var}{\varphi}
\newcommand{\varo}{\mbox{\boldmath $\varphi$}}
\newcommand{\vare}{\varepsilon}
\newcommand{\ven}{{\bf v}}
\newcommand{\vs}{\vspace*}

\newcommand{\we}{\wedge}

\newcommand{\wid}{\widetilde}
\newcommand{\wide}{\widehat}

\newcommand{\xis}{\raisebox{2.2pt}{$\chi$}}

\newcommand{\zo}{\mbox{\boldmath $z$}}

\newcommand{\clif}{{\cal C}\! \ell}
\newcommand{\dirac}{\mbox{\boldmath $\partial$}}

\begin{document}

\begin{center}
{\large\bf  On the Existence of
Undistorted Progressive Waves (UPWs) of Arbitrary Speeds $0 \leq v<
\infty$ in Nature}
\end{center}

\vs{5mm}

\cent{\it Waldyr A.\ Rodrigues, Jr.$^{\rm (a)}$ \ {\it and} \ Jian-Yu
Lu$^{\rm (b)}$}

\vs{5mm}

\begin{tabular}{ll}
(a) & Instituto de Matem\'atica, Estat\'\i stica e 
Computa\c c\~ao  Cient\'{\i}fica \\
 & IMECC-UNICAMP; CP 6065, 13081-970, Campinas, SP, Brasil \\
 &   e-mail: walrod@ime.unicamp.br\\
(b) & Biodynamics Research Unit, Department of Physiology and
Biophysics \\
& Mayo Clinic and Foundation, Rochester, MN55905, USA \\ 
& e-mail: jian@us0.mayo.edu 
\end{tabular}

\begin{abstract}
We present the theory, the experimental evidence and fundamental
physical consequences concerning the existence of families of
undistorted progressive waves (UPWs) of arbitrary speeds $0\leq v <
\infty$, which are solutions of the homogeneous wave equation, Maxwell
equations, Dirac, Weyl and Klein-Gordon equations.
\end{abstract}


\bigbreak
\n {\large\bf 1. Introduction} \\
\nobreak

In this paper we present the theory, the experimental evidence, and the
fundamental physical consequences concerning the existence of families
of undistorted progressive waves (UPWs)$^{(*)}$\footnotetext{$^{(*)}$UPW
is used for the singular, {\em i.e.\/}, for undistorted progressive
wave.} moving with arbitrary speeds$^{(**)}$\footnotetext{$^{(**)}$We
use units where $c=1$, $c$ being the so called velocity of light in
vacuum.} $0 \leq v < \infi$. We show that the main equations of
theoretical physics, namely: the scalar {\it homogeneous wave equation}
(HWE); the {\it Klein-Gordon equation} (KGE); the {\it Maxwell
equations}, the {\it Dirac and Weyl equations} have UPWs solutions in a
homogeneous medium, including the vacuum. By UPW, following Courant and
Hilbert$^{[1]}$ we mean that the UPW waves are distortion free, {\em
i.e.\/} they are translationally invariant and thus do not spread, or
they reconstruct their original form after a certain period of time.
Explicit examples of how to construct the UPWs solutions for the HWE are
found in Appendix A.\ The UPWs solutions to any field equations have
infinite energy. However, using the finite aperture approximation (FAA)
for diffraction (Appendix~A), we can project quasi undistorted
progressive waves (QUPWs) for any field equation which have finite
energy and can then in principle be launched in physical space.

In section 2 we show results of a recent experiment proposed and
realized by us where the measurement of the speeds of a FAA to a {\it
subluminal}$^{(*)}$\footnotetext{$^{(*)}$In this experiment the
waves are sound waves in water and, of course, the meaning of the
words subluminal, luminal and superluminal in this case is that the
waves travel with speed less, equal or greater than $c_s$, the so
called velocity of sound in water.} Bessel pulse [eq.(2.1)] and of the
FAA to a {\it superluminal} $X$-wave [eq.(2.5)] are done. The results
are in excellent agreement with the theory.

In section 3 we discuss some examples of UPWs solutions of Maxwell
equations; (i) subluminal solutions which  are interesting concerning
some recent attempts appearing in the literature$^{[2, 3, 4]}$ of
construction of {\it purely electromagnetic particles\/} (PEP) and (ii)
a superluminal UPW solution of Maxwell equations called the superluminal
electromagnetic {\em X-wave}$^{[5]}$ (SEXW). We briefly discuss how to
launch a FAA to SEXW.\ In view of the experimental results presented in
section 2 we are confident that such electromagnetic waves will be
produced in the next few years. In section 4 we discuss the important
question concerning the speed of propagation of the energy carried by
superluminal UPWs solutions of Maxwell equations, clearing some
misconceptions found in the literature. In section 5 we show that the
experimental production of a superluminal electromagnetic wave implies
in a breakdown of the Principle of Relativity.  In section 6 we present
our conclusions.

Appendix B presents a {\it unified theory} of how to construct UPWs of
arbitrary speeds \protect{$0 \leq v < \infty$} which are solutions of
Maxwell, Dirac and Weyl equations. Our unified theory is based on the
Clifford bundle formalism$^{[6,7,8,9,10]}$ where all fields quoted above
are represented by objects of the same mathematical nature. We take the
care of translating all results in the standard mathematical formalisms
used by physicists in order for our work to be usefull for a larger
audience.

Before starting the technical discussions it is worth to briefly recall
the history of the UPWs of arbitrary speeds $0 \leq v < \infty$, which
are solutions of the main equations of theoretical physics.

To the best of our knowledge H.\ Bateman$^{[11]}$ in 1913 was the first
person to present a subluminal UPW solution of the HWE.\ This solution
corresponds to what we called the subluminal {\it spherical Bessel beam}
in Appendix A [see eq.(A.31)]. Apparently this solution has been
rediscovered and used in diverse contexts many times in the literature.
It appears, {\em e.g.\/}, in the papers of Mackinnon$^{[12]}$ of 1978
and of Gueret and Vigier$^{[13]}$ and more recently in the papers of
Barut and collaborators$^{[14,15]}$. In particular in$^{[14]}$ Barut
also shows that the HWE has superluminal solutions. In 1987 Durnin and
collaborators rediscovered a subluminal UPW solution of the HWE in
cylindrical coordinates$^{[16,17,18]}$. These are the Bessel beams of
section A4 [see eq.(A.41)]. We said rediscovered because these solutions
are known at least since 1941, as they are explicitly written down in
Stratton's book$^{[19]}$. The important point here is that
Durnin$^{[16]}$ and collaborators constructed an optical subluminal
Bessel beam. At that time they didn't have the idea of measuring the
speed of the beams, since they were interested in the fact that the FAA
to these beams were quasi UPWs and could be very usefull for optical
devices. Indeed they used the term ``diffraction-free beams" which has
been adopted by some other authors later. Other authors still use for
UPWs the term {\it non-dispersive beams}. We quote also that Hsu and
collaborators$^{[20]}$ realized a FAA to the $J_0$ Bessel beam
[eq.(A.41)] with a narrow band PZT ultrasonic transducer of non-uniform
poling. Lu and Greenleaf$^{[21]}$ produced the first $J_0$
nondiffracting annular array transducers with PZT ceramic/polymer
composite and applied it to medical acoustic imaging and tissue
characterization$^{[22,23]}$.  Also Campbell et al$^{[24]}$ used an
annular array to realize a FAA to a $J_0$ Bessel beam and compared the
$J_0$ beam to the so called {\it axicon beam\/}$^{[25]}$. For more on
this topic see$^{[26]}$.

Luminal solutions of a new kind for the HWE and Maxwell equations, also
known as focus wave mode [FWM] (see Appendix A), have been discovered by
Brittingham~$^{[27]}$ (1983) and his work inspired many interesting and
important studies as, {\em e.g.\/},$^{[29-40]}$.

To our knowledge the first person to write about the possibility of a
{\it superluminal} UPW solution of HWE and, more important, of Maxwell
equations was Band$^{[41,42]}$. He constructed a superluminal
electromagnetic UPW from the modified Bessel beam [eq.(A.42)] which was
used to generate in an appropriate way an electromagnetic potential in
the Lorentz gauge. He suggested that his solution could be used to
eventually launch a superluminal wave in the exterior of a conductor
with cylindrical symmetry with appropriate charge density. We discuss
more some of Band's statements in section 4.

In 1992 Lu and Greenleaf$^{[43]}$ presented the first superluminal UPW
solution of the HWE for acoustic waves which could be launched by a
physical device$^{[44]}$. They discovered the so called $X$-waves, a
name due to their shape (see Fig. 3). In the same  year Donnelly and
Ziolkowski$^{[45]}$ presented a thoughtfull method for generating UPWs
solutions of homogeneous partial equations. In particular they studied
also UPW solutions for the wave equation in a lossy infinite medium and
to the KGE.\ They clearly stated also how to use these solutions to
obtain through the Hertz potential method (see appendix B, section B3)
UPWs solutions of Maxwell equations.

In 1993 Donnely and Ziolkowski$^{[46]}$ reinterpreted their study
of$^{[45]}$ and obtained subluminal, luminal and superluminal UPWs
solutions of the HWE and of the KGE.\ In Appendix A we make use of the
methods of this important paper in order to obtain some UPWs solutions.
Also in 1992 Barut and Chandola$^{[47]}$ found superluminal UPWs
solutions of the HWE.\ In 1995 Rodrigues and Vaz$^{[48]}$ discovered in
quite an independent way$^{(*)}$\footnotetext{$^{(*)}$Rodrigues and Vaz
are interested in obtaining solutions of Maxwell equations characterized
by non-null field invariants, since solutions of this kind
are$^{[49,50]}$ necessary in proving a surprising relationship between
Maxwell and Dirac equations.} subluminal and superluminal UPWs solutions
of Maxwell equations and the Weyl equation. At that time Lu and
Greenleaf$^{[5]}$ proposed also to launch a superluminal electromagnetic
$X$-wave.$^{(**)}$\footnotetext{$^{(**)}$A version of [5] was submitted
to {\em IEEE Trans. Antennas Propag.\/} in 1991. See reference 40
of$^{[43]}$.}

In September 1995 Professor Ziolkowski took knowledge of$^{[48]}$ and
informed  one of the authors [WAR] of his publications and also of Lu's
contributions. Soon a collaboration with Lu started which produced this
paper. To end this introduction we must call to the reader's attention
that in the last few years several important experiments concerning the
superluminal tunneling of electromagnetic waves appeared in the
literature$^{[51,52]}$. Particularly interesting is Nimtz's
paper$^{[53]}$  announcing that he transmitted Mozart's Symphony \# 40
at 4.7c through a retangular waveguide. The solutions of Maxwell
equations in a waveguide lead to solutions of Maxwell equations that
propagate with subluminal or superluminal speeds.  These solutions can
be obtained with the methods discussed in this paper  and will be
discussed in another publication.

\vs{1cm}

\n {\large\bf 2. Experimental Determination of the Speeds of Acoustic
Finite Aperture Bessel Pulses and $X$-Waves.} \\

In appendix A we show the existence of several UPWs solutions to the
HWE, in particular the subluminal UPWs Bessel beams [eq.(A.36)] and the
superluminal UPWs $X$-waves [eq.(A.52)].  Theoretically the UPWs
$X$-waves, both the broad-band and band limited [see eq.(2.4)] travel
with speed $v = c_s/ \cos \eta > 1$. Since only FAA to these $X$-waves
can be launched with appropriate devices, the question arises if these
FAA $X$-waves travel also with speed  greater than $c_s$, what can be
answered only by experiment.  Here we present the results of
measurements of the speeds of a FAA to a broad band Bessel beam, called
a Bessel pulse (see below) and of a FAA to a band limited $X$-wave, both
moving in water. We write the formulas for these beams inserting into
the HWE the parameter $c_s$ known as the speed of sound in water. In
this way the dispersion relation [eq.(A.37)] must read
\renewcommand{\theequation}{2.\arabic{equation}}
\setcounter{equation}{0}
\begin{equation}
\frac{\om^2}{c_s^2} - k^2 = \alpha^2 \, . 
\end{equation}
Then we write for the Bessel beams
\begin{equation}
\Phi^<_{J_n} (t, \vec x) = J_n(\al\rho) e^{i(kz-\om t + n\te)}, \ \
n=0,1,2,\ld
\end{equation}
Bessel pulses are obtained from eq.(2.2) by weighting it with a
transmitting transfer function, $T(\om)$ and then linearly superposing
the result over angular frequency $\om$, {\em i.e.\/}, we have
\begin{equation} 
\Phi^<_{JBB_n} (t, \vec x) = 2\pi e^{in\te}
J_n(\al\rho) \Fe^{-1} [T(\om)e^{ikz}], 
\end{equation} 
where $\Fe^{-1}$ is the inverse Fourier transform. The FAA to
$\Phi^<_{JBB_n}$ will be denoted by FAA$\Phi^<_{JBB_n}$ (or
$\Phi^<_{FAJ_n}$).

We recall that the $X$-waves are given by eq.(A.52), {\em i.e.\/},
\begin{equation}
\Phi^>_{X_n} (t, \vec x) = e^{in\te} \int^\infi_0 B(\ov k)J_n (\ov k
\rho\sin\eta) e^{-\ov k [a_0-i(z\cos\eta-c_s t)]} d \ov k \, ,
\end{equation}
where $\ov k = k /\cos \eta, \ \ov k=\om/c_s$. By choosing $B(\ov
k)=a_0$ we have the infinite aperture broad bandwidth $X$-wave
[eq.(A.53)] given by
\begin{eqnarray}
&& \Phi^>_{XBB_n} (t,\vec x) = \frac{a_0(\rho\sin\eta)^n
e^{in\te}}{\sqrt{M} (\tau+\sqrt{M})^n}, \nb \\
&& \\
&& M = (\rho\sin\eta)^2 + \tau^2, \ \ \tau = [a_0-i(z\cos\eta-c_s t)].
\nb
\end{eqnarray}
A FAA to $\Phi^>_{XBB_n}$ will be denoted by FAA$\Phi^>_{XBB_n}$.  When
$B(\ov k)$ in eq.(2.4) is different from a constant, {\em e.g.\/}, if
$B(\ov k)$ is the Blackman window function we denote the $X$-wave by
$\Phi^>_{XBL_n}$, where $BL$ means band limited. A FAA to
$\Phi^>_{XBL_n}$ will be denoted FAA$\Phi_{XBL_n}$. Also when
$T(\omega)$ in eq.(2.3) is the Blackman window function we denote the
respective wave by $\Phi_{JBL_n}$.

As discussed in Appendix A and detailed in$^{[26,44]}$ to produce a FAA
to a given beam the aperture of the transducer used must be finite.  In
this case the beams produced, in our case FAA$\Phi_{JBL_0}$ and
FAA$\Phi_{XBB_0}$, have a finite depth of field$^{[26]}$
(DF)$^{(*)}$\footnotetext{$^{(*)}$DF is the distance where the field
maximum drops to half the value at the surface of the transducer.} and
can be approximately produced by truncating the infinite aperture beams
$\Phi_{JBL_0}$ and $\Phi_{XBB_0}$ (or $\Phi_{XBL_0}$) at the transducer
surface $(z=0)$. Broad band pulses for $z > 0$ can be obtained by first
calculating the fields at all frequencies with eq.(A.28), {\em i.e.\/},
\begin{eqnarray}
 \wid{\wid\Phi}_{FA} (\om, \vec x) & = & \frac{1}{i\la} \int^a_0 \!
\int^\pi_{-\pi} \rho' d\rho' d\te' \wid{\wid\Phi} (\om, \vec{x}')
\frac{e^{i\ov k R}}{R^2} z \nb \\
&&  \\
&+& \frac{1}{2\pi} \int^a_0 \! \int^\pi_{-\pi} \rho' d \rho' d\theta'
\wid{\wid\Phi} (\om, \vec{x}') \frac{e^{i\ov kR}}{R^3} z , \nb
\end{eqnarray}
where the aperture weighting function $\wid{\wid\Phi} (\om, \vec x')$ is
obtained from the temporal Fourier transform of eqs.(2.3) and (2.4). If
the aperture is circular of radius $a$ [as in eq.(2.6)], the depth of
field of the FAA$\Phi_{JBL_0}$ pulse, denoted $BZ_{\max}$ and the depth
of field of the FAA$\Phi_{XBB_0}$ or FAA $\Phi_{XBL_0}$ denoted by
$XZ_{\max}$ are given by$^{[26]}$
\begin{equation}
BZ_{\max} = a\, \sqrt{\left(\frac{\om_0}{c_s\al}\right)^2-1} \, ; \ \ \
XZ_{\max} = a \cot\eta.
\end{equation}

For the FAA$\Phi_{JBL_0}$ pulse we choose $T(\om)$ as the Blackman
window function$^{[54]}$ that is peaked at the central frequency
$f_0=2.5 MHz$ with a relative bandwidth of about 81\% ($-6\,dB$
bandwidth divided by the central frequency). We have
\begin{equation}
B(\ov k) = \left\{\begin{array}{l}
a_0 \left[ 0.42 - 0.5 {\dis\frac{\pi\ov k}{\ov k_0} + 0.08 \cos
\frac{2\pi\ov k}{\ov k_0}} \right]\, , \ 0 \leq \ov k \leq 2 \ov k_0 ; \\
0 \ \ {\rm otherwise} .
\end{array}\right.
\end{equation}
The ``scaling factor" in the experiment is $\al = 1202.45m^{-1}$ and the
weighting function $\wid{\wid\Phi}_{JBB_0} (\om,\vec x)$ in eq.(2.6) is
approximated with stepwise functions.  Practically this is done with the
10-element annular array transfer built by Lu and Greenleaf$^{[26,44]}$.
The diameter of the array is 50mm.
Fig.~1$^{(**)}$\footnotetext{$^{(**)}$Reprinted with permission from
fig.\ 2 of$^{[44]}$.} shows the block diagram for the production of FAA
$\Phi^>_{XBL_0}$ and FAA$\Phi^<_{JBL_0}$.  The measurement of the speed
of the FAA Bessel pulse has been done by comparing the speed  with which
the peak of the FAA Bessel pulse travels  with the speed of the peak of
a pulse produced by a small circular element of the array (about 4mm or
$6.67\la$ in diameter, where $\la$ is 0.6mm in water). This pulse
travels with speed $c_s=1.5\,mm/\mu s$. The distance between the peaks
and the surface of the transducer are 104.33(9)mm and 103.70(5)mm for
the single-element wave and the Bessel pulse, respectively, at the same
instant $t$ of measurement. The results can be seen in the pictures
taken from of the experiment in Fig.~2. As predicted by the theory
developed in Appendix A the speed of the Bessel pulse is 0.611(3)\%
slower than the speed $c_s$ of the usual sound wave produced by the
single element.

The measurement of the speed of the central peak of the FAA
$\Phi^>_{XBL_0}$ wave obtained from eq.(2.4) with a Blackman window
function [eq.(2.8)] has been done in the same way as for the Bessel
pulse. The FAA$\Phi_{XBL_0}$ wave has been produced by the 10-element
array transducer of 50mm of diameter with the techniques developed by Lu
and Greenleaf$^{[26,44]}$. The distances traveled at the same instant
$t$ by the single element wave and the $X$-wave are respectively
173.48(9)mm and 173.77(3)mm. Fig. 3 shows the pictures taken from the
experiment. In this experiment the axicon angle is $\eta = 4^0$.  The
theoretical speed of the infinite aperture $X$-wave is predicted to be
0.2242\% greater then $c_s$. We found that the FAA$\Phi_{XBB_0}$ wave
traveled with speed 0.267(6)\% greater then $c_s$~!

These results, which we believe are the first experimental determination
of the speeds of subluminal and superluminal quasi-UPWs
FAA$\Phi^>_{JBL_0}$ and FAA$\Phi^<_{JBB_0}$ solutions of the HWE,
together with the fact that, as already quoted, Durnin$^{[16]}$ produced
subluminal {\it optical} Bessel beams, give us confidence  that
electromagnetic subluminal and superluminal waves may be physically
launched with appropriate devices. In the next section we study in
particular the superluminal electromagnetic $X$-wave (SEXW).

It is important to observe here the following crucial points: (i)~The
FAA $\Phi_{XBB_n}$ is produced by the source (transducer) in a short
period of time $\Delta t$. However, different parts of the transducer
are activated at different times, from 0 to $\Delta t$, calculated from
eqs.(A.9) and (A.28). As a result the wave is born as an integral object
for time $\Delta t$ and propagates with the same speed as the peak.
This is exactly what has been seen in the experiments and is
corroborated by the computer simulations we did for the superluminal
electromagnetic waves (see section 3). (ii)~One can find in almost all
textbooks that the velocity of transport of energy for waves obeying the
scalar wave equation
\begin{equation}
\left( \frac{1}{c^2} \frac{\partial^2}{\partial t^2} - \nabla^2 \right)
\phi = 0
\end{equation}
is given by 
\begin{equation} 
\vec{v}_\varepsilon = \frac{\vec{S}}{u} , 
\end{equation} 
where $\vec{S}$ is the flux of momentum and $u$ is the energy density,
given by
\begin{equation}
\vec{S} = \nabla \phi \frac{\partial \phi}{\partial t} \, , \quad u =
\frac{1}{2} \left[ (\nabla\phi)^2 + \frac{1}{c^2} \left( \frac{\partial
\phi}{\partial t} \right)^2 \right] \, ,
\end{equation}
from which it follows that
\begin{equation}
v_\varepsilon = \frac{|\vec{S}|}{u} \leq c_s \, .
\end{equation}
Our acoustic experiment shows that for the $X$-waves the speed of
transport of energy is $c_s/\cos\eta$, since it is the energy of the
wave that activates the detector (hydro-phone). This shows explicitly 
that the definition of $v_\varepsilon$ is meaningless. This fundamental 
experimental result must be kept in mind when we discuss the meaning of
the velocity of transport of electromagnetic waves in section 4.

\begin{figure}[hb] 
\vspace{8cm} 
\caption{Block diagram of acoustic production of Bessel
pulse and $X$-Waves.} 
\end{figure}  
 
\newpage 

\mbox{} 

\vfill 

\begin{figure}[hb] 
\vspace{10cm} 
\caption{Propagation speed of the peak of Bessel pulse and
its comparison with that of a pulse produced by a small
circular element (about 4~mm or 6.67~$\lambda$ in
diameter, where $\lambda$ is 0.6~mm in water). The Bessel
pulse was produced by a 50~mm diameter transducer. The
distances between the peaks and the surface of the
transducer are 104.339~mm and 103.705~mm for the single-element
wave and the Bessel pulse, respectively. The time used by
these pulses is the same. Therefore, the speed of the peak
of the Bessel pulse is 0.611(3)\% slower than that of the
single-element wave.} 
\end{figure} 
  

\mbox{} 

\vfill 

\begin{figure}[hb] 
\vspace{10cm} 
\caption{Propagation speed of peak of $X$-wave and its
comparison with that of a pulse produced by small circular
element (about 4~mm or 6.67~$\lambda$, where $\lambda$ is
0.6~mm in water). The $X$-wave was produced by a 50~mm
diameter transducer. The distance between the peaks and
the surface of the transducer are 173.489~mm and
173.773~mm for the single-element wave and the $X$-wave,
respectively. The time used by these pulses is the same.
Therefore, the speed of the peak of the $X$-wave is
0.2441(8)\% faster than that of the single-element wave.
The theoretical ratio for $X$-waves and the speed of sound
is $\displaystyle \frac{(c_s/\cos \eta - c_s)}{c_s} =
0.2442\%$ for $\eta = 4^{\mbox{o}}$.} 
\end{figure}  


\clearpage 

\n {\large \bf 3. Subluminal and Superluminal UPWs Solutions of Maxwell
Equations(ME)} \\

In this section we make full use of the Clifford bundle formalism (CBF)
resumed in Appendix B, but we give translation of all the main results
in the standard vector formalism used by physicists. We start by
reanalyzing in section 3.1 the plane wave solutions (PWS) of ME with the
CBF.\ We clarify some misconceptions and explain the fundamental role of
the duality operator $\ga_5$ and the meaning of $i=\sqrt{-1}$ in
standard formulations of electromagnetic theory.  Next in section 3.2 we
discuss subluminal UPWs solutions of ME and an unexpected relation
between these solutions and the possible existence of purely
electromagnetic particles (PEPs) envisaged by Einstein$^{[55]}$,
Poincar\'e$^{[56]}$, Ehrenfest$^{[57]}$ and recently discussed by Waite,
Barut and Zeni$^{[2,3]}$.  In section 3.3 we discuss in detail the
theory of superluminal electromagnetic $X$-waves (SEXWs) and how to
produce these waves by appropriate physical devices. \\

\n {\bf 3.1 Plane Wave Solutions of Maxwell Equations} \\

We recall that Maxwell equations in vacuum can be written as [eq.(B.6)]

\renewcommand{\theequation}{3.\arabic{equation}}
\setcounter{equation}{0}
\begin{equation} \label{3e1}
\pa F = 0,
\end{equation}
where $F \sec \bwe^2(M) \sub \sec \Ca\ell(M)$. The well known PWS of
eq.(\ref{3e1}) are obtained as follows. We write in a given Lorentzian
chart $\lan x^\mu\ran$ of the maximal atlas of $M$ (section B2) a PWS
moving in the $z$-direction
\begin{equation} \label{3e2}
F = f e^{\ga_5 kx} \, , 
\end{equation}
\begin{equation}
k = k^\mu \ga_\mu, \,  k^1=k^2=0, \, x = x^\mu\ga_\mu ,
\end{equation}
where $k,\, x \in \sec \bwe^1(M) \sub \sec \Ca\ell(M)$ and where $f$
is a constant 2-form. From eqs.(\ref{3e1}) and (\ref{3e2}) we obtain
\begin{equation}
kF=0
\end{equation}
Multiplying eq.(3.4) by $k$ we get
\begin{equation}
k^2F=0
\end{equation}
and since $k \in \sec \bwe^1(M) \sub \sec \Ca\ell(M)$ then
\begin{equation}
k^2 = 0 \ \leri \ k^0 = \pm |\vec k|= \pm k^3 ,
\end{equation}
{\em i.e.\/}, the propagation vector is light-like. Also
\begin{equation}
F^2 = F.\ F + F \we F = 0
\end{equation}
as can be easily seen by multiplying both members of eq.(3.4) by $F$ and
taking into account that $k \neq 0$. Eq(3.7) says that the field
invariants are null.

It is interesting to understand the fundamental role of the volume
element $\ga_5$ (duality operator) in electromagnetic theory. In
particular since $e^{\ga_5 kx} = \cos kx + \ga_5 \sin kx$, $\gamma_5
\equiv \ia$, writing $F= \vec{E} + \ia \vec{B}$ (see eq.(B.17)), 
$f = \vec{e}_1 + \ia \vec{e}_2$,  we see that
\begin{equation} 
\vec{E} + \ia \vec{B} = \vec{e}_1 \cos kx - \vec{e}_2 \sin kx + \ia ( \vec{e}_1 \sin kx + \vec{e}_2 \cos kx) \, . 
\end{equation}  
From this equation, using $\partial F =0$, it follows that $\vec{e}_1 .
\vec{e}_2 =0$, $\vec{k} . \vec{e}_1 = \vec{k} . \vec{e}_2 = 0$ and then
\begin{equation} 
\vec{E} . \vec{B} =0 \, . 
\end{equation}  
This equation is important because it shows that we must take care with
the $i=\sqrt{-1}$ that appears in usual formulations of Maxwell Theory
using complex electric and magnetic fields. The $i=\sqrt{-1}$ in many
cases unfolds a secret that can only be known through eq.(3.8).  It also
follows that $\vec k . \vec E = \vec k .  \vec B = 0$, {\em i.e.\/}, PWS
of ME are {\it transverse} waves.  We can rewrite eq.(3.4) as
\begin{equation}
k\ga_0 \ga_0 F \ga_0 = 0
\end{equation}
and since $k \ga_0 = k_0 + \vec k, \ \ga_0 F \ga_0 = - \vec E + \ia
\vec B$ we have
\begin{equation}
\vec k f = k_0 f .
\end{equation}

Now, we recall that in $\Ca\ell^+(M)$ (where, as we say in Appendix B,
the typical fiber is isomorphic to the Pauli algebra $\Ca\ell_{3,0}$) we
can introduce the operator of space conjugation denoted by $*$ such that
writing $f = \vec e + \ia \vec b$ we have
\begin{equation}
f^* = - \vec{e} + \ia \vec{b} \ \ ; \ \ k^*_0 = k_0 \ \ ; \ \ \vec k^* = - \vec k .
\end{equation}
We can now interpret the two solutions of $k^2 = 0$, {\em i.e.\/}
$k_0=|\vec k|$ and $k_0 = -|\vec k|$ as corresponding to the solutions
$k_0 f = \vec kf$ and $k_0 f^* = - \vec kf^*$; $f$ and $f^*$ correspond
in quantum theory to ``photons" of positive or negative helicities. We
can interpret $k_0 = |\vec k|$ as a particle and $k_0 = -|\vec k|$ as an
antiparticle.

Summarizing we have the following  important facts concerning PWS of ME:
(i) the propagation vector is light-like, $k^2=0$; (ii) the field
invariants are null, $F^2=0$; (iii) the PWS are transverse waves, {\em
i.e.\/}, $\vec k . \vec E = \vec k . \vec B = 0$. \\

\n {\bf 3.2 Subluminal Solutions of Maxwell Equations and Purely
Electromagnetic Particles.} \\

We take $\Phi \in \sec (\bwe^0(M) \oplus \bwe^4(M)) \sub \sec
\Ca\ell(M)$ and consider the following Hertz potential $\pi \in \sec
\bwe^2(M) \sub \sec \Ca\ell(M)$ [eq.(\ref{be23})]
\begin{equation}
\pi = \Phi \ga^1 \ga^2 .
\end{equation}
We now write
\begin{equation}
\Phi(t, \vec x) = \phi(\vec x) e^{\ga_5 \Om t} .
\end{equation}
Since $\pi$ satisfies the wave equation, we have
\begin{equation}
\nab^2 \phi(\vec x) + \Om^2 \phi(\vec x) = 0\, . 
\end{equation}

Solutions of eq.(3.15) (the Helmholtz equation) are well known. Here we
consider the simplest solution in spherical coordinates,

\begin{equation}
\phi(\vec x) = C \frac{\sin\Om r}{r} \ \ , \ \ r =
\sqrt{x^2+y^2+z^2},
\end{equation}
where $C$ is an arbitrary real constant. From the results of Appendix B
we obtain the following stationary electromagnetic field, which is at
rest in the reference frame $Z$ where $\lan x^\mu\ran$ are naturally
adapted coordinates (section B2):

\begin{eqnarray}
F_0 & = & \frac{C}{r^3} [\sin \Om t(\al\Om r\sin\te \sin\varphi - \be
\sin \te \cos\te \cos\varphi) \ga_0 \ga_1 \nb  \\ 
& - & \sin \Om t (\al\Om r\sin\te \cos\varphi + \be \sin \te \cos\te
\sin\varphi) \ga_0 \ga_2 \nb  \\ 
& + &  \sin \Om t(\be\sin^2 \te - 2\al)\ga_0 \ga_3 + \cos\Om t
(\be\sin^2 \te -  2\al)  \ga_1 \ga_2 \\
& + & \cos\Om t(\be\sin\te \cos\te \sin\varphi + \al\Om r \sin \te
\cos\varphi)\ga_1 \ga_3 \nb \\ 
& + & \cos\Om t(-\be\sin\te \cos\te
\cos\varphi + \al\Om r \sin \te \sin\varphi)\ga_2 \ga_3 ] \nb
\end{eqnarray}
with $\al=\Om r \cos\Om r - \sin\Om r$ and $\be = 3\al + \Om^2 r^2
\sin \Om r$. Observe that $F_0$ is regular at the origin and vanishes
at infinity. Let us rewrite the solution using the Pauli-algebra in
$\Ca\ell^+(M)$. Writing $(\ia \equiv \ga_5)$
\begin{equation}
F_0 = \vec E_0 + \ia \vec B_0
\end{equation}
we get

\begin{equation}
\vec E_0 =  \vec W \sin \Om t , \ \ \vec B_0=\vec W \cos\Om t , 
\end{equation}
with

\begin{equation}
\vec W = - C \left( \frac{\al\Om y}{r^3} - \frac{\be x z}{r^5} , -
\frac{\al\Om x}{r^3} - \frac{\be y z}{r^5}, \frac{\be(x^2+y^2)}{r^5}
- \frac{2\al}{r^3}\right) .
\end{equation}
We verify that $\divi \vec W = 0$, $\divi \vec E_0 = \divi \vec B_0 =
0$, $\rot \vec E_0 + \pa \vec B_0/\pa t = 0$, $\rot \vec B_0 - \pa \vec
E_0/\pa t = 0$, and
\begin{equation}
\rot \vec W =  \Om \vec W.
\end{equation}

Now, from eq.(B.88) we know that $T_0 = {\dis\frac{1}2} \wid F \ga_0 F$
is the 1-form representing the energy density and the Poynting vector.
It follows that $\vec E_0 \times \vec B_0 = 0$, {\em i.e.\/}, the
solution has zero angular momentum. The energy density $u = S^{00}$ is
given by
\begin{equation}
u = \frac{1}{r^6} [\sin^2 \te (\Om^2 r^2 \al^2 + \be^2 \cos^2 \te ) +
(\be\sin^2 \te - 2\al)^2]
\end{equation}
Then $\int\!\int\!\int_{\R^3} u\,d\ven = \infi$.  As explained in
section A.6 a finite energy solution can be constructed by considering
``wave packets" with a distribution of intrinsic frequencies $F(\Om)$
satisfying appropriate conditions. Many possibilities exist, but they
will not be discussed here. Instead, we prefer to direct our attention
to eq.(3.21). As it is well known, this is a very important equation
(called the force free equation$^{[2]}$) that appears {\em e.g.\/} in
hydrodynamics and in several different situations in plasma
physics$^{[58]}$. The following considerations are more important.

Einstein$^{[55]}$ among others (see$^{[3]}$ for a review) studied the
possibility of constructing purely electromagnetic particles (PEPs). He
started from Maxwell equations for a PEP configuration described by an
electromagnetic field $F_p$ and a current density $J_p$, where
\begin{equation}
\pa F_p = J_p
\end{equation}
and rightly concluded that the condition for existence of PEPs is
\begin{equation}
J_p . F_p = 0.
\end{equation}
This condition implies in vector notation
\begin{equation}
\rho_p \vec E_p = 0 , \ \ \vec j_p . \vec E_p = 0  , \ \ \vec
j_p \times \vec B_p = 0 . 
\end{equation}
From eq.(3.24) Einstein concluded that the only possible solution of
eq.(3.22) with the subsidiary condition given by eq.(3.23) is $J_p = 0$.
However, this conclusion is correct, as pointed in$^{[2,3]}$, only if
$J^2_p > 0$, {\em i.e.\/}, if $J_p$ is a time-like current density.
However, if we suppose that $J_p$ can be spacelike, {\em i.e.\/}, $J^2_p
< 0$, there exists a reference frame where $\rho_p = 0$ and a possible
solution of eq.(3.24) is
\begin{equation}
\rho_p = 0  , \ \ \vec E_p . \vec B_p = 0  , \ \ \vec j_p =
KC\vec B_p ,
\end{equation}
where $K = \pm 1$ is called the chirality of the solution and $C$ is a
real constant. In$^{[2,3]}$ static solutions of eqs.(3.22) and (3.23)
are exhibited where $\vec E_p = 0$. In this case we can verify that
$\vec B_p$ satisfies 
\begin{equation}
\nab \times \vec B_p = KC\vec B_p .
\end{equation}
Now, if we choose $F_0 \in \sec \bwe^2 (M) \sub \sec \Ca\ell(M)$ such
that
\begin{equation}
\begin{array}{c}
F_0 = \vec E_0 + \ia \vec B_0 , \\
\vec E_0 = \vec B_p \cos \Om t , \ \ \vec B_0 = \vec B_p \sin
\Om t
\end{array}
\end{equation}
and $\Omega = KC >0$, we immediately realize that
\begin{equation}
\pa F_0 = 0 . 
\end{equation}
This is an amazing result, since it means that the free Maxwell
equations may have stationary solutions that may be used to model PEPs.
In such solutions the structure of the field $F_0$ is such that we can
write
\begin{equation}
\begin{array}{c}
F_0 = F_{p}^{'} + \ov F = \ia \vec{W} \cos \Omega t - \vec{W} \sin
\Omega t ,  \\
\pa F_{p}^{'} = -\pa \ov F = J_{p}^{'} , 
\end{array}
\end{equation}
{\em i.e.\/}, $\pa F_0=0$ is equivalent to a field plus a current. This
fact opens several interesting possibilities for modeling PEPs (see
also$^{[4]}$) and we discuss more this issue in another publication.

We observe that moving subluminal solutions of ME can be easily obtained
choosing as Hertz potential, {\em e.g.\/},

\begin{eqnarray}
&& \h \pi^<(t,\vec x) = C \frac{\sin\Om\xi_<}{\xi_<} \exp [\ga_5(\om_<
t-k_< z)]\ga_1 \ga_2 , \\
&& \h\h \om^2_< - k^2_< = \Om^{2}_{<} ; \nb \\
&& \xi_< = [x^2+y^2+ \ga^2_< (z-v_< t)^2] , \\
&& \h \ga_< = \frac{1}{\sqrt{1-v^2_<}} , \ \ v_< = d\om_</ dk_< . \nb
\end{eqnarray}
We are not going to write explicitly the expression for $F^<$
corresponding to $\pi^<$ because it is very long and will not be used in
what follows.

We end this section with the following observations: (i) In general for
subluminal solutions of ME (SSME) the propagation vector satisfies an
equation like eq.(3.30).  (ii) As can be easily verified, for a SSME the
field invariants are non-null. (iii) A SSME is not a transverse wave.
This can be seen explicitly from eq.(3.21).  Conditions (i), (ii) and
(iii) are in contrast with the case of the PWS of ME.\ In$^{[49,50]}$
Rodrigues and Vaz showed that for free electromagnetic fields $(\pa
F=0)$ such that $F^2 \neq 0$, there exists a Dirac-Hestenes equation
(see section A.8) for $\psi \in \sec (\bwe^0(M) + \bwe^2(M) + \bwe^4(M))
\sub \sec \Ca\ell(M)$ where $F = \psi \ga_1 \ga_2 \wid \psi$. This was
 the reason why Rodrigues and Vaz discovered subluminal and superluminal
solutions of Maxwell equations (and also of Weyl equation)$^{[48]}$
which solve the Dirac-Hestenes equation [eq.(\ref{be33})]. \\

\n {\bf 3.3 The Superluminal Electromagnetic $X$-Wave (SEXW)} \\

To simplify the matter in what follows we now suppose that the functions
$\Phi_{X_n}$ [eq.(A.52)] and $\Phi_{XBB_n}$ [eq.(A.53)] which are
superluminal solutions of the scalar wave equation are 0-forms sections
of the complexified Clifford bundle $\Ca\ell_C(M) = \C \ot \Ca\ell(M)$
(see section B4). We rewrite eqs.(A.52) and (A.53)
as$^{(*)}$\footnotetext{$^{(*)}$In what follows $n=0,1,2,\ld$}

\begin{equation}
\Phi_{X_n} (t, \vec x) = e^{in\te} \int^\infi_0 B(\ov k) J_n (\ov k
\rho\sin\eta) e^{-\ov k[a_0 - i(z \cos\eta-t)]} d\ov k
\end{equation}
and choosing $B(\ov k) = a_0$, we have
\begin{eqnarray}
&& \h \Phi_{XBB_n} (t,\vec x) = \frac{a_0(\rho\sin \eta)^n
e^{in\te}}{\sqrt{M}(\tau+\sqrt{M})^n} \\
&& M = (\rho\sin\eta)^2 + \tau^2 ; \ \ \  \tau = [a_0 -
i(z\cos\eta-t)].
\end{eqnarray}
As in section 2, when a finite broadband $X$-wave is obtained from
eq.(3.31) with $B(\ov k)$ given by the Blackman spectral function
[eq.(2.8)] we denote the resulting $X$-wave by $\Phi_{XBL_n}$ ($BL$
means band limited wave). The finite aperture approximation (FAA)
obtained with eq.(A.28) to $\Phi_{XBL_n}$ will be denoted
FAA$\Phi_{XBL_n}$ and the FAA to $\Phi_{XBB_n}$ will be denoted by
FAA$\Phi_{XBB_n}$. We use the same nomenclature for the electromagnetic
fields derived from these functions. Further, we suppose now that the
Hertz potential $\pi$, the vector potential A and the corresponding
electromagnetic field $F$ are appropriate sections of $\Ca\ell_C(M)$. We
take
\begin{equation}
\pi=\Phi \ga_1 \ga_2 \in \sec \C \ot \bwe^2(M) \sub \sec
\Ca\ell_C(M),
\end{equation}
where $\Phi$ can be $\Phi_{X_n}, \Phi_{XBB_n}, \Phi_{XBL_n}$, FAA
$\Phi_{XBB_n}$ or FAA$\Phi_{XBL_n}$. Let us start by giving the explicit
form of the $F_{XBB_n}$, {\em i.e.\/}, the SEXWs. In this case eq.(B.81)
gives $\pi=\vec \pi_m$ and
\begin{equation}
\vec \pi_m = \Phi_{XBB_n} \zo
\end{equation}
where ${\bf z}$ is the versor of the $z$-axis. Also, let $\ro$, $\teo$
be respectively the versors of the $\rho$ and $\te$ directions where
$(\rho, \te, z)$ are the usual cylindrical coordinates.  Writing
\begin{equation}
F_{XBB_n} = \vec E_{XBB_n} + \ga_5 \vec B_{XBB_n}
\end{equation}
we obtain from equations (A.53) and (B.25):
{\small
\begin{equation}
\vec E_{XBB_n} = -\frac{\ro}{\rho} \frac{\pa^2}{\pa t\pa\te}
\Phi_{XBB_n} + \teo \frac{\pa^2}{\pa t\pa \rho} \Phi_{XBB_n}\, ; 
\end{equation} 
\begin{equation} 
\vec B_{XBB_n} = \ro \frac{\pa^2}{\pa \rho\pa z}
\Phi_{XBB_n} + \teo \frac{1}{\rho} \frac{\pa^2}{\pa \theta \pa z}
\Phi_{XBB_n} +
\zo \left(\frac{\pa^2}{\pa z^2} \Phi_{XBB_n} - \frac{\pa^2}{\pa
t^2} \Phi_{XBB_n} \right) ; 
\end{equation} 
}

\noindent Explicitly we get for the components in cylindrical
coordinates:\\

\h $(\vec E_{XBB_n})_\rho = -{\dis\frac{1}{\rho} n\frac{M_3}{\sqrt{M}}}
\Phi_{XBB_n}$; \hfill (3.41a)

\h $(\vec E_{XBB_n})_\te = {\dis\frac{1}{\rho} i\frac{M_6}{\sqrt{M}
M_2}}
\Phi_{XBB_n}$; \hfill (3.41b)

\h $(\vec B_{XBB_n})_\rho = \cos\eta (\vec E_{XBB_n})_\te$; \hfill
(3.41c)

\h $(\vec B_{XBB_n})_\te = -\cos\eta (\vec E_{XBB_n})_\rho$; \hfill
(3.41d)

\h $(\vec B_{XBB_n})_z = -\sin^2\eta {\dis\frac{M_7}{\sqrt{M}}
 \Phi_{XBB_n}}$. \hfill
(3.41e) \\ 

\noindent The functions $M_i$, $(i=2,\ld,7)$ in (3.41) are: \\ 

\h $M_2 = \tau + \sqrt{M}$; \hfill (3.42a)

\h $M_3 = n + {\dis\frac{1}{\sqrt{M}}} \tau$; \hfill (3.42b)

\h $M_4 = 2n + {\dis\frac{3}{\sqrt{M}}}\tau$; \hfill (3.42c)

\h $M_5 = \tau + n\sqrt{M}$;  \hfill (3.42d)

\h $M_6 = (\rho^2 \sin^2\eta {\dis\frac{M_4}{M}} - n M_3)M_2 + n \rho^2
 {\dis\frac{M_5}{M}} \sin^2\eta$;  \hfill (3.42e)

\h $M_7 = (n^2-1) {\dis\frac{1}{\sqrt{M}} + 3 n \frac{1}{M}} \tau +
3  {\dis\frac{1}{\sqrt{M^3}}}\tau^2$.  \hfill (3.42f)

We immediately see from eqs.(3.41) that the $F_{XBB_n}$ are indeed
superluminal UPWs solutions of ME, propagating with speed $1/\cos\eta$
in the $z$-direction. That $F_{XBB_n}$ are UPWs is trivial and that they
propagate with speed $c_1=1/\cos\eta$ follows because $F_{XBB_n}$
depends only on the combination of variables $(z-c_1t)$ and any
derivatives of $\Phi_{XBB_n}$ will keep the $(z-c_1t)$ dependence
structure.

Now, the Poynting vector $\vec P_{XBB_n}$ and the energy density
$u_{XBB_n}$ for $F_{XBB_n}$ are obtained by considering the real
parts of $\vec E_{XBB_n}$ and $\vec B_{XBB_n}$. We have \\

\h $(\vec P_{XBB_n})_\rho = - Re\{(\vec E_{XBB_n})_\te\} Re\{(\vec
B_{XBB_n})_z\}$; \hfill (3.43a)

\h $(\vec P_{XBB_n})_\te =  Re\{(\vec E_{XBB_n})_\rho\} Re\{(\vec
B_{XBB_n})_z\}$; \hfill (3.43b)

\h $(\vec P_{XBB_n})_z = \cos\eta \left[|Re\{(\vec E_{XBB_n})_\rho\}|^2 + |
Re\{(\vec E_{XBB_n})_\te\}|^2 \right]$; \hfill (3.43c)

\setcounter{equation}{43}

\begin{equation}
u_{XBB_n} = (1+\cos^2 \eta) \left[|Re\{(\vec E_{XBB_n})_\rho\}|^2 + |Re
\{(\vec E_{XBB_n})_\te\}|^2\right] + |Re\{(\vec B_{XBB_n})_z\}|^2 . 
\end{equation}

The total energy of $F_{XBB_n}$ is then
\begin{equation}
\vare_{XBB_n} = \int^\pi_{-\pi} d\te \int^{+\infi}_{-\infi} dz
\int^\infi_0 \rho \, d\rho \, u_{XBB_n}
\end{equation}
Since as $z \rig \infi$, $\vec E_{XBB_n}$ decreases as
$1/|z-t\cos\eta|^{1/2}$, what occurs for the $X$-branches of
$F_{XBB_n}$, $\vare_{XBB_n}$ may not be finite. Nevertheless, as in the
case of the acoustic $X$-waves discussed in section 2, we are quite sure
that a FAA$F_{XBL_n}$ can be launched over a large distance. Obviously
in this case the total energy of the FAA$F_{XBL_n}$ is finite.

We now restrict our attention to $F_{XBB_0}$. In this case from
eq.(3.40)  and eqs.(3.43) we see that $(\vec E_{XBB_0})_\rho = (\vec
B_{XBB_0})_\te = (\vec P_{XBB_0})_\te = 0$. In Fig.\
4$^{(*)}$\footnotetext{${(*)}$Figures 4, 5 and 6 were reprinted with
permission from$^{[5]}$.} we see the amplitudes of $Re \{\Phi_{XBB_0}\}$
[4(1)],  $Re\{(\vec E_{XBB_0})_\te \}$ [4(2)], $Re\{(\vec
B_{XBB_0})_\rho\}$ [4(3)] and $Re \{(\vec B_{XBB_0})_z\}$ [4(4)]. Fig.\
5 shows respectively  $(\vec P_{XBB_0})_\rho$ [5(1)], $(\vec
P_{XBB_0})_z$ [5(2)]  and $u_{XBB_0}$ [5(3)].  The size of each panel in
Figures 4 and 5 is 4m~($\rho$-direction)~$\times$ 2mm~($z$-direction)
and the maxima and minima of the images in Figures 4 and 5 (before
scaling) are shown in Table 1, in MKSA
units$^{(**)}$\footnotetext{$^{(**)}$Reprinted with permission from
Table I of$^{[5]}$.}.

\begin{center}
\begin{tabular}{|c|c|c|c|c|}
\hline
 & $Re\{\Phi_{XBB_0}\}$ & $Re\{(\vec{E}_{XBB_0})_\theta\}$
 & $Re\{(\vec{B}_{XBB_0})_\rho\}$ & $Re\{(\vec{B}_{XBB_0})_z\}$ \\
\hline
\ max \ & 1.0 & $9.5\times 10^6$ & $2.5\times 10^4$ & 6.1 \\
\hline
\ min \ & 0.0 &$-9.5\times 10^6$ &$-2.5\times 10^4$ &$-1.5$ \\
\hline
\end{tabular}

\vs{5mm}

\begin{tabular}{|c|c|c|c|}
\hline
     & $(\vec{P}_{XBB_0})_\rho$  &
$(\vec{P}_{XBB_0})_z$  & $U_{XBB_0}$ \\
\hline
\ max \ & $2.4\times 10^7$ & $2.4\times 10^{11}$ & $1.6\times 10^3$ \\
\hline
\ min \ &$-2.4\times 10^7$ & 0.0 & 0.0 \\
\hline
\end{tabular}

\vs{5mm}

{\small\bf Table 1: Maxima and Minima of the zeroth-order
nondiffracting electromagnetic $X$ waves (units: MKSA).}
\end{center}

\noindent Fig.\ 6 shows the beam plots of $F_{XBB_0}$ in Fig.\ 4 along
one of the $X$-branches (from left to right). Fig.\ 6(1) represents the
beam plots of $Re\{\Phi_{XBB_0}\}$ (full line), $Re\{(\vec
E_{XBB_0})_\te\}$ (dotted line), $Re\{(\vec B_{XBB_0})_\rho\}$ (dashed
line) and $Re\{(\vec B_{XBB_0})_z\}$ (long dashed line). Fig.\ 6(2)
represents the beam plots of $(\vec P_{XBB_0})_\rho$ (full line), $(\vec
P_{XBB_0})_z$ (dotted line) and $u_{XBB_0}$ (dashed line). \\

\n {\bf 3.4 Finite Aperture Approximation to $F_{XBB_0}$ and
$F_{XBL_0}$} \\

From eqs.(3.40), (3.43) and (3.44) we see that $\vec E_{XBB_0}$, $\vec
B_{XBB_0}$, $\vec P_{XBB_0}$ and $u_{XBB_0}$ are related to the scalar
field $\Phi_{XBB_0}$. It follows that the depth of the field$^{[5]}$ (or
non diffracting distance --- see section 2) of the FAA$F_{XBB_0}$ and of
the FAA$F_{XBL_0}$, which of course are to be produced by a finite
aperture radiator, are equal and given by
\begin{equation}
Z_{\max} = D/2 \cot \eta ,
\end{equation}
where $D$ is the diameter of the radiator and $\eta$ is the axicon
angle.  It can be proved also$^{[5]}$ that for $\Phi_{XBL_0}$ (and more
generally for $\Phi_{XBL_n}$), that $Z_{\max}$ is independent of the
central frequency of the  spectrum $B(\ov k)$ in eq.(3.1). Then if we
want, {\em e.g.\/}, that $F_{XBB_0}$ or $F_{XBL_0}$ travel 115~km with a
20~m diameter radiator, we need $\eta = 0.005^{\mbox{o}}$.

Figure 7 shows the envelope of $Re\{{\rm FAA}\Phi_{XBB_0}\}$ obtained
with the finite aperture approximation (FAA) given by eq.(A.28), with $D
= 20\ \mbox{m}$, $a_0 = 0.05\ \mbox{mm}$ and $\eta = 0.005^{\mbox{o}}$,
for distances $z = 10\ \mbox{km}$ [6(1)] and $z=100\ \mbox{km}$ [6(2)],
respectively, from the radiator which is located at the plane $z=0$.
Figures 7(3) and 7(4) show the envelope of $Re\{{\rm FAA}\Phi_{XBL_0}\}$
for the same distances and the same parameters $(D,a_0$ and $\eta$)
where $B(\ov k)$ is the following Blackman window function, peaked at
the frequency $f_0 = 700\ \mbox{GHz}$ with a $6\ \mbox{dB}$ bandwidth
about $576~\mbox{GHz}$:

\begin{equation}
B(\ov k) = \left\{\begin{array}{l}
a_0 [0.42-0.5 \cos\frac{\pi\ov k}{\ov k_0} + 0.08 \cos \frac{2\pi \ov
k}{\ov k_0}], \ 0 \leq \ov k \leq 2 \ov k_0 ;\\
0 \ \ {\rm otherwise}; 
\end{array}\right.
\end{equation}
where $\ov k_0 = 2\pi f_0/c \ \ (c=300,000km/s)$. From eq.(3.46) it
follows that for the above choice of $D$, $a_0$ and $\eta$
\begin{equation}
Z_{\max} = 115\, km
\end{equation}
Figs. 8(1) and 8(2) show the lateral beam plots and Figs. 8(3) and 8(4)
show the axial beam plots respectively for $Re\{{\rm FAA}\Phi_{XBB_0}\}$
and for $Re\{{\rm FAA}\Phi_{XBL_0}\}$ used to calculate $F_{XBB_0}$ and
$F_{XBL_0}$. The full and dotted lines represent $X$-waves at distances
$z = 10~\mbox{km}$ and $z=100~\mbox{km}$. Fig. 9 shows the peak values
of $Re\{{\rm FAA}\Phi_{XBB_0}\}$ (full line) and $Re\{{\rm
FAA}\Phi_{XBL_0}\}$ (dotted line) along the $z$-axis from
$z=3.45~\mbox{km}$ to $z=230~\mbox{km}$.  The dashed line represents the
result of the exact $\Phi_{XBB_0}$ solution. The $6~\mbox{dB}$ lateral
and axial beam widths of $\Phi_{XBB_0}$, which can be measured in Fig
7(1) and 7(2), are about 1.96~m and 0.17~mm respectively, and those of
the FAA$\Phi_{XBL_0}$ are about 2.5~m and 0.48~mm as can be measured
from 7(3) and 7(4).  For $\Phi_{XBB_0}$ we can calculate$^{[43,26]}$ the
theoretical values of the $6~\mbox{dB}$ lateral ($BW_L$) and axial
($BW_A$) beam widths, which are given by
\begin{equation}
BW_L = \frac{2 \sqrt{3} a_0}{|\sin\eta|} ; \ \ \ BW_A =
\frac{2\sqrt{3}a_0}{|\cos\eta|} \; .
\end{equation}
With the values of $D$, $a_0$ and $\eta$ given above, we have $BW_L =
1.98\ \mbox{m}$ and $BW_A = 0.17\ \mbox{mm}$. These are to be compared
with the values of these quantities for the FAA$\Phi_{XBL_0}$.

We remark also that eq.(3.46) says that $Z_{\max}$ does not depend on
$a_0$. Then we can choose an arbitrarily small $a_0$ to increase the
localization (reduced $BW_L$ and $BW_A$) of the $X$-wave without
altering $Z_{\max}$. Smaller $a_0$ requires that the FAA$\Phi_{XBL_0}$
be transmitted with broader bandwidth. The depths of field of
$\Phi_{XBB_0}$ and of $\Phi_{XBL_0}$ that we can measure in Fig. 9 are
approximately 109~km and 110~km, very close to the value given by
eq.(3.46) which is 115~km.

We conclude this section with the following observations.

\begin{itemize}
\item[(i)] In general both subluminal and superluminal UPWs solutions of
ME have non null field invariants and are not transverse waves. In
particular our solutions have a longitudinal component along the
$z$-axis. This result is important because it shows that, contrary to
the speculations of Evans$^{[59]}$, we do not need an electromagnetic
theory with a non zero  photon-mass, {\em i.e.\/}, with $F$ satisfying
Proca  equation in order to have an electromagnetic wave with a
longitudinal component.  Since Evans presents evidence$^{[59]}$ of the
existence on longitudinal magnetic fields in many different physical
situations, we conclude that the theoretical and experimental study of
subluminal and superluminal UPW solutions of ME must be continued.

\item[(ii)] We recall that in microwave and optics, as it is well known,
the electromagnetic intensity is approximately represented by the
magnitude of a scalar field solution of the HWE.\ We already quoted in
the introduction that Durnin$^{[16]}$ produced an optical $J_0$-beam,
which as seen from eq.(3.1) is related to $\Phi_{XBB_0}$ \linebreak
$(\Phi_{XBL_0})$. If we take into account this fact together with the
results of the acoustic experiments described in section 2, we arrive at
the conclusion that subluminal electromagnetic pulses $J_0$ and also
superluminal $X$-waves can be launched with appropriate antennas using
present technology.

\item[(iii)] If we take a look at the structure of {\em e.g.\/} the
FAA$\Phi_{XBB_0}$ [eq.(3.40)] plus eq.(A.28) we see that it is a
``packet" of wavelets, each one traveling with speed $c$.  Nevertheless,
the electromagnetic $X$-wave wave that is an interference pattern is
such that its peak travels with speed $c/\cos \eta > 1$. (This indeed
happens in the acoustic experiment with $c \mapsto c_s$, see section 2).
Since as discussed above we can project an experiment to launch the peak
of the FAA$\Phi_{XBB_0}$ from a point $z_1$ to a point $z_2$, the
question arises: Is the existence of superluminal electromagnetic waves
in conflict with Einstein's Special Relativity? We give our answer to
this fundamental issue in section 5, but first we discuss in section 4
the speed of propagation of the energy associated with a superluminal
electromagnetic wave.  \end{itemize}

\newpage 

\begin{figure}[hb] 
\vspace{16cm} 
\caption{ Real part of field components of the exact solution
superluminal electromagnetic $X$-wave at distance $z=ct/ \cos
\eta$ ($\eta = 0.005^{\mbox{o}}$, $a_0 = 0.05\ \mbox{mm}$, $n=0$).}
\end{figure} 

\newpage 

\begin{figure}[hb] 
\vspace{16cm} 
\caption{Poynting flux and energy density of the exact
solution superluminal electromagnetic $X$-wave at distance
$z = ct/\cos\eta$, ($\eta = 0.005^{\mbox{o}}$, $a_0 =
0.05\ \mbox{mm}$, $n=0$).} 
\end{figure} 

\newpage 

\begin{figure}[hb] 
\vspace{16cm} 
\caption{(6.1)~Beam plots along the $X$-branches of
$F_{XBB_0}$ for Re$\{\Phi_{XBB_0}\}$ or Hertz potential,
Re$\{(\vec{E}_{XBB_0})_\theta \}$,
Re$\{(\vec{B}_{XBB_0})_\rho \}$, and Re$\{ ( \vec{B}_{XBB_0}
)_z \}$. (6.2)~Beam plots for $( \vec{P}_{XBB_0})_\rho$
(full line), $(\vec{P}_{XBB_0})_z$ (dotted line) and
$u_{XBB_0}$ (dashed line).} 
\end{figure} 

\newpage 

\begin{figure}[hb] 
\vspace{16cm} 
\caption{7(1) and 7(2) show the real part of
FAA$\Phi_{XBB_0}$ at distances $z=10~\mbox{km}$ and
$z=100~\mbox{km}$ from the radiator located at the plane
$z=0$ with $D=20~\mbox{m}$ and $\eta = 0.005^{\mbox{o}}$.
7(3) and 7(4) show the real parts of FAA$\Phi_{XBL_0}$ for
the same distances.} 
\end{figure} 

\newpage 

\begin{figure}[hb] 
\vspace{16cm} 
\caption{Beam plots of scalar $X$-waves (finite aperture).} 
\end{figure} 

\newpage 

\begin{figure}[hb] 
\vspace{16cm} 
\caption{Peak magnitude of $X$-waves along the $z$ axis.} 
\end{figure} 

\clearpage

\n {\large\bf 4. The Velocity of Transport of Energy of the UPWs
Solutions of Maxwell Equations }\\

Motivated by the fact that the acoustic experiment of section 2 shows
that the energy of the FAA $X$-wave travels with speed greater than
$c_s$ and since we found in this paper UPWs solutions of Maxwell
equations with speeds $0 \leq v < \infi$, the following question arises
naturally: Which is the velocity of transport of the energy of a
superluminal UPW (or quasi UPW) solution of ME?

We can find in many physics textbooks ({\em e.g.\/}$^{[10]}$) and in
scientific papers$^{[41]}$ the following argument. Consider  an
arbitrary solution of ME in vacuum, $\pa F = 0$. Then if $F = \vec E +
\ia \vec B$ (see eq.(B.17)) it follows that the Poynting vector and the
energy density of the field are
\renewcommand{\theequation}{4.\arabic{equation}}
\setcounter{equation}{0}
\begin{equation}
\vec P = \vec E \times \vec B , \ \ u=\frac{1}2 (\vec E^2 + \vec
B^2) . 
\end{equation}
It is obvious that the following inequality always holds:
\begin{equation}
v_\vare = \frac{|\vec P|}{u} \leq 1 . 
\end{equation}
Now,  the conservation of energy-momentum reads, in integral form over a
finite volume $V$ with boundary $S = \pa V$
\begin{equation}
\frac{\pa}{\pa t} \left\{ \int\!\!\int\!\!\int_V d{\mbox{\bf v}}
\frac{1}2 (\vec E^2 + \vec
B^2)\right\} = \oint_S d\vec S . \vec P
\end{equation}
Eq.(4.3) is interpreted saying that $\oint_S d\vec S . \vec P$ is the
 field energy flux across the surface $S = \pa V$, so that $\vec P$ is
the flux density --- the amount of field energy passing through a unit
area of the surface in unit time. For plane wave solutions of Maxwell
equations,
\begin{equation}
v_\vare = 1
\end{equation}
and this result gives origin to the ``dogma'' that free electromagnetic
fields transport energy at speed $v_\vare=c=1$.

However $v_\vare \leq 1$ is true even for subluminal and superluminal
solutions of ME, as the ones discussed in section 3.  The same is true
for the superluminal modified Bessel beam found by Band$^{[41]}$ in
1987. There he claims that since $v_\vare \leq 1$ there is no conflict
between superluminal solutions of ME and Relativity Theory since what
Relativity forbids is the propagation of energy with speed greater than
$c$.

Here we challenge this conclusion. The fact is that as is well known
$\vec P$ is not uniquely defined. Eq(4.3) continues to hold true if we
substitute $\vec P \mapsto \vec P + \vec P'$ with $\nab . \vec P' = 0$.
But of course we can easily find for subluminal, luminal or superluminal
solutions of Maxwell equations a $\vec P'$ such that
\begin{equation}
\frac{|\vec P + \vec P'|}{u} \geq 1 .
\end{equation}
We come to the conclusion that the question of the transport of energy
in superluminal UPWs solutions of ME is an experimental question. For
the acoustic superluminal $X$-solution of the HWE (see section 2) the
energy around the peak area flows together with the wave, {\em i.e.\/},
with speed $c_1 = c_s/\cos \eta$ (although the ``canonical'' formula
[eq.(2.10)] predicts that the energy flows with $v_\varepsilon < c_s$).
Since we can see no possibility for the field energy of the superluminal
electromagnetic wave to travel outside the wave we are confident to
state that the velocity of energy transport of superluminal
electromagnetic waves is superluminal.

Before ending we give another example to illustrate that eq.(4.2) (as is
the case of eq.(2.10)) is devoid of physical meaning. Consider a
spherical conductor in electrostatic equilibrium with uniform
superficial charge density (total charge $Q$) and with a dipole magnetic
moment. Then, we have

\begin{equation}
\vec E = Q\frac{\mbox{\bf r}}{r^2}  ; \ \ \vec B = \frac{C}{r^3}
(2\cos\te \, {\bf r} + \sin \te \, \teo)
\end{equation}
and
\begin{equation}
\vec P = \vec E \times \vec B = \frac{CQ}{r^5} \sin\te \, \varo \, ,
\ \
\ \ u = \frac{1}{2} \left( \frac{Q^2}{r^4} + \frac{C^2}{r^6}  (3\cos^2
\te + 1) \right).
\end{equation}
Thus
\begin{equation}
\frac{|\vec P|}{u} = \frac{2CQ r \sin\te }{r^2 Q^2+C^2(3\cos^2 \te+1)}
\neq 0 \  \mbox{for $r\neq 0$.}
\end{equation}
Since the fields are static the conservation law eq.(4.3) continues to
hold true, as there is no motion of charges and for any closed surface
containing the spherical conductor we have
\begin{equation}
\oint_S d\vec S . \vec P = 0.
\end{equation}
But {\it nothing\/} is in motion! In view of these results we must
investigate whether the existence of superluminal UPWs solutions of ME
is compatible or not with the Principle of Relativity. We analyze this
question in  detail in the next section.

To end this section we recall that in section~2.19 of his book
Stratton$^{[19]}$ presents a discussion of the Poynting vector and
energy transfer which essentially agrees with the view presented above.
Indeed he finished that section with the words: ``By this standard there
is every reason to retain the Poyinting-Heaviside viewpoint until a
clash with new experimental evidence shall call for its
revision.''$^{(*)}$\footnotetext{$^{(*)}$Thanks are due to the referee
for calling our attention to this point.} \\

\n {\large\bf 5. Superluminal Solutions of Maxwell Equations and the
Principle of Relativity} \\

In section 3 we showed that it seems possible with present technology to
launch in free space superluminal electromagnetic waves (SEXWs).  We
show in the following that the physical existence of SEXWs implies a
breakdown of the Principle of Relativity (PR). Since this is a
fundamental issue, with implications for all branches of theoretical
physics, we will examine the problem with great care. In section 5.1 we
give a rigorous mathematical definition of the PR and in section 5.2 we
present the proof of the above statement.\\

\n {\bf 5.1 \  Mathematical Formulation of the Principle of Relativity
and Its Physical Meaning}\\

In Appendix B we define Minkowski spacetime as the triple
$\lan M,g,D\ran$, where $M \simeq \R^4, g$ is a Lorentzian metric and
$D$ is the Levi-Civita connection of $g$.

Consider now $G_M$, the group of all diffeomorphisms of $M$, called the
manifold mapping group. Let $\T$ be a geometrical object defined in
$A\subq M$. The diffeomorphism $h\in G_M$ induces a deforming mapping $
h_*: \T \mapsto h_* \T = \T $ such that: 

\renewcommand{\theequation}{5.\arabic{equation}}
\setcounter{equation}{0}

\begin{description}
\item[(i)] If $f: M\sups A \rig \R$, then $h_*f = f\circ h^{-1}: h(A)
\rig \R$

\item[(ii)] If $\T \in \sec T^{(r,s)}(A) \subset \sec T(M)$, where
$T^{(r,s)}(A)$ is the sub-bundle of tensors of type $(r,s)$ of the
tensor bundle $T(M)$, then
\[ 
(h_*\T)_{h_e} (h_* \omega_1,\ld, h_* \omega_r,
h_* X_1,\ld,h_*X_s)
= \T_e(\omega_1,\ld,\omega_r, X_1,\ld,X_s)
\] 
$\for X_i \in T_eA,\, i=1,\ld,s,\, \for \om_j \in T^*_eA,\, j=1,\ld,r,
\, \for e \in A.$

\item[(iii)] If $D$ is the Levi-Civita connection and $X,Y \in \sec
TM$, then
\begin{equation}
(h_*D_{h*X} h_* Y)_{he} h_{*f} = (D_XY)_e f \ \ \ \for e \in M . 
\end{equation}
\end{description}

If $\{f_\mu = \pa/\pa x^\mu\}$ is a coordinate basis for $TA$ and
$\{\te^\mu =dx^\mu\}$ is the corresponding dual basis for $T^*A$ and if
\begin{equation}
\T = T^{\mu_1\ld \mu_r}_{\nu_1\ld\nu_s} \te^{\nu_1} \ot\ld
\ot \te^{\nu_s} \ot f_{\mu_1}\ot\ld\ot f_{\mu_r} ,
\end{equation}
then
\begin{equation}
h_*\T = [T^{\mu_1\ld \mu_r}_{\nu_1\ld\nu_s}\circ h^{-1}] h_*
\te^{\nu_1} \ot\ld \ot h_* \te^{\nu_s} \ot h_* f_{\mu_1} \ot\ld\ot
h_*f_{\mu_r} . 
\end{equation}
Suppose now that $A$ and $h(A)$ can be covered by the local chart
$(U,\eta)$ of the maximal atlas of $M$, and $A \subq U, h(A) \subq
U$. Let $\lan x^\mu\ran$ be the coordinate functions associated with
$(U,\eta)$. The mapping
\begin{equation}
x^{'\mu} = x^{\mu} \circ h^{-1} \ : \ h(U) \rig \R
\end{equation}
defines a coordinate transformation $\lan x^\mu\ran \mapsto \lan
x^{'\mu}\ran$ if $h(U)\sups A \cup h(A)$. Indeed $\lan x^{'\mu}\ran$ are
the coordinate functions associated with the local chart $(V,\var)$
where $h(U)\subq V$ and $U\cap V \neq \phi$. Now, since it is well known
that under the above conditions $h_* \pa/\pa x^\mu \equiv \pa/\pa
x^{'\mu}$ and $h_*dx^\mu \equiv dx^{'\mu}$, eqs.(5.3) and (5.4) imply
that
\begin{equation}
(h_*\T)_{\lan x^{'\mu} \ran}(he) = \T_{\lan x^\mu \ran}(e) , 
\end{equation}
where $\T_{\lan x^\mu\ran}(e)$ means the components of $\T$ in the chart
$\lan x^\mu\ran$ at the event $e \in M$, {\em i.e.\/}, $\T_{\lan x^\mu
\ran}(e) = T^{\mu_1\ld \mu_r}_{\nu_1\ld\nu_s} (x^\mu(e))$ and where
$\bar{T}^{'\mu_1 \ldots \mu_r}_{\nu_1 \ldots \nu_s}(x^{'\mu}(he))$ are
the components of $\bar{\T} = h_{*}\T$ in the basis $\{h_* \partial /
\partial x^\mu = \partial/\partial x^{'\mu}\}$, $\{h_* dx^\mu =
dx^{'\mu}\}$, at the point $h(e)$. Then eq.(5.6) reads
\begin{equation}
{\ov{T}}^{'\mu_1\ld \mu_r}_{\nu_1\ld\nu_s} (x^{'\mu} (he)) = T^{\mu_1
\ld \mu_r}_{\nu_1\ld\nu_s} (x^\mu(e)) , 
\end{equation}
or using eq.(5.5)
\begin{equation}
\ov T^{'\mu_1\ld \mu_r}_{\nu_1\ld\nu_s}(x^{'\mu}(e)) =
(\Lambda^{-1})^{\mu_1}_{\alpha_1} \ldots \Lambda^{\beta_s}_{\nu_s}
T^{'\alpha_1 \ldots \alpha_r}_{\beta_1 \ldots \beta_s}
(x^{'\mu}(h^{-1}e)) ,
\end{equation}
where $\Lambda^{\mu}_{\alpha}= \partial x^{'\mu}/\partial
x^{\alpha}$, etc.

In appendix B we introduce the concept of inertial reference
frames $I \in \sec TU$, $U\subq M$ by
\begin{equation}
g(I,I) =1 \ \ {\rm and} \ \ DI =0 . 
\end{equation}
A general frame $Z$ satisfies $g(Z,Z)=1$, with $DZ \neq 0$. If $\alpha =
g(Z,\ ) \in \sec  T^*U$, it holds
\begin{equation}
(D\al)_e = a_e \ot \al_e + \sig_e + \om_e + \frac{1}{3} \te_e h_e, e \in
U \subq M ,
\end{equation}
where $a = g(\mbox{A},\ )$, $\mbox{A}=D_Z Z$ is the acceleration and
where $\om_e$ is the rotation tensor, $\sig_e$ is the shear tensor,
$\te_e$ is the expansion and $h_e = g_{|H_e}$ where
\begin{equation}
T_eM = [Z_e]\oplus [H_e] . 
\end{equation}
$H_e$ is the rest space of an {\it instantaneous  observer} at $e$, {\em
i.e.\/} the pair $(e, Z_e)$. Also $h_e(X,Y) = g_e(pX, pY)$, $\for X,Y
\in T_e M$ and $p: T_eM \rig H_e$. (For the explicit form of $\om, \sig,
\te$, see$^{[60]})$. From eqs.(5.9) and (5.10) we see that an inertial
reference frame has no acceleration, no rotation, no shear and no
expansion.

We introduce also in Appendix B the concept of a (nacs/$I$). A
(nacs/$I$)  $\lan x^\mu\ran$ is said to be in the Lorentz gauge if
$x^\mu$, $\mu=0,1,2,3$ are the usual Lorentz coordinates and $I =\pa/\pa
x^0 \in \sec TM$.  We recall that it is a theorem that putting $I=e_0
=\pa/\pa x^0$, there exist three other fields $e_i \in \sec TM$ such
that $g(e_i, e_i) =-1, \ i=1,2,3$, and $e_i =\pa/\pa x^i$.

Now, let $\lan x^\mu \ran$ be Lorentz coordinate functions as above. We
say that
$\ell \in G_M$ is a {\it Lorentz mapping} if and only if
\begin{equation}
x^{'\mu}(e) = \La^\mu_\nu x^\mu(e),
\end{equation}
where $\La^\mu_\nu \in \Le^\upa_+$ is a Lorentz transformation. For
abuse of notation we denote the subset $\{\ell\}$ of $G_M$ such that
eq.(5.12) holds true also by $\Le^\upa_+ \sub G_M$.

When $\lan x^{\mu} \ran$ are Lorentz coordinate functions, $\lan
x^{'\mu} \ran$ are also Lorentz coordinate functions. In this case we
denote
\begin{equation}
e_\mu = \pa/\pa x^\mu, \ e'_\mu = \pa/\pa x^{'\mu} , \
\ga_\mu = dx^\mu , \ \ga'_\mu = dx^{'\mu} \; ;
\end{equation}
when $\ell \in \Le^\upa_+ \sub G_M$ we say that $\ell_*\T$ is the {\it
Lorentz deformed version} of $\T$. 

Let $h \in G_M$. If for a geometrical object $\T$ we have
\begin{equation}
h_* \T =\T,
\end{equation}
then $h$ is said to be a symmetry of $\T$ and the set of all $\{h \in
G_M\}$ such that eq.(5.13) holds is said to be the symmetry group of
$\T$.  We can immediately verify that for $\ell \in \Le^\upa_+ \sub G_M$
\begin{equation}
\ell_* g = g , \ \ell_*D = D,
\end{equation}
{\em i.e.\/}, the special restricted orthochronous Lorentz group
$\Le^\upa_+ $ is a symmetry group of $g$ and $D$.

In$^{[62]}$ we maintain that a physical theory $\tau$ is
characterized by: 

\begin{description}
\item{(i)} the theory of a certain ``species of structure" in the
sense of Boubarki$^{[63]}$;
\item{(ii)} its physical interpretation;
\item{(iii)} its present meaning and present applications.
\end{description}

We recall that in the mathematical exposition of a given physical theory
$\tau$, the postulates or basic axioms are presented as definitions.
Such definitions mean that the physical phenomena described by $\tau$
behave in a certain way. Then, the definitions require more motivation
than the pure mathematical definitions. We call coordinative definitions
the physical definitions, a term introduced by Reichenbach$^{[64]}$. It
is necessary also to make clear that completely convincing and genuine
motivations for the coordinative definitions cannot be given, since they
refer to nature as a whole and to the physical theory as a whole.

The theoretical approach to physics behind (i), (ii) and (iii) above is
then to admit the mathematical concepts of the ``species of structure"
defining $\tau$ as primitives, and define coordinatively the observation
entities from them.  Reichenbach assumes that ``{\it physical knowledge}
is characterized by the fact that concepts are not only defined by other
concepts, but are also coordinated to real objects". However, in our
approach, each physical theory, when characterized as a species of
structure, contains some implicit geometric objects, like some of the
reference frame fields defined above, that cannot in general  be
coordinated to real objects. Indeed it would be an absurd to suppose
that all the infinity of IRF that exist in $M$ must have a material
support.

We define a {\it spacetime} theory as a theory of a species of structure
such that, if Mod $\tau$ is the class of models of $\tau$, then each
$\Upsilon \in$ Mod $\tau$ contains a substructure called spacetime (ST).
More precisely, we have
\begin{equation}
\Upsilon = (\mbox{ST}, \T_1\ld \T_m\}\; ,
\end{equation}
where ST can be a very general structure$^{[62]}$.  For what follows we
suppose that $\mbox{ST} = \Me = (M, g, D)$, {\em i.e.\/} that ST is
Minkowski spacetime. The $\T_i$, $i=1,\ld,m$ are (explicit) geometrical
objects defined in $U \subq M$ characterizing the physical fields and
particle trajectories that cannot be geometrized in $\Upsilon$. Here, to
be geometrizable means to be a metric field or a connection on $M$ or
objects derived from these concepts as, {\em e.g.\/}, the Riemann tensor
or the torsion tensor.

The reference frame fields will be called the {\it implicit} geometrical
objects of $\tau$, since they are mathematical objects that do not
necessarily correspond to properties of a physical system described by
$\tau$.

Now, with the Clifford bundle formalism we can formulate in $\Ca\ell(M)$
all modern physical theories (see Appendix B) including Einstein's
gravitational theory$^{[6]}$. We introduce now the Lorentz-Maxwell
electrodynamics (LME) in $\Ca\ell(M)$ as a theory of a species of
structure. We say that LME has as model
\begin{equation}
\Upsilon_{LME} = \lan M,g,D,F,J,\{\varphi_i,m_i,e_i\}\ran, 
\end{equation}
where $(M,g,D)$ is Minkowski spacetime, $\{\varphi_i,m_i,e_i\}$,
$i=1,2,\ld,N$ is the set of all charged particles, $m_i$ and $e_i$ being
the masses and charges of the  particles and $\var_i: \R \supset I \rig
M$ being the world lines of the particles characterized by the fact that
if $\varphi_{i\ast} \in \sec TM$ is the velocity vector, then
$\check\varphi_i = g(\var_{i*},\ ) \in \sec \La^1(M) \sub \sec
\Ca\ell(M)$ and $\check\var_i.\check\var_i =1$. $F \in \sec \La^2(M)
\sub\sec \Ca\ell(M)$ is the electromagnetic field and $J \in \sec
\La^1(M)\sub \sec \Ca\ell(M)$ is the current density. The proper axioms
of the theory are
\begin{equation}
\begin{array}{c} 
\pa F = J \\ 
m_i D_{\var_{i*}} \check\varphi_i = e_i \check\varphi_i \cdot F
\end{array} 
\end{equation}

From a mathematical point of view it is a trivial result that
$\tau_{LME}$ has the following property: If $h \in G_M$ and if
eqs.(5.16) have a solution $\lan F,J, (\var_i,m_i,e_i)\ran$ in $U \subq
M$ then $\lan h_*F, h_*J, (h_* \var_i, m_i, e_i) \ran $ is also a
solution of eqs.(5.16) in $h(U)$. Since the result is true for any $h
\in G_M$ it is true for $\ell \in \Le^\upa_+ \sub G_M$, {\em i.e.\/},
for any Lorentz mapping.

We must now make it clear that $\lan F,J, \{\var_i,m_i,e_i\}\ran$ which
is a solution of eq.(5.16) in $U$ can be obtained only by imposing {\it
mathematical boundary conditions} which we denote by $BU$. The
solution will be realizable in nature if and only if the mathematical
boundary conditions can be physically realizable. This is indeed a
nontrivial point$^{[62]}$ for in particular it says to us that
even if $\lan h_*F, h_*J, \{h_*\var_i,m_i,e_i\}\ran$ can be a solution
of eqs.(5.16) with mathematical boundary conditions $Bh(U)$, it may
happen that $Bh(U)$ cannot be physically realizable in nature. The
following statement, denoted $PR_1$,  is usually presented$^{[62]}$
as the Principle of (Special) Relativity in active form:


\begin{description} 
\item{$PR_1$:}Let $\ell \in \Le^\upa_+ \sub G_M$. If for a physical
theory $\tau$ and $\Upsilon \in \mbox{Mod}\, \tau $, $\Upsilon = \lan
M,g,D,\T_1,\ld,\T_m\ran$ is a possible physical phenomenon, then $\ell_*
\Upsilon = \lan M, g, D, \linebreak l_*\T_1, \ld, l_*\T_m\ran$ is  also a possible
physical phenomenon.  
\end{description}


\noindent It is clear that hidden in $PR_1$ is the assumption that the
boundary conditions that determine $\ell_*\Upsilon$ are physically
realizable.  Before we continue we introduce the statement denoted
$PR_2$, known as the Principle of (Special) Relativity in passive
form$^{[62]}$

\begin{description} 
\item{$PR_2$:}``All inertial reference frames are physically equivalent
or indistinguishable".
\end{description} 


\noindent We now give a precise mathematical meaning to the above
statement.

Let $\tau$ be a spacetime theory and let $\mbox{ST} = \lan M,g,D\ran$ be
a substructure of Mod~$\tau$ representing spacetime. Let $I \in \sec TU$
and $I' \in \sec TV$, $U, V \subq M$, be two inertial reference frames.
Let $(U,\eta)$ and $(V,\var)$ be two Lorentz charts of the maximal atlas
of $M$ that are naturally adapted respectively to $I$ and $I'$. If $\lan
x^\mu\ran$ and $\lan x^{'\mu}\ran$ are the coordinate functions
associated with $(U,\eta)$ and $(V,\var)$, we have $I =\pa/\pa x^0, I' =
\pa/\pa x^{'0}$. 

\medskip 

{\bf Definition:} Two inertial reference frames $I$ and $I'$ as above
are said to be physically equivalent according to $\tau$ if and only if
the following conditions are satisfied:

(i) $G_M \supset \Le^\upa_+ \ni \ell: U \rig \ell(U) \subq V, \
x^{'\mu} = x^{\mu} \circ \ell^{-1} \Rig I' = \ell_* I$

When $\Upsilon \in Mod \tau$, $\Upsilon = \lan M,g,D, \T_1,\ld
\T_m\ran$, is such that $g$ and $D$ are defined over all $M$ and $\T_i
\in \sec \Ca\ell (U) \subset \sec \Ca\ell (M)$, calling $o=\lan
g,D,\T_1,\ld \T_m\ran$, $o$ solves a set of differential equations in
 $\eta(U) \sub \R^4$ with a given set of boundary conditions denoted
$b^{o\lan x^{\mu}\ran}$, which we write as

\begin{equation} \label{5e19} 
D^\al_{\lan x^\mu \ran} (o_{\lan x^\mu \ran})_e =0 \ ; \
b^{o\lan x^{\mu}\ran} \ ; \ e \in U
\end{equation}
and we must have:

(ii) If $\Upsilon \in$ Mod $\tau \Lfr \ell_* \Upsilon \in $ Mod $\tau$,
then necessarily
\begin{equation} \label{5e20} 
\ell_*\Upsilon = \lan M,g,D, \ell_*\T_1,\ld \ell_* \T_m\ran
\end{equation}
is defined in $\ell(U) \subq V$ and calling $\ell_* o\equiv
\{g,D,\ell_* \T_1,\ld, \ell_* \T_m\}$ we must have
\begin{equation} \label{5e21} 
D^\al_{\lan x^{'\mu} \ran} (\ell_* o_{\lan x^{'\mu}
\ran})_{|\ell e} =0 \ ; \  b^{\ell_* o\lan x^{'\mu}\ran} \ \ \ell e \in
\ell (U) \subq V .
\end{equation}
In eqs.(\ref{5e19}) and (\ref{5e21}) $D^{\alpha}_{\langle x^\mu
\rangle}$ and $D^{\alpha}_{\langle x^{'\mu} \rangle}$ mean $\alpha =
1, 2, \ldots,m$ sets of differential equations in $\R^4$. The system of
differential equations (5.19) must have the same functional form as  the
system of differential equations (5.17) and $b^{\ell_* o\lan
x^{'\mu}\ran}$ must be relative to $\lan x^{'\mu}\ran$ the same  as
$b^{o\lan x^{\mu}\ran}$ is relative to $\lan x^{\mu}\ran$ and if
$b^{o\lan x^{\mu}\ran}$ is physically realizable then $b^{\ell_* o\lan
x^{'\mu}\ran}$ must also be physically realizable.  We say under these
conditions that $I \sim I'$ and that $\ell_* o$ is the Lorentz deformed
version of  the phenomena described by $o$.

\medskip 

Since in the above definition $\ell_* \Upsilon= \lan M,g,D, \ell_*
T_1,\ld,\ell_*T_m\ran$, it follows that when $I \sim I'$, then $\ell_*
g=g, \ell_* D = D$ (as we already know) and this means that the
spacetime structure does not give a preferred status to $I$ or $I'$
according to $\tau$.\\

\n {\bf 5.2 \ Proof that the Existence of SEXWs Implies a Breakdown of
{\boldmath $PR_1$ and $PR_2$}}\\

We are now able to prove the statement presented at the beginning of
this section, that the existence of SEXWs implies a breakdown of the
Principle of Relativity in both  its active ($PR_1$) and passive
($PR_2$) versions.

Let  $\ell \in \Le^\upa_+ \sub G_M$ and let $F$,  $\ov{F} \in \sec
\La^2(M) \sub \sec \Ca\ell(M)$, $\ov{F} = \ell_* F$.  Let $\ov F =
\ell_* F=R\check{F}R^{-1}$, where $\check{F}_e = (1/2) F_{\mu\nu}
(x^\delta (\ell^{-1}e))\gamma^\mu \gamma^\nu$ and where $R \in \sec
\mbox{Spin}_+(1,3) \sub \sec \Ca\ell(M)$ is a Lorentz mapping, such that
$\ga^{'\mu} =R\ga^\mu R^{-1} = \La^\mu_\al \ga^\al, \La^\mu_\al \in
\Le^\upa_+$ and let $\lan x^\mu\ran$ and $\lan x^{'\mu}\ran$ be Lorentz
coordinate functions as before such that $\ga^\mu = dx^\mu, \ga^{'\mu} =
dx^{'\mu}$ and $x^{'\mu} = x^\mu \circ \ell^{-1}$. We write

$$F_e = {\dis\frac{1}{2}} F_{\mu\nu}(x^\delta(e))
\ga^\mu\ga^\nu ; \eqno{(5.22\mbox{a})}$$

$$F_e = {\dis\frac{1}{2}} F'_{\mu\nu}(x^{'\delta}(e))
\ga^{'\mu}\ga^{'\nu} ; \eqno{(5.22\mbox{b})}$$

$$\ov F_e = {\dis\frac{1}{2}} \ov F_{\mu\nu}(x^\delta(e))
\ga^\mu\ga^\nu ; \eqno{(5.23\mbox{a})}$$

$$\ov F_e = {\dis\frac{1}{2}} \ov F'_{\mu\nu}(x^{'\delta}(e))
\ga^{'\mu}\ga^{'\nu} . \eqno{(5.23\mbox{b})}$$

From (5.22a) and (5.22b) we get that
\setcounter{equation}{23}
\begin{equation}
F'_{\al\be} (x^{'\delta}(e)) = (\La^{-1})^\mu_\al(\La^{-1})^\nu_\be
F_{\mu\nu} (x^\delta(e)) .
\end{equation}
From (5.22a) and (5.23b) we also get
\begin{equation}
\ov F_{\al\be} (x^{\delta}(e)) = \La^\mu_\al \La^\nu_{\be} F_{\mu\nu}
(x^\delta(\ell^{-1} e))
\end{equation}

Now, suppose that $F$ is a superluminal solution of Maxwell equation, in
particular a SEXW as discussed in section 3.  Suppose that $F$ has been
produced in the inertial frame $I$ with $\lan x^\mu\ran$ as  (nacs/$I$),
with the physical device described in section 3. $F$ is generated in the
plane $z=0$ and is traveling with speed $c_1=1/\cos \eta$ in the
negative $z$-direction. It will then travel to the future in spacetime,
according to the observers in $I$.  Now, there exists $\ell \in
\Le^\upa_+$ such that $\ell_* F = \ov F = RFR^{-1}$ will be a solution
of Maxwell equations and such that if the velocity 1-form of $F$ is
$v_F=(c_1^2 -1)^{-1/2}(1,0,0,-c_1)$, then the velocity 1-form of $\ov F$
 is $v_{\ov{F}} = (c^{'2}_{1} -1)^{-1/2} (-1,0,0,-c'_1)$, with $c'_1
>1$, {\em i.e.\/}, $v_{\ov{F}}$ is pointing to the past. As its is well
known $\ov{F}$ carries negative energy according to the observers in the
$I$ frame.

We then arrive at the conclusion that to assume the  validity of $PR_1$
is to assume the physical possibility of sending to the past waves
carrying negative energy.  This seems to the authors an impossible task,
and the reason is that there do no exist physically realizable boundary
conditions that would allow the observers in $I$ to launch $\bar{F}$ in
spacetime and such that it traveled to its own past.

We now show that there is also a breakdown of $PR_2$, {\em i.e.\/}, that
it is not true that all inertial frames are physically equivalent.
Suppose we have two inertial frames $I$ and $I'$ as above, {\em i.e.\/},
$I =\pa/\pa x^0$, $I' = \pa/\pa x^{'0}$.

Suppose that $F$ is a SEXW which can be launched in $I$ with velocity
1-form as above and suppose $\ov F$ is a SEXW built in $I'$ at the plane
$z'=0$ and with velocity 1-form relative to $\lan x^{'\mu} \ran$ given
by $v_{\ov F} = v^{'\mu} \ga'_{\mu}$ and
\begin{equation}
v_{\ov F} = \biggl(\frac{1}{\sqrt{c^{2}_1-1}}, 0,0, -
\frac{c_1}{\sqrt{c^{2}_1-1}}\biggr)
\end{equation}
If $F$ and $\ov F$ are related as above we see (See Fig.10) that $\ov
F$, which has positive energy and is traveling to the future according
to $I'$, {\it can be sent} to the past of the observers at rest in the
$I$ frame. Obviously this is impossible and we conclude that $\ov F$ is
not a physically realizable phenomenon in nature. It cannot be realized
in $I'$ but $F$ can be realized in $I$. It follows that $PR_2$ does not
hold.

If the elements of the set of inertial reference frames are not
equivalent then there must exist a fundamental reference frame. Let $I
\in \sec TM$ be that fundamental frame. If $I'$ is moving with speed $V$
relative to $I$, {\em i.e.\/},
\begin{equation}
I' = \frac{1}{\sqrt{1-V^2}} \frac{\pa}{\pa t} -
\frac{V}{\sqrt{1-V^2}}\pa/\pa z \; ,
\end{equation}
then, if observers in $I'$ are equipped with a generator of SEXWs and if
they prepare their apparatus in order to send SEXWs with different
velocity 1-forms in all possible directions in spacetime, they will find
a particular velocity 1-form in a given spacetime direction in which the
device stops working. A simple calculation yields then, for the observes
in $I'$, the value of $V$~! 

In $^{[65]}$ Recami argued that the Principle of Relativity continues to
hold true even though superluminal phenomena exist in nature. In this
theory of tachyons there exists, of course, a situation completely
analogous to the one described  above (called the Tolman-Regge paradox),
and according to Recami's view $PR_2$ is valid because $I'$ must
interpret $\ov F$ a being an anti-SEXW carrying positive energy and
going into the future according to him.  In his theory of tachyons
Recami was able to show that  the dynamics of tachyons implies that no
detector {\it at rest} in  $I$ can detect a tachyon (the same would be
valid for a SEXW like $\ov{F}$) sent by $I'$ with velocity 1-form given
by eq.(4.26). Thus he claimed that $PR_2$ is true. At first sight the
argument seems good, but it is at least incomplete. Indeed, a detector
in $I$ does not need to be at rest in $I$.  We can imagine a detector in
periodic motion in $I$ which could absorb the $\ov F$ wave generated by
$I'$ if this was indeed possible. It is enough for the detector to have
relative to $I$ the speed $V$ of the $I'$ frame in the appropriate
direction at the moment of absorption.  This simple argument shows that
there is no salvation for $PR_2$ (and for $PR_1$) if superluminal
phenomena exist in nature.

The attentive reader at this point probably has the following
question in his/her mind: How could the authors start with Minkowski
spacetime, with equations carrying the Lorentz symmetry and yet
arrive at the conclusion that $PR_1$ and $PR_2$ do not hold?

The reason is that the Lorentzian structure of $\lan M,g,D\ran$ can be
seen to exist directly from the Newtonian spacetime structure as proved
in $^{[66]}$. In that paper Rodrigues and  collaborators show that even
if $\Le^\upa_+$ is not a symmetry group of Newtonian dynamics it is a
symmetry group of the only possible coherent formulation of
Lorentz-Maxwell electrodynamic theory compatible with experimental
results that is possible to formulate in the Newtonian
spacetime$^{(*)}$. \footnotetext{$(*)$ We recall that Maxwell equations
have, as  is well known, many symmetry groups besides $\Le^\upa_+$.}

We finish calling to the reader's attention that there are some
experiments reported in the literature which suggest also a breakdown of
$PR_2$ for the roto-translational motion of solid bodies. A discussion
and references can be found in$^{[67]}$.

\begin{figure}[htb] 
\vspace*{8cm} 
\caption{$\bar{F}$ cannot be launched by $I'$.} 
\end{figure}

\bigskip 

\n {\large\bf 6. Conclusions}
\nobreak

\bigskip

In this paper we presented a unified theory showing that the homogeneous
wave equation, the Klein-Gordon equation, Maxwell equations and the
Dirac and Weyl equations have solutions with the form of  undistorted
progressive waves (UPWs) of arbitrary speeds $0 \leq v <\infty$.

We present also the results of an experiment which confirms that finite
aperture approximations to a Bessel pulse and to an $X$-wave in water
move as predicted by our theory, {\em i.e.\/}, the Bessel pulse moves
with speed less than $c_s$ and the $X$-wave moves with speed greater
than $c_s$, $c_s$ being the sound velocity in  water.

We exhibit also some subluminal and superluminal solutions of Maxwell
equations. We showed that subluminal solutions can in principle be used
to model purely electromagnetic particles. A detailed discussion is
given about the superluminal electromagnetic $X$-wave solution of Maxwell
equations and we showed that it can in principle be launched with
available technology. Here a point must be clear, the $X$-waves, both
acoustic and electromagnetic, are signals in the sense defined by
Nimtz$^{[74]}$. It is a widespread misunderstanding that signals must
have a front. A front can be defined only mathematically because it
implies an infinite frequency spectrum. Every real signal does not have
a well defined front. 

The existence of superluminal electromagnetic waves implies in the
breakdown of the Principle of
Relativity.$^{(*)}$\footnotetext{$^{(*)}$It is important to recall that
there exists the possibility of propagation of superluminal signals
inside the hadronic matter. In this case the ingenious construction of
Santilli's isominkowskian spaces (see$^{[68-73]}$) is useful.} We
observe that besides its fundamental theoretical implications, the
practical implications of the existence of UPWs solutions of the main
field equations of theoretical physics (and their finite aperture
realizations) are very important.  This practical importance ranges from
applications in ultrasound medical imaging to the project of
electromagnetic bullets and new communication devices$^{[33]}$. Also we
would like to conjecture that the existence of subluminal and
superluminal solutions of the Weyl equation may be important to solve
some of the mysteries associated with neutrinos.  Indeed, if neutrinos
can be produced in subluminal or superluminal modes --- see$^{[75,76]}$
for some experimental evidence concerning superluminal neutrinos ---
they can eventually escape detection on earth after leaving the sun.
Moreover, for neutrinos in a subluminal or superluminal mode it would be
possible to define a kind of ``effective mass". Recently some
cosmological evidences that neutrinos have a non-vanishing mass have
been discussed by {\em e.g.\/} Primack et al$^{[77]}$. One such
``effective mass" could be responsible for those cosmological evidences,
and in such a way that we can still have a left-handed neutrino since it
would satisfy the Weyl equation. We discuss more this issue in another
publication.

\bigbreak

\n {\large\bf Acknowledgments}
\nobreak

\bigskip
The authors are grateful to CNPq, FAPESP and FINEP for partial financial
support. We would like also to thank Professor V.\ Barashenkov,
Professor G.\ Nimtz, Professor E.\ Recami, Dr.  E.\ C.\ de Oliveira, Dr.
Q.\ A.\ G.\ de Souza, Dr. J.\ Vaz Jr.\ and Dr. W.\ Vieira for many
valuable discussions, and J.\ E.\  Maiorino for collaboration and a
critical reading of the manuscript. WAR recognizes specially the
invaluable help of his wife Maria de F\'atima and his sons, whom with
love supported his variations of mood during the very hot summer of 96
while he was preparing this paper.  We are also grateful to the referees
for many useful criticisms and suggestions and for calling our attention
to the excellent discussion concerning the Poynting vector in the books
by Stratton$^{[19]}$ and Whittaker$^{[78]}$.  

\bigskip 

\n {\large\bf Appendix A.\ Solutions of the (Scalar) Homogeneous Wave
Equation and Their Finite Aperture Realizations} \\

In this appendix we first recall briefly  some well known results
concerning the fundamental  (Green's functions) and the general
solutions of the (scalar) {\it homogeneous wave equation} (HWE) and the
theory of their {\it finite aperture approximation} (FAA). FAA is based
on the {\it Rayleigh-Sommerfeld formulation of diffraction} (RSFD) by a
plane screen. We show that under certain conditions the RSFD is useful
for designing physical devices to launch waves that travel with the
characteristic velocity in a homogeneous medium ({\em i.e.\/}, the speed
$c$ that appears in the wave equation). More important, RSFD is also
useful for projecting physical devices to launch some of the subluminal
and superluminal solutions of the HWE ({\em i.e.\/}, waves that
propagate in an homogeneous medium with speeds respectively less and
greater than $c$) that we present in this appendix. We use units such
that $c=1$ and $\hbar =1$, where $c$ is the so called velocity of light
in vacuum and $\hbar$ is Planck's constant divided by $2\pi$. \\

\n {\bf A1. Green's Functions and the General Solution of the (Scalar)
HWE} \\

Let $\Phi$ in what follows be a complex function in Minkowski spacetime
$M$:
\renewcommand{\theequation}{A.\arabic{equation}}
\setcounter{equation}{0}
\begin{equation}
\Phi : M  \ni x \mapsto \Phi(x) \in \C \; .
\end{equation}
The inhomogeneous wave equation for $\Phi$ is
\begin{equation}
\Box\Phi = \left(\frac{\pa^2}{\pa t^2} -
\nab^2\right) \Phi = 4\pi\rho \; ,
\end{equation}
where $\rho$ is a complex function in Minkowski spacetime. We define a
two-point Green's function for the wave equation (A.2) as a solution of
\begin{equation}
\Box G(x-x') = 4\pi\de(x-x') \; .
\end{equation}

As it is well known, the fundamental solutions of (A.3) are: \\

\hfill {\sl Retarded Green's function:} $G_R(x-x')=2H(x-x')\de[(x-x')^2]$;
\hfill (A.4a)

\hfill {\sl Advanced Green's function:}
$G_A(x-x')=2H[-(x-x')]\de[(x-x')^2]$;  \hfill (A.4b)
\\

\n where $(x-x')^2 \equiv (x^0-x^{'0})^2 - (\vec x - \vec x')^2$,
$H(x) = H(x^0)$ is the step function and $x^0=t,x^{'0}=t'$.

We can rewrite eqs.(A.4) as ($R = |\vec x-\vec x'|$):

$$ G_R(x^0-x^{'0}; \vec x-\vec x') = \frac{1}R \de(x^0-x^{'0}-R) \; ;
\eqno{(\mbox{A.4c})} $$

$$ G_A(x^0-x^{'0}; \vec x-\vec x') = \frac{1}R \de(x^0-x^{'0}-R) \; .
\eqno{(\mbox{A.4d})} $$

We define the Schwinger function by
\setcounter{equation}{4}
\begin{equation}
G_S = G_R - G_A = 2 \vare(x) \de(x^2); \ \vare(x)=H(x) - H(-x) \; .
\end{equation}
It has the properties
$$
\Box G_S = 0; \ G_S(x) = - G_S (-x); \ G_S(x) = 0 \ \ {\rm if} \ \
x^2 < 0 \; ; \eqno{(\mbox{A.6a})}
$$
$$
G_S (0, \vec x) = 0;   \ {\dis\frac{\pa G_S}{\pa x^i}}
\biggl|_{x^i=0} = 0; \ {\dis\frac{\pa G_s}{\pa x^0}}\biggl|_{x^0=0} =
\de (\vec x) \; . \eqno{(\mbox{A.6b})}
$$

For the reader who is familiar with the material presented in Appendix
B, we observe that these equations can be rewritten in a very elegant
way in $\Ca\ell_C(M)$. (If you haven't read Appendix B, go to
eq.(A.8$'$).) We have
\setcounter{equation}{6}
\begin{equation}
\int_\sig \star dG_S (x-y) = -\int_\sig d G_S(x-y)\ga^5 = 1, \ \ {\rm
if} \ \ y \in \sig,
\end{equation}
where $\sig$ is any spacelike surface. Then if $f \in \sec \C \ot
\bwe^0(M) \sub \sec \Ca\ell_C (M)$ is any function defined on a
spacelike surface $\sig$, we can write

\begin{equation}
\int_\sig [\star dG_S (x-y)]f(x) = - \int d G_s (x-y) f(x) \ga^5 =
f(y) \; .
\end{equation}
Eqs.(A.7) and (A.8) appear written in textbooks on field theory as
$$
{\dis\int_\sig} \pa_\mu G_S (x-y)d\sig^\mu (x) = 1 \ ; \
{\dis\int_\sig} f(x) \pa_\mu G_S(x-y) d\sig^\mu (x) =
f(y) \; . \eqno{(\mbox{A.8'})}
$$
We now express the general solution of eq.(A.2), including the initial
conditions, in a bounded constant time  spacelike hypersurface $\sig$
characterized by $\ga^1 \we \ga^2  \we \ga^3$ in terms of $G_R$. We
write the solution in the {\it standard vector notation}. Let
the constant time hypersurface $\sig$ be the volume $V \sub \R^3$ and
$\pa V = S$ its boundary. We have, 
\begin{eqnarray} 
\Phi(t,\vec x) & = & \int^{t_+}_0 dt' \int\!\!\int\!\!\int_V d\ven' G_R
(t-t', \vec x-\vec x') \rho (t', \vec x') \nb \\ 
&+& \frac{1}{4\pi}
\int\!\!\int\!\!\int_V d\ven' \left[ G_R |_{t'=0} \frac{\pa\Phi}{\pa t'}
(t',\vec x')|_{t'=0} -\Phi(t', \vec x')|_{t'=0} \frac{\pa}{\pa t'} 
G_R |_{t'=0} \right] \nb \\ 
&+& \frac{1}{4\pi} \int^{t_+}_0 dt' \int\!\!\int_S d\vec
S' . (G_R \grad' \Phi-\Phi \grad' G_R),
\end{eqnarray} 
where grad$'$ means that the gradient operator acts on $\vec x'$, and
where $t_+$ means that the integral over $t'$ must end on $t' = t+\vare$
in order to avoid ending the integral exactly at the peak of the
$\de$-function. The {\it first term} in eq.(A.9) represents the effects
of the sources, the {\it second term} represents the effects of the
initial conditions (Cauchy problem) and the {\it third term} represents
the effects of the boundary conditions on the space boundaries $\pa V =
S$.This term is essential for the theory of diffraction and in
particular for the RSFD.\ 

 \smallskip  

\n {\bf Cauchy problem:} Suppose that $\Phi(0, \vec x)$ and
${\dis\frac{\pa}{\pa t}} \Phi(t, \vec x)|_{t=0}$ are known at every
point in space, and assume that there are no sources present, {\em
i.e.\/}, $\rho = 0$.  Then the solution of the HWE becomes
\begin{equation}
\Phi(t,\vec x) = \frac{1}{4\pi} \int\!\!\int\!\!\int d\ven'
\left[G_R|_{t'=0} \frac{\pa}{\pa t}\Phi(t', \vec x')|_{t'=0} -
\frac{\pa}{\pa t} G_R |_{t'=0} \Phi(0,\vec x')\right] \; .
\end{equation}
The integration extends over all space and we explicitly assume that
the third term in eq.(A.9) vanishes at infinity.

\smallskip 

We can give an intrinsic formulation of eq.(A.10). Let  $x \in \sig$,
where $\sig$ is a spacelike surface {\it without} boundary. Then the
solution of the HWE can be written
\begin{eqnarray}
\Phi(x) & = & \frac{1}{4\pi} \int_\sig \{G_S (x-x') [\star d\Phi(x')] -
[\star dG_S(x-x')] \Phi (x')\} \nb \\
& & \\
& = & \frac{1}{4\pi} \int_\sig d\sig^\mu(x) [G_S(x-x')\pa_\mu
\Phi(x') - \pa_\mu G_S (x-x')\Phi(x')] \nb
\end{eqnarray}
where $G_S$ is the Schwinger function [see eqs.(A.7, A.8)]. $\Phi(x)$
given by eq.(A.11) corresponds to ``causal propagation" in the usual
Einstein sense, {\em i.e.\/}, $\Phi(x)$ is influenced only by points of
$\sig$ which lie in the backward (forward) light cone of $x'$, depending
on whether $x$ is ``later" (``earlier") than $\sig$.\\

\n {\bf A2. Huygen's Principle; the Kirchhoff and Rayleigh-Sommerfeld
Formulations of Diffraction by a Plane Screen$^{[79]}$} \\

Huygen's principle is essential for  understanding  Kirchhoff's
formulation and the Rayleigh-Sommerfeld formulation (RSF) of diffraction
by a plane screen. Consider again the general solution [eq.(A.9)] of the
HWE which is non-null in the surface $S = \pa V$ and suppose also that
$\Phi (0, \vec x)$ and ${\dis\frac{\pa}{\pa t}} \Phi(t, \vec x)|_{t=0}$
are null for all $\vec x \in V$. Then eq.(A.9) gives
\begin{equation}
\Phi(t,\vec x) = \frac{1}{4\pi} \int\!\!\int_S d\vec S' . \left[
\frac{1}R \grad' \Phi(t', \vec x') + \frac{\vec R}{R^3} \Phi (t',\vec x')
- \frac{\vec R}{R^2} \frac{\pa}{\pa t'} \Phi (t', \vec
x')\right]_{t'=t-R} . 
\end{equation}
From eq.(A.12) we see that if $S$ is along a wavefront and the rest of
it is at infinity or where $\Phi$ is zero, we can say that the field
value $\Phi$ at $(t,\vec x)$ is caused by the field $\Phi$ in the wave
front at time $(t-R)$ earlier. This is Huygen's principle. \\ 

{\bf Kirchhoff's theory}: Now, consider a screen with a hole like in
Fig.11. 


\begin{figure}[htb]
\vspace*{7cm}
\caption{Diffraction from a finite aperture.} 
\end{figure}


Suppose that we have an {\it exact solution} of the HWE that can be
written as
\begin{equation}
\Phi(t,\vec x) = F(\vec x) e^{i\om t} ,
\end{equation}
where we define also
\begin{equation}
\om =\ov k
\end{equation}
and $\ov k$ is not necessarily the propagation vector (see bellow).  We
want to find the field at $\vec x \in V$, with $\pa V = S_1 + S_2$
(Fig.11), with $\rho=0 \ \for \vec x \in V$. Kirchhoff proposed to use
eq.(A.12) to give an approximate solution for the problem.  Under the so
called Sommerfeld radiation condition,
\begin{equation}
\lim_{\und r \rig \infi} \und r \left(\frac{\pa F}{\pa n} - i\ov k F
\right) = 0 ,
\end{equation}
where $\und r = |\und{\vec r}| = \vec x-\vec x'$, $\vec x'$ being a
point of $S_2$, the integral in eq.(A.12) is null over $S_2$. Then, we
get
\begin{eqnarray}
 F(\vec x) &=&  \frac{1}{4\pi} \int\!\!\int_{S_1} dS' \left(\frac{\pa
F}{\pa n} G_K - F\frac{\pa G_K}{\pa n}\right) ; \\
 \h\h G_K &=& \frac{e^{-i \ov k R}}{R}, \ R=|\vec x-\vec x'|, \vec x'
\in S_1 \; .
\end{eqnarray}

Now, the ``source" is opaque, except for the aperture which is denoted
by $\Sig$ in Fig.11. It is reasonable to suppose that the major
contribution to the integral arises from points of $S_1$ in the aperture
$\Sig \sub S_1$. Kirchhoff then proposed the conditions:

\begin{itemize}
\item[(i)] Across $\Sig$, the fields $F$ and $\pa F/\pa n$ are exactly
the same as they would be in the absence of sources. 

\item[(ii)] Over the portion of $S_1$ that lies in the geometrical
shadow of the screen the field $F$ and $\pa F/\pa n$ are null.
\end{itemize}

Conditions (i) plus (ii) are called {\it Kirchhoff boundary
conditions\/}, and we end with

\begin{equation}
F_K(\vec x) = \int\!\!\int_\Sig dS' \left(\frac{\pa F}{\pa n} G_K - F
\frac{\pa}{\pa n} G_K\right),
\end{equation}
where $F_K(\vec x)$ is the Kirchhoff approximation to the problem. As is
well known, $F_K$ gives results that agree very well with experiments,
if the dimensions of the aperture are large compared with the wave
length. Nevertheless, Kirchhoff's solution is inconsistent, since under
the hypothesis given by eq.(A.13), $F(\vec x)$ becomes a solution of the
Helmholtz equation
\begin{equation}
\nab^2 F + \om^2 F = 0 \; ,
\end{equation}
and as is well known it is illicit for this equation to impose
simultaneously arbitrary boundary conditions for both $F$ and $\pa F/\pa
n$.

A further shortcoming of $F_K$ is that it fails to reproduce the assumed
boundary conditions when $\vec x \in \Sig \sub S_1$. To avoid such
inconsistencies Sommerfeld proposed to eliminate the necessity of
imposing boundary conditions on both $F$ and $\pa F/\pa n$
simultaneously. This gives the so called {\it Rayleigh-Sommerfeld
formulation of  diffraction by a plane screen\/} (RSFD). RSFD is
obtained as follows. Consider again  a solution of eq.(A.18) under
Sommerfeld radiation condition [eq.(A.15)]
\begin{equation}
F(\vec x) = \frac{1}4 \int\!\!\int_{S_1} \left(\frac{\pa F}{\pa n}
G_{RS} - F\frac{\pa G_{RS}}{\pa n}\right) dS' ,
\end{equation}
where now $G_{RS}$ is a Green function for eq.(A.19) different from
$G_K$. $G_{RS}$ must provide an exact solution of eq.(A.19) but we want
in
addition that $G_{RS}$ or $\pa G_{RS}/\pa n$ vanish over the entire
surface $S_1$, since as we already said we cannot impose the values of
$F$ and $\pa F/\pa n$ simultaneously.

A solution for this problem is to take $G_{RS}$ as a three-point
function, {\em i.e.\/}, as a solution of
\begin{equation}
(\nab^2+\om^2)G^-_{RS} (\vec x, \vec x', \vec x'') = 4\pi \de(\vec
x-\vec x') - 4\pi\de(\vec x-\vec x'').
\end{equation}
We get
\begin{equation}
G^-_{RS} (\vec x,\vec x', \vec x'') = \frac{e^{i\ov k R}}{R} -
\frac{e^{i\ov k R'}}{R'} ,
\end{equation} 
\begin{equation}  
R = |\vec x-\vec x'|; \ R'=|\vec x-\vec x''| ,
\end{equation}
where $\vec x \in S_1$ and $\vec x' = - \vec x''$ are mirror image
points relative to $S_1$.  This solution gives $G^-_{RS}\biggl|_{S_1} =
0$ and $\pa G^-_{RS}/\pa n\biggr|_{S_1} \neq 0$.

Another solution for our problem such that $G^+_{RS}\biggl|_{S_1} \neq
0$ and $\pa G^+_{RS} / \pa n\biggr|_{S_1} = 0$ is realized for
$G^+_{RS}$ satisfying
\begin{equation}
(\nab^2+\om^2)G^+_{RS} (\vec x, \vec x', \vec x'') = 4\pi \de(\vec
x-\vec x') + 4\pi \de (\vec x - \vec x'') . 
\end{equation}
Then 
\begin{equation}
G^+_{RS} (\vec x, \vec x', \vec x'') = \frac{e^{i\ov k R}}{R} +
\frac{e^{i\ov kR'}}{R'} \; ,
\end{equation}
with $R$ and $R'$ as in eq.(A.23).

We now use $G^+_{RS}$ in eq.(A.25) and take $S_1$ as being the $z=0$
plane. In this case $\vec n = -\wide k$, $\wide k$ being the versor of
the $z$ direction, $\vec R= \vec x - \vec x'$, $\vec R . \vec n = z'-z
\cos (\vec n, \vec R) = (z'-z)/R$ and we get
\begin{equation}
F(\vec x) = -\frac{1}{2\pi} \int\int_{S_1} dS' F(x',y', 0) \left[i\ov kz
\frac{e^{i\ov kR}}{R^2} - \frac{e^{i\ov kR}}{R^3} z \right] \; .
\end{equation}

\bigskip 

\noindent {\bf A3. Finite Aperture Approximation for Waves Satisfying
\mbox{\boldmath $\Phi(t,\vec x)=F(\vec x)e^{-i\om t}$}} \nopagebreak 

\bigskip 

The finite aperture approximation to  eq.(A.26) consists in integrating
only over $\Sig \sub S_1$, {\em i.e.\/}, we suppose $F(\vec x) = 0 \
\for \vec x \in (S_1\backslash \Sig)$. Taking into account that
\begin{equation}
\ov k = 2\pi/\la, \ \om = \ov k,
\end{equation}
we get
\begin{equation}
F_{FAA} = \frac{1}{\la} \int\!\!\int_{\Sig} dS' F(x',y', 0)
\frac{e^{i\ov kR}}{R^2}z + \frac{1}{2\pi} \int\!\!\int_\Sig dS'
F(x',y',0) \frac{e^{i\ov kR}}{R^3} z  .
\end{equation}

In section A4 we show some subluminal and superluminal solutions of the
HWE and then discuss for which solutions the FAA is valid. We show that
there are indeed subluminal and superluminal solutions of the HWE for
which (A.28) can be used. {\it Even more important}, we describe in
section 2 the results of recent experiments, conducted by us, that
confirm the predictions of the theory for acoustic waves in water.\\

\n {\bf A4. Subluminal and Superluminal Solutions of the HWE} \\

Consider the HWE ($c=1$)
$$
{\dis\frac{\pa^2}{\pa t^2}} \Phi - \nab^2 \Phi = 0 \;. \qquad
 \eqno{(\mbox{A.2}')}
$$
We now present some subluminal and superluminal solutions of
eq.(A.$2'$).$^{[80]}$
\\

\noindent {\it Subluminal and Superluminal Spherical Bessel Beams.} To
introduce these beams we define the variables
$$\xi_< = [x^2+y^2+\ga^2_< (z-v_<t)^2]^{1/2} \; ;
\eqno{(\mbox{A.29a})}$$
$$\ga_< = {\dis\frac{1}{\sqrt{1-v^2_<}}} \; ; \ \ \om^2_< - k^2_< =
\Om^2_<  \; ; \ \ v_< = {\dis\frac{d\om_<}{dk_<}} \;
;\eqno{(\mbox{A.29b})}$$
$$\xi_> = [-x^2-y^2+\ga^2_> (z-v_> t)^2]^{1/2} \; ;
\eqno{(\mbox{A.29c})}$$
$$\ga_> = {\dis\frac{1}{\sqrt{v^2_> - 1}}} \; ; \ \ \om^2_> - k^2_> =
-\Om^2_> \; ; \ \ v_> = d\om_>/dk_> \; . \eqno{(\mbox{A.29d})}$$

We can now easily verify that the functions $\Phi^{\ell_m}_<$ and
$\Phi^{\ell_m}_>$ below are respectively subluminal and superluminal
solutions of the HWE (see example 3 below for how to obtain these
solutions). We have

\setcounter{equation}{29}
\begin{equation}
\Phi^{\ell m}_p (t, \vec x) = C_\ell \, j_\ell (\Om_p \xi_p) \, P^\ell_m
(\cos \te) e^{im\theta} e^{i (\om_pt - k_p z)}
\end{equation}
where the index $p = < , \; > $, $C_\ell$ are constants, $j_\ell$ are
the spherical Bessel functions, $P^\ell_m$ are the Legendre functions
and $(r,\te, \var)$ are the usual spherical coordinates. $\Phi^{\ell
m}_<$ $[\Phi^{\ell m}_>]$ has phase velocity $(w_</k_<) < 1$ $[(w_>/k_>)
> 1]$ and the modulation function $j_\ell (\Om_< \xi_<)$ $[j_\ell (\Om_>
\xi_>)]$ moves with group velocity $v_<$ $[v_>]$, where $0 \leq v_< < 1$
$[1 < v_> < \infi]$. Both $\Phi^{\ell m}_<$ and $\Phi^{\ell m}_>$ are
{\it undistorted progressive waves} (UPWs). This term has been
introduced by Courant and Hilbert$^{[1]}$; however they didn't suspect
of UPWs moving with speeds greater than $c=1$. For use in the main text
we write the explicit  form of $\Phi^{00}_<$ and $\Phi^{00}_>$, which we
denote simply by $\Phi_<$ and $\Phi_>$:

\begin{equation}
\Phi_p(t,\vec x) = C \frac{\sin (\Om_p \xi_p)}{\xi_p}
e^{i(\om_pt-k_pz)}  ; \ \ p=< \; \mbox{or} \; > \; .
\end{equation}
When $v_< = 0$, we have $\Phi_< \rig \Phi_0$,

$$
\Phi_0(t,\vec x) = C {\dis\frac{\sin \Om_< r}{r}} e^{i\Om_< t}, \;  r =
(x^2+y^2+z^2)^{1/2} \; .  \eqno{(\mbox{A.32})}
$$
When $v_>=\infi$, $\om_>=0$ and $\Phi^0_> \rig \Phi_\infi$,

\setcounter{equation}{32}
\begin{equation}
\Phi_\infi(t,\vec x) = C_\infi {\dis\frac{\sinh \rho}{\rho}} e^{i\Om_>
z}  ,
\ \  \rho = (x^2+y^2)^{1/2} \; .
\end{equation}

We observe that if our interpretation of phase and group velocities is
correct, then there must be a Lorentz frame where $\Phi_<$ is at rest.
It is trivial to verify that in the coordinate chart $\langle x^{'\mu}
\rangle$ which is a (nacs/$I'$), where $I' = (1-v_{<}^{2})^{-1/2}
\partial/\partial t + (v_</\sqrt{1-v_{<}^{2}}) \partial/\partial z$ is a
Lorentz frame moving with speed $v_<$ in the $z$ direction relative to
$I = \partial/\partial t$, $\Phi_p$ goes in $\Phi_0 (t',\vec{x}')$ given
by eq.(A.32) with $t \mapsto t'$, $\vec{x} \mapsto \vec{x}'$. \\

\noindent {\it Subluminal and Superluminal Bessel Beams.} The solutions
of the HWE in cylindrical coordinates are well known$^{[19]}$. Here we
recall how these solutions are obtained in order to present new
subluminal and superluminal solutions of the HWE.\ In what follows the
cylindrical coordinate functions are denoted by $(\rho,\te, z)$, $\rho =
(x^2+y^2)^{1/2}$, $x=\rho\cos\te$, $y=\rho\sin\te$. We write for $\Phi$:
\begin{equation}
\Phi(t,\rho,\te,z)=f_1(\rho)f_2(\te)f_3(t,z) \; .
\end{equation}
Inserting (A.34) in (A.$2'$) gives

\hfill $\rho^2 {\dis\frac{d^2}{d\rho^2}} f_1 + \rho{\dis\frac{d}{d\rho}}
f_1 + (B\rho^2-\nu^2)f_1 = 0$;\hfill (A.35a)

\hfill $\left({\dis\frac{d^2}{d\theta^2}} + \nu^2\right)f_2 = 0$; \hfill
(A.35b)

\hfill $\left({\dis\frac{d^2}{dt^2}} - {\dis\frac{\pa^2}{\pa z^2}}  +
B\right)f_3=0$. \hfill (A.35c)

\noindent In these equations $B$ and $\nu$ are separation constants.
Since we want $\Phi$ to be periodic in $\te$ we choose $\nu=n$ an
integer. For $B$ we consider two cases:

\medskip 

(i) {\it Subluminal Bessel solution}, $B = \Om^2_< > 0$

\noindent In this case (A.35a) is a Bessel equation and we have
\setcounter{equation}{35}
\begin{equation}
\Phi^<_{J_n} (t,\rho,\te,z) = C_n J_n (\rho\Om_<) e^{i(k_< z - w_< t + n
\te)}, \ \ n=0,1,2,\ld ,
\end{equation}
where $C_n$ is a constant, $J_n$ is the $n$-th order Bessel function and
\begin{equation}
\omega^2_< - k^2_< = \Om^2_< \; .
\end{equation}
In$^{[43]}$ the $\Phi^<_{J_n}$ are called the $nth$-order
non-diffracting Bessel beams$^{(*)}$\footnotetext{$^{(*)}$The only
difference is that $k_<$ is denoted by $\be = \sqrt{\om^2_<-\Om^{2}_<}$
and $\om_<$ is denoted by $k'=\om/c>0$. (We use units where $c=1$). }. 

Bessel beams are examples of undistorted progressive waves (UPWs). They
are ``subluminal" waves. Indeed, the group velocity for each wave is
\begin{equation}
v_< = d\om_< / dk_< \, , \ 0 < v_< < 1 \; ,
\end{equation}
but the phase velocity of the wave is $(\om_</k_<) > 1$. That this
interpretation is correct follows from the results of the acoustic
experiment described in section 2.

It is convenient for what follows to define the variable $\eta$,
called the axicon angle$^{[26]}$,
\begin{equation}
k_< = \ov k_< \cos \eta\;  , \ \ \Om_< = \ov k_< \sin \eta \; , \ \
0 < \eta < \pi/2 \; .
\end{equation}
Then
\begin{equation}
\ov k_< = \om_< > 0
\end{equation}
and eq.(A.36) can be rewritten as $\Phi^<_{A_n} \equiv \Phi^<_{J_n}$,
with
\begin{equation}
\Phi^<_{A_n} = C_n J_n (\ov k_< \rho \sin \eta) e^{i (\ov k_< z
\cos\eta
- \omega_< t + n\te)} .
\end{equation}
In this form the solution is called in$^{[43]}$ the $n$-th order
non-diffracting portion of the  {\it Axicon Beam\/}.  The phase velocity
$v^{ph} = 1/\cos \eta$ is independent of $\ov k_<$, but, of course, it
is dependent on $k_<$.  We shall show below that waves constructed from
the $\Phi^<_{J_n}$ beams can be {\it subluminal} or {\it superluminal} !

\medskip 

(ii) {\it Superluminal (Modified) Bessel Solution}, $B = -\Om^2_> <
0$

\noindent In this case (A.35a) is the modified Bessel equation and we
denote the solutions by
\begin{equation}
\Phi^>_{K_n} (t,\rho,\te,z) = C_n K_n (\Om_> \rho) e^{i(k_> z - \omega_>
t + n\te)}, \ n=0,1, \ld ,
\end{equation}
where $K_n$ are the modified Bessel functions, $C_n$ are constants
and
\begin{equation}
\omega^2_> - k^2_> = -\Om^2_> \;.
\end{equation}
We see that $\Phi^>_{K_n}$ are also examples of UPWs, each of which has
group velocity $v_> = d\omega_> /dk_>$ such that $1 < v_> < \infi$ and
phase velocity  $0 < (\omega_>/k_>) < 1$. As in the case of the
spherical Bessel beam [eq.(A.31)] we see again that our interpretation
of phase and group velocities is correct. Indeed, for the superluminal
(modified) Bessel beam there is no Lorentz frame where the wave is
stationary.

The $\Phi^>_{K_0}$ beam was discussed by Band$^{[41]}$ in 1988 as an
example of superluminal motion. Band proposed to launch the
$\Phi^>_{K_0}$ beam in the exterior of a cylinder of \linebreak radius
$r_1$ on which there is an appropriate superficial charge density.
Since $K_0 (\Om_> r_1)$ is non singular, his solution works. In section
3 we discuss some of Band's statements.

We are now prepared to present some other very interesting solutions of
the HWE, in particular the so called $X$-waves$^{[43]}$, which are
superluminal, as proved by the acoustic experiments described in section
2.  

\medskip 

\n {\bf Theorem [{\em Lu and Greenleaf\/}]}$^{[43]}$: The three
functions below are families of exact solutions of the HWE [eq.(A.2$'$)]
in cylindrical coordinates:
\begin{eqnarray}
&& \Phi_\eta(s) = \int^\infi_0 T(\ov k_<) \left[ \frac{1}{2\pi}
\int^\pi_{-\pi} A(\phi)f(s)d\phi \right] d\ov k_< \; ; \\
&& \Phi_K(s) = \int^{\pi}_{-\pi} D(\eta) \left[ \frac{1}{2\pi}
\int^{\pi}_{-\pi} A(\phi)f(s)d\phi \right] d\eta \; ; \\
&& \Phi_L (\rho,\te,z-t) = \Phi_1(\rho,\te) \Phi_2(z-t) \; ;
\end{eqnarray}
where
\begin{equation}
s = \al_0 (\ov k_<, \eta)\rho \cos (\te-\phi) + b(\ov k_<, \eta) [z\pm
c_1(\ov k_<,\eta)t]
\end{equation}
and
\begin{equation}
c_1(\ov k_<, \eta) = \sqrt{1+[\al_0 (\ov k_<, \eta)/b (\ov
k_<,\eta)]^2} \; .
\end{equation}
In these formulas $T(\ov k_<)$ is any complex function (well behaved) of
$\ov k_<$ and could include the {\it temporal frequency transfer
function} of a radiator system, $A(\phi)$ is any complex function (well
behaved) of $\phi$ and represents a {\it weighting function} of the
integration with respect to $\phi, f(s)$ is  any complex function (well
behaved) of $s$ (solution of eq.(A.29)), $D(\eta)$ is any complex
function (well behaved) of $\eta$ and represents a weighting function of
the integration with respect to $\eta$, called the axicon angle (see
eq.(A.39)), $\al_0(\ov k_<, \eta)$ is any complex function of $\ov k_<$
and $\eta, b(\ov k_<, \eta)$ is any complex function of $\ov k_<$ and
$\eta$.

\medskip 

As in the previous solutions, we take $c = 1$. Note that $\ov k_<$,
$\eta$ and the wave vector $k_<$ of the $f(s)$ solution of eq.(A.29) are
related by eq.(A.39). Also $\Phi_2(z-t)$ is any complex function of
$(z-t)$ and $\Phi_1(\rho,\te)$ is any solution of the transverse Laplace
equation, {\em i.e.\/},
\begin{equation}
\left[\frac{1}\rho \frac{\pa}{\pa \rho} \left(\rho \frac{\pa}{\pa
\rho}\right) +
\frac{1}{\rho^2} \frac{\pa^2}{\pa\te^2} \right] \Phi_1 (\rho,\te)=0.
\end{equation}
The proof is obtained by direct substitution of $\Phi_\eta$, $\Phi_K$
and $\Phi_L$ in the HWE.\ Obviously,  the exact solution $\Phi_L$ is an
example of a luminal UPW, because if one ``travels"  with the speed $c =
1$, {\em i.e.\/}, with $z-t = {\rm const.}$, both the lateral and axial
components $\Phi_1 (\rho, \te)$ and $\Phi_2(z-t)$ will be the same for
all time $t$ and distance $z$. When $c_1(\ov k, \eta)$ in eq.(A.47) is
real, ($\pm$) represent respectively backward and forward propagating
waves.

We recall that $\Phi_\eta (s)$ and $\Phi_K(s)$ represent families of
UPWs if $c_1(\ov k_<, \eta)$ is independent of $\ov k_<$ and $\eta$
respectively. These waves travel to infinity at speed $c_1$.
$\Phi_\eta(s)$ is a generalized function that contains some of the UPWs
solutions of the HWE derived previously. In particular, if $T(\ov k_<) =
\de(\ov k_< - \ov k_<')$,  $\ov k_<' = \om > 0$ is a constant and if
$f(s)=e^s$, $\al_0(\ov k_<, \eta) = -i\Om_<$, $b(\ov k_<, \eta) =i\be =
i\om/c_1$, one obtains  Durnin's UPW beam$^{[16]}$
\begin{equation}
\Phi_{Durnin} (s) = \left[ \frac{1}{2\pi} \int^\pi_{-\pi} A(\phi)
e^{-i\Om_< \rho \cos (\te-\phi)} d\phi \right] e^{i(\be z-\om t)}.
\end{equation}
If $A(\phi) = i^n e^{in\phi}$ we obtain the $n$-th order UPW Bessel
beam  $\Phi^<_{J_n}$ given by eq.(A.36). $\Phi^<_{A_n}(s)$ is obtained
in the same way with the transformation $k_< = \ov k_< \cos \eta$;
$\Om_< = \ov k_< \sin \eta$. \\

\n {\bf The {\em X}-waves}. We now present a superluminal UPW  wave
which, as discussed in section 2, is physically realizable in an
approximate way (FAA) in the acoustic case and  can be used to generate
Hertz potentials for the electromagnetic field (see section 3). We take
in eq.(A.44):

\begin{eqnarray}
&& T(\ov k_<) = B(\ov k_<) e^{-a_0 \ov k_<}; \ \ A(\phi) = i^n
e^{in\phi};
\ \ \al_0 (\ov k_<, \eta) = -i\ov k_< \sin \eta ; \nb \\
&& b(\ov k_<, \eta) = ik \cos\eta; \ \  f(s)=e^s .
\end{eqnarray}
We then get
\begin{equation}
\Phi^>_{X_n} = e^{in\te} \int^\infi_0 B(\ov k_<) J_n (\ov k_< \rho
\sin\eta) e^{-\ov k_< [a_0 - i(z\cos\eta-t)]} d\ov k_< \, .
\end{equation}
In eq.(A.52) $B(\ov k_<)$ is any well behaved complex function of $\ov
k_<$ and represents a {\it transfer function of practical radiator},
$\ov k_< = \om$ and $a_0$ is a constant, and $\eta$ is again called the
axicon angle$^{[26]}$. Eq.(A.52) shows that $\Phi^>_{X_n}$ is
represented by a Laplace transform of the function $B(\ov k_<) J_n (\ov
k_< \rho \sin \eta)$ and an azimuthal phase term $e^{in \te}$.  The
name {\it X-waves} for the $\Phi^>_{X_n}$ comes from the fact that these
waves have an $X$-like shape in a plane containing the axis of symmetry
of the waves (the $z$-axis, see Fig.4(1) in section 3). \\ 

\noindent {\bf The $\Phi^>_{XBB_n}$ waves}. This wave is obtained from
eq.(A.44) putting $B(\ov k_<) = a_0$. It is called the $X$-wave produced
by an {\it infinite} aperture and {\it broad bandwidth}. We use in this
case the notation $\Phi^>_{XBB_n}$. Under these conditions we get

\begin{equation}
\Phi^>_{XBB_n} = \frac{a_0(\rho\sin \eta )^n
e^{in\te}}{\sqrt{M}(\tau+\sqrt{M})^n} \ \ , \ \ (n=0,1,2,\ld)
\end{equation}
where the subscript denotes ``broadband". Also
\begin{equation}
M = (\rho\sin \eta)^2+\tau^2 \; ;
\end{equation}
\begin{equation}
\tau = [a_0 - i(z\cos\eta-t)]
\end{equation}

For $n=0$ we get $\Phi^>_{XBB_0}$: 
\begin{equation} \label{ae56}
\Phi^>_{XBB_0} = \frac{a_0}{\sqrt{(\rho\sin\eta)^2 +
[a_0-i(z\cos\eta-t)]^2}} \; .
\end{equation}
It is clear that all $\Phi^>_{XBB_n}$ are UPWs which propagate with
speed $c_1 = 1/\cos\eta > 1$ in the $z$-direction. Our statement is
justified for as can be easily seen (as in the modified superluminal
Bessel beam) there is no Lorentz frame where $\Phi_{XBB_n}^{>}$ is at
rest.  Observe that this is the real speed of the wave; phase and group
velocity concepts are not applicable here. Eq(\ref{ae56}) does not give
any dispersion relation.

The $\Phi^>_{XBB_n}$ waves cannot be produced in practice as they have
infinite energy (see section A7), but as we shall show  a good
approximation for them can be realized with {\it finite aperture}
radiators. \\

\n {\bf A5. Construction of $\Phi^<_{J_n}$ and $X$-Waves with Finite
Aperture Radiators} \\

In section A3 we study the condition under which the Rayleigh-Sommerfeld
solution to HWE [eq.(A.24)] can be derived. The condition is just that
the wave $\Phi$ must be written as $\Phi(t,\vec x) = F(\vec x) e^{-i\om
t}$; which is true for the Bessel beams $\Phi^<_{J_n}$. In section 2 we
show that a finite aperture approximation (FAA) to a broad band Bessel
beam or Bessel pulse denoted FAA$\Phi_{BBJ_n}$ or $\Phi_{FAJ_n}$ [see
eq.(2.3)] can be physically realized and moves as predicted by the
theory.

At first sight it is not obvious that for the $\Phi_{X_n}$ waves  we can
use eq.(A.26), but actually we can. This happens because we can write,
\begin{eqnarray} \Phi^>_{X_n} (t, \vec x) &=& \frac{1}{2\pi}
\int^{+\infi}_{-\infi} \wid{\wid \Phi}^>_{X_n} (\om, \vec x) e^{-i\om t}
d \om  \\ \wid{\wid \Phi}^>_{X_n} (\om , \vec x) &=& 2\pi e^{in\te}
 B(\om) J_n (\om \rho\sin \eta) H(\om) e^{-\om(a_0-iz\cos\eta)},
\hs{3mm} n=0,1,2,\ld \end{eqnarray} where $H(\om)$ is the step function
and each $\wid{\wid \Phi}_{X_n} (\om, \vec x)$ is a solution of the
transverse Helmholtz  equation. Then the Rayleigh-Sommerfeld
approximation can be written and  the FAA can be used. Denoting the FAA
to $\Phi_{X_n}^>$ by $\Phi_{FAX_n}^>$ and using eq.(A.28) we get

\begin{eqnarray}
&& \Phi^>_{FAX_n} (\ov k_<, \vec x) = \frac{1}{i\la} \int^{2\pi}_0
d\te' \int^{D/2}_0 \rho' d\rho' \wid{\wid \Phi}_{X_n} (\ov k_<, \rho',
\phi')
\frac{e^{i\ov k_< R}}{R^2} z \nb \\
&& \ \ + \frac{1}{2\pi} \int^{2\pi}_0 d\te' \int^{D/2}_0 \rho'd\rho'
\wid{\wid \Phi}_{X_n} (\ov k_<,\rho', \te') \frac{e^{i\ov k_<
R}}{R^3}z \, ; \\
&& \Phi^>_{FAX_n} (t, \vec x) = \Fe^{-1} [\Phi^>_{FAX_n} (\om,\vec x)],
\ n=0,1,2,\ld,
\end{eqnarray}
where $\la$ is the wave length and $R = |\vec x - \vec x'|$.  $\Fe^{-1}$
represents the inverse Fourier transform. The first and second terms in
eq.(A.59) represent respectively the contributions from high and low
frequency components. We attached  the symbol $>$ to $\Phi^>_{FAX_n}$
meaning  as before that the wave is superluminal.  This is justified
from the results of the experiment described in section 2. \\

\n {\bf A6. The Donnelly-Ziolkowski Method (DZM) for Designing
Subluminal, Luminal
and Superluminal UPWs Solutions of the HWE and the Klein-Gordon
Equation (KGE)} \\

Consider first the HWE for $\Phi$ [eq.(A.2$'$)] in a homogeneous medium.
Let $\wid \Phi(\om,\vec k)$ be the Fourier transform of $\Phi(t,\vec
x)$, {\em i.e.\/},

\h\h $\wid \Phi(\om,\vec k) = {\dis\int_{R^3} d^3 x
\int^{+\infi}_{-\infi}} dt \, \Phi(t,\vec x) e^{-i(\vec k \vec x - \om t)}$
, \hfill (A.61a) \\

\h\h $\Phi(t,\vec x) = {\dis\frac{1}{(2\pi)^4} \int_{R^3} d^3 \vec k
\int^{+\infi}_{-\infi}} d\om \, \wid \Phi (\om,\vec k) e^{i(\vec k \vec x -
\om t)}$. \hfill (A.61b)

\noindent Inserting (A.61a) in the HWE we get
\setcounter{equation}{61}
\begin{equation}
(\om^2-\vec k^2) \wid \Phi (\om, \vec k) = 0
\end{equation}
and we are going to look for solutions of the HWE and eq.(A.62) in the
sense of distributions. We rewrite eq.(A.62) as
\begin{equation}
(\om^2 - k^2_z - \Om^2) \wid \Phi (\om, \vec k) = 0 .
\end{equation}
It is then obvious that any $\Phi(\om, \vec k)$ of the form
\begin{equation}
\wid \Phi(\omega,\vec k) = \Xi (\Om,\be) \, \de [\om - (\be+\Om^2/4\be)]
\, \de[k_z - (\be-\Om^2/4\be)] \; ,
\end{equation}
where $\Xi(\Om, \be)$ is an arbitrary weighting function,  is a
solution of eq.(A.63) since the $\de$-functions imply that
\begin{equation}
\om^2-k^2_z = \Om^2 \; .
\end{equation}

In 1985$^{[30]}$ Ziolkowski found a {\it luminal} solution of the HWE
called the Focus Wave Mode. To obtain this solution we choose, {\em
e.g.\/},
\begin{equation}
\Xi_{FWM} (\Om, \be) = \frac{\pi^2}{i\be} \exp (-\Om^2 z_0/4\be) ,
\end{equation}
whence we get, assuming $\be > 0$ and $z_0 > 0$,
\begin{equation}
\Phi_{FWM} (t,\vec x) = e^{i\be(z+t)} \frac{\exp\{-\rho^2\be/[z_0 +
i(z-t)]\}}{4\pi i [z_0+i(z-t)]} .
\end{equation}
Despite the velocities $v_1 = + 1$ and $v_2 = -1$ appearing in the
phase, the modulation function of $\Phi_{FWM}$ has very interesting
properties, as discussed in details in $^{[46]}$. It remains to observe
that eq.(A.67) is a special case of Brittingham's formula.$^{[26]}$

Returning to eq.(A.64) we see that the $\de$-functions make any function
of the Fourier transform variables $\om, k_z$ and $\Om$ to lie in a line
on the surface $\om^2 - k^2_z - \Om^2 = 0$ [eq.(A.63)]. Then, the
support of the $\de$-functions is the line

\begin{equation}
\om= \be + \Om^2/4\be ; \ k_z = \be - \Om^2/4\be\; .
\end{equation}
The projection of this line in the $(\om, k_z)$ plane is a straight line
of slope $-1$ ending at the point $(\be, \be)$. When $\be = 0$ we must
have $\Om=0$, and in this case the line is $\om = k_z$ and $\Phi(t,\vec
x)$ is simply a superposition of plane waves, each one having frequency
$\om$ and traveling with speed $c=1$ in the positive $z$ direction.

Luminal UPWs solutions can be easily constructed by the ZM$^{[46]}$,
but will not be discussed here. Instead, we now show how to use
ZM to construct subluminal and superluminal solutions of the HWE.\ \\

\n {\bf First Example:} Reconstruction of the subluminal Bessel Beams
$\Phi^<_{J_0}$ and the superluminal $\Phi^>_{XBB_0}$ ($X$-wave) \\

Starting from the ``dispersion relation" $\om^2-k^2_z - \Om^2 = 0$, we
define
\begin{equation}
\wid \Phi (\om, \ov k) = \Xi (\ov k, \eta) \de (k_z - \ov k \cos\eta)
\de (\om-\ov k).
\end{equation}
This implies that
\begin{equation}
k_z = \ov k \cos\eta; \quad \cos \eta = k_z / \om ,
\ \ \om > 0, \ \ -1 < \cos\eta < 1\, .
\end{equation}
We take moreover
\begin{equation}
\Om = \ov k \sin \eta ; \ \ \ \ov k > 0 \; .
\end{equation}

We recall that $\vec\Om = (k_x, k_y)$, $\vec\rho = (x,y)$ and we choose
$\vec \Om . \vec \rho = \Om \rho \cos \te$. Now, putting eq.(A.69) in
eq.(A.61b) we get

{\small
\begin{equation}
\Phi(t, \vec x) = \frac{1}{(2\pi)^4} \int^\infi_0 d\ov k \; \ov k \sin^2
\eta \left[\int^{2\pi}_0 d\te \, \Xi (\ov k, \eta) \, e^{i\ov k \rho
\sin \eta \cos \te} \right] e^{i(\ov k \cos\eta z - \ov k t)} .
\end{equation} 
}
Choosing
\begin{equation}
\Xi (\ov k, \eta) = (2\pi)^3 \frac{z_0 e^{-\ov k z_0 \sin\eta}}{\ov k
\sin \eta },
\end{equation}
where $z_0 > 0$ is a constant, we obtain
\[
\Phi(t, \vec x) = z_0 \sin \eta \int^\infi_0 d\ov k \, e^{-\ov k z_0 \sin
\eta} \left[\frac{1}{2\pi} \int^{2\pi}_0 d\te \, e^{i\ov k \rho
\sin\eta\cos\te} \right] e^{i \ov k (\cos\eta \, z - t)} \; .
\]
Calling $z_0 \sin\eta = a_0 > 0$, the last equation becomes

\begin{equation}
\Phi^>_{X_0}(t,\vec x) = a_0 \int^\infi_0 d\ov
k \, e^{-\ov k a_0} J_0 (\ov k
\rho \sin \eta) e^{i\ov k(\cos\eta \, z-t)} .
\end{equation}
Writing $\ov k = \ov k_<$ and taking into account eq.(A.41) we see that
\begin{equation}
J_0 (\ov k_< \rho\sin\eta) e^{i\ov k_< (z\cos\eta-t)}
\end{equation}
is a subluminal Bessel beam, a solution of the HWE moving in the
positive $z$ direction. Moreover, a comparison of eq.(A.74) with
eq.(A.52) shows that (A.74) is a particular superluminal $X$-wave, with
$B(\ov k_<)=e^{-a_0\ov k_<}$. In fact it is the $\Phi^>_{XBB_0}$ UPW
given by eq.(A.56). \\

\n {\bf Second Example:} Choosing in (A.72)

\begin{equation}
\Xi (\ov k, \eta) = (2\pi)^3 e^{-z_0|\cos\eta|\ov k} \cot \eta
\end{equation}
gives

\vs{4mm}

$\Phi^> (t,\vec x) = \cos^2 \eta {\dis\int^\infi_0} d\ov k \;\ov k
e^{-z_0|\cos\eta|\ov k} J_0(\ov k \rho\sin\eta) e^{-i\ov k(\cos\eta
z-t)}$ \hfill (A.77a)

\vs{4mm}

\hs{13.95mm} $=
{\dis\frac{[z_0-i\mbox{sgn}(\cos\eta)(z-t/\cos\eta)]}{[\rho^2
\tan^2\eta +
[z_0 + i\mbox{sgn}(\cos\eta)(z-t/\cos\eta)]^2]^{3/2}} }$. \hfill
(A.77b)

\vs{4mm}

Comparing eq.(A.77a) with eq.(A.52) we discover that the ZM produced in
this example a more general $\Phi^>_{X_0}$ wave where $B(\ov k_<) =
e^{-z_0|\cos\eta|\ov k_<}$. Obviously $\Phi^> (t,\vec x)$ given by
eq.(A.77b) moves with superluminal speed $(1/\!\cos\eta)$ in the
positive or negative $z$-direction depending on the sign of $\cos \eta$,
denoted $\mbox{sgn}(\cos \eta)$.

In both examples studied above we see that the projection of the
supporting line of eq.(A.69) in the $(\om,k_z)$ plane is the straight
line $k_z/\om = \cos\eta$, and $\cos\eta$ is its reciprocal slope. This
line is inside the ``light cone" in the $(\om,k_z)$ plane. \\

\n {\bf Third Example:} Consider two arbitrary lines with the same
reciprocal slope that we denote by $v>1$, both running between the lines
$\om=\pm k_z$ in the upper half plane $\om > 0$ and each one cutting the
$\om$-axis at  different  values $\be_1$ and $\be_2$ (Fig.(12)).  The
two lines  are projections of members of a family of HWE solution lines
and each one can be represented as a portion of the straight lines
(between the lines $\om=\pm k_z)$
\setcounter{equation}{77}
\begin{equation}
k_z = v (\om-\be_1) , \ k_z=v(\om-\be_2).
\end{equation}
It is clear that on the solution line of the HWE, $\Om$ takes values
from zero up to a maximum value that depends on $v$ and $\be$ and then
back to zero.

We see also that the  maximum value of $\Om$, given by $\be v /
\sqrt{v^2-1}$, on any HWE solution line occurs for those values of $\om$
and $k_z$ where the corresponding projection lines cut the line $\om=v
k_z$. It is clear that there are two points on any HWE solution line
with the same value of $\Om$ in the interval
\begin{equation}
0 < \Om < v \be / \sqrt{v^2-1} = \Om_0.
\end{equation}
It follows that in this case the HWE solution line breaks into two
segments, as is the case of the projection lines. We can then associate
two different weighting functions, one for each segment. We write
\begin{eqnarray}
\wid \Phi (\Om, \om, k_z) & = & \Xi_1 (\Om,v,\be) \, \de \left[k_z -
\frac{v[\be+\sqrt{\be^2 v^2 - \Om^2(v^2-1)]}}{(v^2-1)} \right] \times
\nb \\
&& \times  \de \left[ \om - \frac{[\be v^2 + \sqrt{v^2
\be^2-\Om^2(v^2-1)]}}{(v^2-1)} \right] + \nb \\
&& + \  \Xi_2 (\Om,v,\be) \, \de \left[ k_z - \frac{v[\be - \sqrt{\be^2
v^2 - \Om^2 (v^2-1)]}}{(v^2-1)} \right] \times \nb \\
&& \times \, \de \left\{ \om - \frac{[\be v^2 - \sqrt{v^2 \be^2
 - \Om^2 (v^2-1)]}}{(v^2-1)} \right\} .
\end{eqnarray}
Now, choosing
\[
\Xi_1 (\Om, v, \be) = \Xi_2 (\Om,v,\be) =
(2\pi)^3/2 \sqrt{\Om^2_0-\Om^2}
\]
we get
\[
\Phi_{v,\be} (t,\rho,z) = \Om_0 \exp \left(\frac{i\be
v(z-vt)}{\sqrt{v^2-1}} \right) \int^\infi_0 d\xis \; \xis J_0 (\Om_0
\rho
\xis) \cos \left\{ \frac{\Om_0 v}{\sqrt{v^2-1}}
\frac{(z-t/v)}{\sqrt{1-\xis^2}}\right\}.
\]
Then  
\begin{equation}
\Phi_{v,\be} (t,\rho,z) = \exp \left[ i \be \frac{v(z-vt)}{\sqrt{v^2-1}}
\right] \frac{\sin \left\{
\Om_0\sqrt{\frac{v^2}{(v^2-1)}(z-t/v)^2+\rho^2}\right\} }{ \left\{
\Om_0\sqrt{\frac{v^2}{(v^2-1)}(z-t/v)^2+\rho^2}\right\} } \, . 
\end{equation}
If we call $v_< = {\dis\frac{1}v} < 1$ and taking into account the value
of $\Om_0$ given by eq.(A.79), we can  write eq.(A.81) as
\begin{eqnarray}
&& \Phi_{v_<} (t,\rho,z) = \frac{\sin(\Om_0 \xi_<)}{\xi_<} e^{i
\Om_0(z-vt)} \; ; \nb \\
&&  \xi_< = \left[x^2+y^2 + \frac{1}{1-v^2_<} (z-v_<
t)^2\right]^{1/2} \; ;
\end{eqnarray}
which we recognize as the subluminal spherical Bessel beam of section A4
[eq.(A.31)]. \\


\begin{figure}[htb] 
\vspace{7cm} 
\caption{Projection of the support lines of the transforms
of two members of a family of subluminal solutions of the HWE.} 
\end{figure}  


\n {\bf Klein-Gordon Equation} (KGE): We show here the existence of
 subluminal, luminal and superluminal UPW solutions of the KGE.\ We
want to solve
\begin{equation}
\left(\frac{\pa^2}{\pa t^2} - \nab^2 + m^2 \right) \Phi^{KG} (t,\vec x)
= 0, \qquad m > 0,
\end{equation}
with the Fourier transform method. We obtain for $\wid
\Phi^{KG}(\om,\vec k)$  (a generalized function) the equation
\begin{equation}
\{\om^2 - k^2_z - (\Om^2+m^2)\} \wid\Phi^{KG} (\om,\vec k)=0 .
\end{equation}

As in the case of the HWE, any solution of the KGE will have a
transform $\wid \Phi (\om,\vec k)$ such that its support line lies on
the surface
\begin{equation}
\omega^2 - k^2_z - (\Om^2 + m^2) = 0 \; .
\end{equation}
From eq.(A.85), calling $\Om^2 + m^2 = K^2$, we see that we are in a
situation identical to the HWE for which we showed the existence of
subluminal, superluminal and luminal solutions. We write down as
examples one solution of each kind. \\

{\it Subluminal UPW solution of the KGE}. To obtain this solution it is
enough to change in eq.(A.81) $\Om_0 = v\be/\sqrt{v^2-1} \rig
\Om^{KG}_0 = \left[\left({\dis\frac{v\be}{\sqrt{v^2-1}}}\right)^2-m^2
\right]^{1/2}$. We have 
\begin{eqnarray}
\Phi^{KG}_< (t,\rho,z) & = & \exp \left\{\frac{i\be
v(z-vt)}{\sqrt{v^2-1}}\right\} \frac{ \sin (\Om^{KG}_0 \xi_< )}{\xi_<}
\; ;
\nb \\
\xi_< & = & \left[ x^2+y^2 + \frac{1}{1-v^2_<} (z-v_< t)^2
\right]^{1/2} , \ v_<
= 1/v .
\end{eqnarray}

{\it Luminal UPW solution of the KGE.} To obtain a solution of this type
it is enough, as in eq.(A.64), to write

\begin{equation}
\wid\Phi^{KG} = \Xi (\Om, \be) \de [k_z - (\Om^2 + (m^2-\be^2)/2\be)]
\de [\om-(\Om^2 + (m^2+\be^2)/2\be)] \; .
\end{equation}
Choosing
\begin{equation}
\Xi (\Om, \be) = \frac{(2\pi)^2}{\be} \exp (-z_0 \Om^2/2\be),  \; z_0 >
0,
\end{equation}
gives
\begin{eqnarray}
&& \!\!\!\!\!\!\!\!\!\!\!\!\!\!\!\!\!\!\!\!\!\!\!\! \Phi^{KG}_\be
(t,\vec x) = \nb \\
&& \!\!\!\!\!\!\!\!\!\!\!\!\!\!\!\!\!\!\!\!\!\!\!\! = \exp
(iz(m^2-\be^2)/2
\be) \exp (-it(m^2+\be)/2\be) \frac{\exp\{-\rho^2
\be/2[z_0-i(z-t)]\}}{[z_0
- i(z-t)]} .
\end{eqnarray}

{\it Superluminal UPW solution of the KGE}. To obtain a solution of this
kind we introduce a parameter $v$ such that $0 < v < 1$ and write for
$\wid \Phi^{KG}$ in (A.84)

\begin{eqnarray}
\wid\Phi^{KG}_{v,\be} (\om,\Om,k_z) & = & \Xi(\Om,v, \be) \,
\de \left[\om-\frac{\left(-\be v^2 + \sqrt{(\Om^2 +
m^2)(1-v^2)+v^2\be^2}\,\right)}{1-v^2} \right] \times  \nb \\
&& \times \de \left[k_z-\frac{v\left(-\be + \sqrt{(\Om^2 +
m^2)(1-v^2)+v^2\be^2}\,\right)}{1-v^2} \right] \; .
\end{eqnarray}

Next we choose

\begin{equation}
\Xi (\Om, v, \be) = \frac{(2\pi)^3
\exp(-z_0\sqrt{\Om^2_0+\Om^2})}{\sqrt{\Om^2_0 + \Om^2}},
\end{equation}
where $z_0 > 0$ is an arbitrary parameter, and where

\begin{equation}
\Om^2_0 = \frac{\be^2 v^2}{1-v^2} + m^2 \; .
\end{equation}
Then introducing $v_> = 1/v > 1$ and $\ga_> = {\dis 
\frac{1}{\sqrt{v^2_>-1}}}$, we get

\begin{equation}
\Phi^{KG_>}_{v,\be} (t,\vec x) = 
\exp \left\{\frac{i(\Om^2_0 - m^2)(z-v t)}{\beta v} \right\}
\frac{\exp\{-\Om_0 \sqrt{[z_0-i\ga_> (z-v_> t)]^2 + x^2 +
y^2}\}}{\sqrt{[z_0-i\ga_> (z-v_>t)]^2+x^2+y^2}} \; ,
\end{equation}
which is a superluminal UPW solution of the KGE moving with speed $v_>$
in the $z$ direction.  From eq.(A.93) it is an easy task to reproduce
the superluminal spherical Bessel beam which is solution of the HWE
[eq.(A.30)]. \\

\n {\bf A7. On the Energy of the UPWs Solutions of the HWE} \\

Let $\Phi_r (t,\vec x)$ be a real solution of the HWE.\ Then, as it is
well known, the energy of the solution is given by
\begin{equation}
\vare = \int\int\int_{\R^3} d\ven \left[\left(\frac{\pa\Phi_r}{\pa t}
\right)^2 - \Phi_r \nab^2 \Phi_r\right] + \lim_{R\rig\infi}
\int\int_{S(R)} dS \; \Phi_r \, \vec n . \nab \Phi_r \; ,
\end{equation}
where $S(R)$ is the 2-sphere of radius $R$ .

We can easily verify that the real or imaginary parts of all UPWs
solutions of the HWE presented above have infinite energy. The question
arises of how to project superluminal waves, solutions of the HWE, with
finite energy. This can be done if we recall that all UPWs discussed
above can be indexed by at least one parameter that here we call $\al$.
Then, calling $\Phi_\al (t,\vec x)$ the real or imaginary parts of a
given UPW solution we may form ``packets" of these solutions as
\begin{equation}
\Phi(t,\vec x) = \int d\al \, F(\al) \Phi_\al (t,\vec x)
\end{equation}

We now may test for a given solution $\Phi_\al$ and for a weighting
function $F(\al)$ if the integral in eq.(A.94) is convergent. We can
explicitly show for some (but not all) of the solutions showed above
(subluminal, luminal and superluminal) that for weighting functions
satisfying certain integrability conditions the energy $\vare$ results
finite.  It is particularly important in this context to quote that the
finite aperture approximations for all UPWs have, of course, finite
energy.  For the case in which $\Phi$ given by eq.(A.95) is used to
generate solutions for, {\em e.g.\/}, Maxwell or Dirac fields (see
Appendix B), the conditions for the energy of these fields to be finite
will in general be different from the condition that gives for $\Phi$ a
finite energy. This problem will be discussed with more details in
another paper.

\vs{1cm}

\n {\large\bf Appendix B.\ A Unified Theory for Construction of UPWs
Solutions of Maxwell, Dirac and Weyl Equations}

\renewcommand{\theequation}{B.\arabic{equation}}
\setcounter{equation}{0}

\vs{5mm}

In this appendix we briefly recall the main results concerning the
theory of Clifford algebras (and bundles) and their relationship with
the Grassmann algebras (and bundles).  Also the concept of
Dirac-Hestenes spinors and their relationship with the usual Dirac
spinors used by physicists is clarified. We introduce moreover the
concepts of the Clifford and Spin-Clifford bundles of spacetime and the
Clifford calculus. As we shall see, this formalism provides a unified
theory for the construction of UPWs subluminal, luminal and superluminal
solutions of Maxwell, Dirac and Weyl equations. More details on the
topics of this appendix can be found in$^{[6-9]}$ and$^{[81]}$.

\bigskip 

\n {\bf B1. Mathematical Preliminaries}

Let ${\cal M}=(M,g,D)$ be Minkowski spacetime. $(M,g)$ is a
four-dimensional time oriented and space oriented Lorentzian manifold,
with $M\simeq {\sl I \!\! R}^4$ and $g \in \sec(T^*M \times T^*M)$
being a Lorentzian metric of signature (1,3). $T^*M$ [$TM$] is the
cotangent [tangent] bundle.  $T^*M = \cup_{x\in M} T^*_xM$ and $TM =
\cup_{x\in M}T_xM$, and $T_xM \simeq T^*_xM \simeq {\sl I \!\!
R}^{1,3}$, where ${\sl I \!\! R}^{1,3}$ is the Minkowski vector
space$^{[60,61,62]}$. $D$ is the Levi-Civita connection of $g$, {\em
i.e.\/}, $Dg=0$, $\mbox{\boldmath $T$}(D) =0$. Also $\mbox{\boldmath
$R$}(D)=0$, $\mbox{\boldmath $T$}$ and $\mbox{\boldmath $R$}$ being
respectively the torsion and curvature tensors. Now, the Clifford bundle
of differential forms ${\cal{C}} \!\ell(M)$ is the bundle of algebras
$\clif(M) = \cup_{x\in M} \clif (T^*_xM)$, where $\forall x\in M,
\clif(T^*_xM) = \clif_{1,3}$, the so called spacetime
algebra$^{[9,81-85]}$. Locally as a linear space over the real field
${\sl I \!\! R}$, $\clif(T^*_x(M))$ is isomorphic to the Cartan algebra
$\bigwedge (T^*_xM)$ of the cotangent space  and $\bigwedge(T^*_x M) =
\sum^4_{k=0} \bigwedge {}^k(T^*_x M)$, where $\bigwedge^k(T^*_x M)$ is
the $4 \choose k$-dimensional space of $k$-forms. The Cartan bundle
$\bigwedge(M) = \cup_{x\in M} \bigwedge(T^*_x M)$ can then be thought as
``embedded" in $\clif(M)$. In this way sections of $\clif(M)$ can be
represented as a sum of inhomogeneous differential forms. Let $\{ e_\mu
= \frac{\partial}{\partial x^\mu}\} \in \sec TM$, $(\mu =
0,1,2,3)$ be an orthonormal basis $g(e_\mu, e_\nu) = \eta_{\mu\nu} =
{\rm diag}(1,-1,-1,-1)$ and let $\{ \gamma^\nu = d x^\nu \} \in {\rm
\sec} \bigwedge^1(M) \subset \sec \clif(M)$ be the dual basis.
Then, the fundamental Clifford product (denoted in what follows by
juxtaposition of symbols) is generated by $\gamma^\mu \gamma^\nu +
\gamma^\nu \gamma^\mu = 2\eta^{\mu\nu}$ and if ${\cal C} \in {\rm
\sec}\clif(M)$ we have

\begin{equation} \label{2e1}
{\cal C} = s+ v_\mu \gamma^\mu + \frac{1}{2!} b_{\mu\nu}
\gamma^\mu\gamma^\nu + \frac{1}{3!} a_{\mu\nu\rho}
\gamma^\mu \gamma^\nu \gamma^\rho + p \gamma^5 \; ,
\end{equation}
where $\gamma^5 = \gamma^0\gamma^1\gamma^2\gamma^3 = dx^0 dx^1 dx^2
dx^3$  is the volume element and $s$, $v_\mu$, $b_{\mu \nu}$, 
$a_{\mu\nu\rho}$, $p\in \sec \bigwedge^0(M) \subset \sec \clif(M)$. For
$A_r \in \sec \bigwedge^r(M)\subset \sec {\cal C}\ell (M)$, $B_s \in
\sec \bigwedge^s(M)$ we define$^{[9,82]}$ $A_r\cdot B_s = \langle A_r
B_s \rangle_{|r-s|}$ and $A_r\wedge B_s = \langle A_r B_s
\rangle_{r+s}$, where $\langle \hspace{1ex} \rangle_k$ is the component
in $\bigwedge^k(M)$ of the Clifford field.

Besides the vector bundle $\clif(M)$ we also need to introduce another
vector bundle $\clif_{{\rm Spin}_+(1,3)}(M)$ $[{\rm Spin}_+(1,3) \simeq
{\rm SL}(2,{\sl I \!\!\!\! C})]$ called the Spin-Clifford
bundle$^{[8,81,84]}$. We can show that $\clif_{{\rm Spin}_+(1,3)} (M)
\simeq \clif(M)/{\cal R}$, {\em i.e.\/} it is a quotient bundle. This
means that sections of $\clif_{{\rm Spin}_+(1,3)}(M)$ are equivalence
classes of sections of the Clifford bundle, {\em i.e.\/}, they are
equivalence sections of non-homogeneous differential forms (see
eqs.(\ref{eq.1},\ref{eq.2}) below).

Now, as is well known, an electromagnetic field is represented by $F \in
\sec \bigwedge^2(M) \subset \sec \clif(M)$. How to represent the Dirac
spinor fields in this formalism~? We can show that the even sections of
$\clif_{{\rm Spin}_+(1,3)}(M)$, called Dirac-Hestenes spinor fields, do
the job. If we fix two orthonormal basis $\Sigma = \{\gamma^\mu\}$ as
before, and $\dot{\Sigma} = \{\dot{\gamma}^\mu = R\gamma^\mu
\widetilde{R} = \Lambda^\mu_\nu \gamma^\nu \}$ with $\Lambda^\mu_\nu \in
{\rm SO}_{+}(1,3)$ and $R (x) \in {\rm Spin}_+(1,3) \; \forall x \in M$,
$R\widetilde{R} = \widetilde{R} R =1$, and where $\widetilde{}$
is the reversion operator in $\clif_{1,3}$, then$^{[8,81]}$ the
representatives of an even section $\mbox{\boldmath $\psi$} \in \sec
\clif_{{\rm Spin}_+(1,3)}(M)$ are the sections $\psi_\Sigma$ and
$\psi_{\dot{\Sigma}}$ of $\clif(M)$ related by
\begin{equation}
\label{eq.1}
\psi_{\dot\Sigma} = \psi_\Sigma R
\end{equation}
and
\begin{equation}
\label{eq.2}
\psi_\Sigma = s + \frac{1}{2!} b_{\mu\nu} \gamma^\mu \gamma^\nu + p
\gamma^5 . 
\end{equation}
Note that $\psi_{\Sigma}$ has the correct number of degrees of
freedom in order to represent a Dirac spinor field, which is not the
case with the so called Dirac-K\"ahler spinor field (see$^{[8,81]}$).

Let $\star$ be the Hodge star operator $\star :\bigwedge^k(M)
\rightarrow \bigwedge^{4-k}(M)$.  We can show that if $A_p \in \sec
\bigwedge^p(M) \subset \sec \clif(M)$ we have $\star A = \widetilde{A}
\gamma^5$.  Let $d$ and $\delta$ be respectively the differential and
Hodge codifferential operators acting on sections of $\bigwedge(M)$. If
$\omega_p \in \sec \bigwedge^p(M)\subset \sec \clif(M)$, then $\delta
\omega_p = (-1)^p \star^{-1} d \star \omega_p$, with $\star^{-1}\star =
{\rm identity}$.

The Dirac operator acting on sections of $\clif(M)$ is the invariant
first order differential operator
\begin{equation}
\dirac = \gamma^\mu D_{e_{\mu}} ,
\end{equation}
and we can show the very important result (see {\em {\em
e.g.\/}\/}$^{[6]}$):
\begin{equation}
\dirac = \dirac \wedge \,  + \, \dirac \cdot = d-\delta .
\end{equation}
With these preliminaries we can write Maxwell and Dirac equations  as
follows$^{[82,85]}$:
\begin{equation} \label{be6}
\dirac F = 0 ,
\end{equation}
\begin{equation} \label{2e7}
\dirac \psi_{\Sigma} \gamma^1\gamma^2 + m \psi_\Sigma \gamma^0 =0 .
\end{equation}
We discuss more this last equation (Dirac-Hestenes equation) in
section B.4. If $m=0$ we have the massless Dirac equation
\begin{equation}
\label{eq.10}
\dirac \psi_\Sigma = 0 ,
\end{equation}
which is Weyl's equation when $\psi_\Sigma$ is reduced to a Weyl spinor
field (see eq.(\ref{2e12n} below).  Note that in this formalism Maxwell
equations condensed in a single equation!  Also, the specification of
$\psi_\Sigma$ depends on the frame $\Sigma$. When no confusion arises we
represent $\psi_\Sigma$ simply by $\psi$.

When $\psi_\Sigma \tilde{\psi}_\Sigma \neq 0$, where $\sim$ is the
reversion operator, we can show that $\psi_\Sigma$ has the following
canonical decomposition:
\begin{equation}
\psi_\Sigma = \sqrt{\rho} \, e^{\beta\gamma_5/2} R \, ,
\end{equation}
where $\rho$, $\beta \in \sec \bwe^0 (M) \subset \sec \clif(M)$ and $R
\in \mbox{Spin}_+(1,3) \subset \clif^{+}_{1,3}$, $\forall x \in M$.
$\beta$ is called the Takabayasi angle$^{[8]}$.

If we want to work in terms of the usual spinor field formalism, we can
translate our results by choosing, for example, the standard matrix
representation of $\{\gamma^\mu\}$, and for $\psi_{\Sigma}$ given by
eq.(\ref{eq.2}) we have the following (standard) matrix
representation$^{[8,49]}$:

\begin{equation}
\Psi = \left( \begin{array}{cc}
	      \phi_1 & -\phi_2^* \\
	      \phi_2 & \phi_1^*
	      \end{array} \right) ,
\end{equation}
where
\begin{equation}
\phi_1 = \left( \begin{array}{cc}
		s - ib_{12} & b_{13}-ib_{23} \\
		-b_{13}-ib_{23} & s+ib_{12}
		\end{array} \right) , \quad
\phi_2 = \left( \begin{array}{cc}
		-b_{03}+i p & -b_{01}+ib_{02} \\
		-b_{01} - ib_{02} & b_{03}+ i p
		\end{array} \right) ,
\end{equation}
with $s, \, b_{12}, \ldots$ real functions; $*$ denotes the 
complex conjugation.  Right multiplication by
$$
\left( \begin{array}{c}
	  1 \\ 0 \\ 0 \\ 0
       \end{array} \right)
$$
gives the usual Dirac spinor field.

We need also the concept of Weyl spinors. By definition, $\psi \in \sec
\clif^+ (M)$ is a Weyl spinor if$^{[83]}$
\begin{equation} \label{2e12n}
\gamma_5 \psi = \pm \psi \gamma_{21} \; .
\end{equation}
The positive [negative] ``eigenstate'' of $\gamma_5$ will be denoted
$\psi_+$ [$\psi_-$]. For a general $\psi \in \sec \clif^+ (M)$ we can
verify that
\begin{equation} \label{be13}
\psi_\pm = \frac{1}{2} [ \psi \mp \gamma_5 \psi \gamma_{21} ]
\end{equation}
are Weyl spinors with eigenvalues $\pm 1$ of eq.(\ref{2e12n}).

We recall that the even subbundle  $\clif^+(M)$ of $\clif(M)$ is such
that its typical fiber is the Pauli algebra $\clif_{3,0} \equiv
\clif_{1,3}^{+}$ (which is isomorphic to $\C (2)$, the algebra of $2
\times 2$ complex matrices). The isomorphism $\clif_{3,0} \equiv
\clif_{1,3}^{+}$ is exhibited by putting $\sigma_i = \gamma_i \gamma_0$,
whence $\sigma_i \sigma_j + \sigma_j \sigma_i = 2 \delta_{ij}$. We
recall also$^{[8,81]}$ that the Dirac algebra is $\clif_{4,1} \equiv \C
(4)$ (see section B4) and $\clif_{4,1} \equiv \C \otimes
\clif_{1,3}$$^{[86]}$.

\bigskip 

\n {\bf B2. Inertial Reference Frames ($I$), Observers and Naturally
Adapted Coordinate Systems} \\

Let $\Me = (M,g,D)$ be Minkowski spacetime. An {\it inertial reference
frame} (irf) $I$ is a timelike vector field $I \in \sec TM$ pointing
into the future such that $g(I,I)=1$ and $DI=0$. Each integral line of
$I$ is called an inertial {\it observer}.  The coordinate functions
$\lan x^\mu\ran, \ \mu = 0,1,2,3$ of the maximal atlas of $M$ are said
to be a naturally adapted coordinate system to $I$ (nacs/$I$)  if $I =
\pa/\pa x^0$ $^{[61,62]}$. Putting $I = e_0$ we can find $e_i = \pa/\pa
x^i, i=1,2,3$ such that $g(e_\mu, e_\nu)=\eta_{\mu\nu}$ and the
coordinate functions $x^{\mu}$ are the usual Einstein-Lorentz ones and
have a precise operational meaning:
$x^0=ct^{(*)}$,\footnotetext{$^{(*)}c$ is the constant called velocity
of light in vacuum. In view of the superluminal and subluminal solutions
of Maxwell equations found in this paper we don't think the terminology
to be still satisfactory.} where $t$ is measured by ``ideal clocks" at
rest on $I$ and synchronized ``\`a la Einstein", $x^i, i=1,2,3$ are
determined with ideal rules$^{[61,62]}$. (We use units where $c=1$.)

\bigskip 

\n {\bf B3. Maxwell Theory in $\clif (M)$ and the Hertz Potential}
\\

Let $e_\mu \in \sec TM$ be an orthonormal basis, $g(e_\mu,e_\nu) =
\eta_{\mu\nu}$ and $e_\mu=\pa/\pa x^\mu$ ($\mu,\nu=0,1,2,3$), such that
$e_0$ determines an IRF.\ Let $\ga^\mu \in \sec \bwe^2(M) \sub \sec
\Ca\ell(M)$ be the dual basis and let $\ga_\mu = \eta_{\mu\nu} \ga^\nu$
be the reciprocal basis to $\ga^\mu$, {\em i.e.\/}, $\ga^\mu .  \ga_\nu
= \de^\mu_\nu$. We have $\ga^\mu = dx^\mu$.

As is well known the electromagnetic field is represented by a two-form
$F \in \sec \bwe^2(M) \sub \sec \Ca\ell(M)$. We have

\begin{equation} \label{be14}
F = \frac{1}2 F^{\mu\nu} \ga_\mu \ga_\nu, \ F^{\mu\nu} =
\left(\begin{array}{cccc}
 0  & -E^1 & -E^2 & -E^3 \\
E^1 &  0   & -B^3 & B^2 \\
E^2 & B^3  & 0    & -B^1 \\
E^3 & -B^2 & B^1  &  0
\end{array}\right) ,
\end{equation}
where $(E^1, E^2, E^3)$ and $(B^1, B^2, B^3)$ are respectively the
Cartesian components of the electric and magnetic fields.  Let $J \in
\sec \bwe^1 (M) \sub \sec \Ca\ell(M)$ be such that

\begin{equation}
J = J^\mu \ga_\mu = \rho \ga_0 + J^1 \ga_1 + J^2 \ga_2 + J^3 \ga_3 ,
\end{equation}
where $\rho$ and $(J^1, J^2, J^3)$ are respectively the Cartesian
components of the charge and of the three-dimensional current densities.

We now write Maxwell equation given by \ref{be6} in $\Ca\ell^+(M)$, the
even sub-algebra of $\Ca\ell(M)$. The typical fiber of $\Ca\ell^+(M)$,
which is a vector bundle, is isomorphic to the Pauli algebra (see
section B1). We put
\begin{equation}
\vec\sig_i = \ga_i\ga_0 , \ \ia = \vec\sig_1 \vec\sig_2 \vec\sig_3 =
\ga_0 \ga_1 \ga_2 \ga_3 = \ga_5 .
\end{equation}

Recall that $\ia$ commutes with bivectors and since $\ia^2=-1$ it acts
like the imaginary unit $i=\sqrt{-1}$ in $\Ca\ell^+(M)$. From
eq.(\ref{be14}),  we get
\begin{equation}
F = \vec E + \ia \vec B
\end{equation}
with $\vec E = E^i \vec\sig_i$, $ \vec B = B^j \vec\sig_j$, $i,j=1,2,3$.
Now, since $\pa=\ga_\mu\pa^\mu$ we get $\pa\ga_0 = \pa/\pa x^0 +
\vec\sig_i \pa^i = \pa/\pa x^0 - \nab$. Multiplying eq.(\ref{be6}) on
the right by $\ga_0$ we have
\[
\pa\ga_0 \ga_0 F \ga_0 = J \ga_0 ,
\]
\begin{equation} \label{be18}
(\pa/\pa x^0 - \nab)(-\vec E+\ia\vec B) = \rho + \vec J,
\end{equation}
where we used $\ga^0 F \ga_0 = -\vec E + \ia \vec B$ and $\vec J = J^i
\vec \sig_i$. From eq.(\ref{be18}) we have
\begin{eqnarray} 
& & - \pa_0 \vec E + \ia \pa_0 \vec B + \nab \vec E -
\ia \nab \vec B = \rho + \vec J \\ 
& & - \pa_0 \vec E + \ia \pa_0 \vec B
+ \nab .\vec E + \nab \we \vec E - \ia \nab . \vec B - \ia\nab \we \vec
B = \rho + \vec J \end{eqnarray}
We have also 
\begin{equation} \label{be19}
-\ia \nab \we \vec A \equiv \nab \times \vec A
\end{equation}
since the usual vector product between two vectors $\vec a = a^i
\vec\sig_i$, $\vec b = b^i \vec\sig_i$ can be identified  with the dual
of the bivector $\vec a \we \vec b$ through the formula $\vec a \times
\vec b = -\ia (\vec a \we \vec b)$. Observe that in this formalism $\vec
a \times \vec b$ is a true vector and not the meaningless pseudovector
of the Gibbs vector calculus. Using eq.(\ref{be19}) and equating the
terms with the same grade we obtain

\begin{equation}
\begin{array}{c}
\nab . \vec E = \rho \; ; \ \ \  \nab \times  \vec B - \pa_0 \vec E =
\vec J \; ;  \\
\\
\nab \times \vec E + \pa_0 \vec B = 0 \; ; \ \ \  \nab . \vec B = 0 \;
;
\end{array}
\end{equation}
which are Maxwell equations in the usual vector notation.

We now introduce the concept of Hertz potential$^{[19]}$ which permits
us to find  nontrivial solutions of the free ``vacuum" Maxwell equation
\begin{equation}
\pa F = 0
\end{equation}
once we know nontrivial solutions of the scalar wave equation,
\begin{equation}
\Box \Phi = (\pa^2/\pa t^2 - \nab^2) \Phi = 0 ; \ \Phi  \in \sec
\mbox{$\bwe^0$} (M) \sub \sec \Ca\ell (M) \; .
\end{equation}

Let $A \in \sec \bwe^1(M) \sub \sec \Ca\ell(M)$ be the vector potential.
We fix the Lorentz gauge, {\em i.e.\/}, $\pa. A = - \de A =0$ such that
$F = \pa A = (d-\de)A = dA$. We have the following important result: 

\medskip 

\n {\bf Theorem:} Let $\pi \in \sec \bwe^2(M) \sub \sec \Ca\ell(M)$ be
the so called Hertz potential. If $\pi$ satisfies the wave equation,
{\em i.e.\/}, $\Box\pi=\pa^2 \pi = (d-\de)(d-\de)\pi=-(d\de+\de d)\pi=0$ and
if we take $A = -\de\pi$, then $F = \pa A$ satisfies the Maxwell
equation $\pa F = 0$. 

\medskip 

The proof is trivial. Indeed, $A = -\de\pi$, implies $\de A = -\de^2
\pi=0$ and $F=\pa A = dA$. Then $\pa F =(d-\de)(d-\de)A = \de d(\de\pi)
= -\de^2 d\pi=0$, since $\de d\pi=-d\de\pi$ from $\pa^2\pi=0$.

From this result we see that if $\Phi \in \sec \bwe^0 (M) \sub \sec
\Ca\ell(M)$ satisfies $\pa^2 \Phi=0$, then we can find non trivial
solution of $\pa F = 0$, using a Hertz potential given, {\em e.g.\/}, by
\begin{equation} \label{be23}
\pi = \Phi \ga_1\ga_2 \; .
\end{equation}
In section 3 this equation is used to generate the superluminal
electromagnetic $X$-wave.

We now express the Hertz potential and its relation with the $\vec E$
and $\vec B$ fields, in order for our reader to see more familiar
formulas. We write $\pi$ as sum of electric and magnetic parts, {\em
i.e.\/},
\begin{equation}
\begin{array}{c}
\pi = \vec\pi_e + \ia \vec\pi_m \\
\\
\vec\pi_e = - \pi^{0i} \vec\sig_i , \ \vec\pi_m = -\pi^{23}
\vec\sig_1 + \pi^{13} \vec\sig_2 - \pi^{12} \vec\sig_3
\end{array}
\end{equation}
Then, since $A = \pa\pi$ we have
\begin{eqnarray}
A & = & \frac{1}2 (\pa\pi-\pi\! \stackrel{\leftarrow}{\pa}) \\
A\ga_0 & = & -\pa_0 \vec\pi_e + \nab . \vec \pi_e - (\nab \times
\vec\pi_m)
\end{eqnarray}
and since $A = A^\mu \ga_\mu$ we also have
\[
A^0 = \nab . \vec\pi_e \; ; \qquad \vec A = A^i \vec\sig_i =
-\frac{\pa}{\pa x^0} \vec\pi_e - \nab \times \vec\pi_m \; .
\]
Since $\vec E = -\nab A^0 - {\dis\frac{\pa}{\pa x^0}} \vec A$ and 
$\vec B = \nab \times \vec A$, we obtain
\begin{eqnarray}
&& \vec E = - \pa_0(\nab \times \vec \pi_m) + \nab \times \nab \times
\vec \pi_e \; ; \\
&& \vec B = \nab \times (-\pa_0 \vec\pi_e - \nab \times \vec \pi_m) =
-\pa_0(\nab \times \vec\pi_e) - \nab \times \nab \times \vec\pi_m \; .
\end{eqnarray}

We define $\vec E_e, \vec B_e, \vec E_m, \vec B_m$ by 

\begin{equation}
\begin{array}{l}
\vec E_e = \nab \times \nab \times \vec\pi_e \ \ ; \ \ \vec B_e =
-\pa_0(\nab \times \vec \pi_e) \ \ ; \\
\vec E_m = -\pa_0(\nab \times \vec\pi_m) \ \ ; \ \ \vec B_m = -\nab
\times
\nab  \times \vec \pi_m \ \ .
\end{array}
\end{equation}

We now introduce the 1-forms of stress-energy. Since $\pa F=0$ we have
$\wid F \wid \pa = 0$. Multiplying the first of these equation on the
left by $\wid F$ and the second on the right by $\wid F$ and summing we
have:
\begin{equation}
(1/2) (\wid F \pa F + \wid F \wid \pa F) = \pa_\mu ((1/2) \wid F \ga^\mu
 F )=  \pa_\mu T^\mu =  0 ,
\end{equation}
where $\wid F \wid\pa \equiv - (\pa_\mu \frac{1}2 F_{\al\be} \ga^\al
\ga^\be) \ga^\mu$. Now,
\begin{equation}
-\frac{1}2 (F\ga^\mu F)\ga^\nu = -\frac{1}2 ( F\ga^\mu
F\ga^\nu ) 
\end{equation}

Since $\ga^\mu . F = {\dis\frac{1}2} (\ga^\mu F - F \ga^\mu) = F .
\ga^\mu$, we have
\begin{eqnarray}
T^{\mu\nu} & = & -\lan(F.\ga^\mu)F\ga^\nu\ran_0 - \frac{1}{2} \lan
\ga^\mu F^2 \ga^\nu\ran_0  \nb \\
& = & -(F.\ga^\mu) . (F.\ga^\nu) - \frac{1}2 (F.\ F)\ga^\mu .
\ga^\nu \\
& = & F^{\mu\alpha} F^{\la\nu} \eta_{\al\la} + \frac{1}4 \eta^{\mu\nu}
F_{\al\be} F^{\al\be}, \nb
\end{eqnarray}
which we recognize as the stress-energy momentum tensor of the
electromagnetic field, and $T^\mu = T^{\mu\nu}\ga_\nu$.

By writing $F = \vec E + \ia\vec B$ as before we can immediately
verify that
\begin{eqnarray}
T_0 & = & -\frac{1}2 F \ga_0 F \nb \\
& = & \left[\frac{1}2 (\vec E^2 + \vec B^2) + (\vec E \times \vec
B)\right] \ga_0.
\end{eqnarray}
We have already shown that $\pa_\mu T^\mu = 0$, and we can easily show
that
\begin{equation}
\pa . T^\mu = 0 . 
\end{equation}

We now define the density of {\it angular momentum}. Choose as before a
Lorentzian chart
$\lan x^\mu\ran$ of the maximal atlas of $M$ and consider the 1-form $x
=
x^\mu \ga_\mu = x_\mu \ga^\mu$. Define
\[
M_\mu = x \we T_\mu = \frac{1}2 (x_\al T_{\mu\nu} - x_\nu
T_{\al\mu}) \ga^\al \we \ga^\nu \; .
\]
It is trivial to verify that as $T_{\mu\nu} = T_{\nu\mu}$ and $\pa_\mu
T^{\mu\nu} = 0$, it holds

\begin{equation}
\pa^\mu M_\mu = 0.
\end{equation}

The {\it invariants} of the electromagnetic field $F$ are $F.\ F$ and $F
\we
F$ and
\[
F^2 = F.\ F + F\we F ;
\]
\begin{equation}
F.\ F = -\frac{1}2 F^{\mu\nu} F_{\mu\nu} \; ; \qquad F \we F = -\ga_5
F^{\mu\nu} F^{\al\be} \vare_{\mu\nu\al\be} \; .
\end{equation}
Writing as before $F = \vec E + \ia \vec B$ we have

\begin{equation}
F^2 = (\vec E^2 - \vec B^2) + 2\ia \vec E . \vec B = F.\ F + F \we F.
\end{equation}

\bigskip 

{\bf B4. Dirac Theory in $\Ca\ell(M)$} \\

Let $\Sig = \{\ga^\mu\} \in \sec \mbox{$\bwe^1$} (M) \sub \sec
\Ca\ell(M)$ be an orthonormal basis. Let $\psi_\Sig \in \sec (\bwe^0
(M)+\bwe^2(M) + \bwe^4(M)) \sub \sec \Ca\ell(M)$ be the representative
of a Dirac-Hestenes Spinor field in the basis $\Sig$. Then, the
representative of Dirac equation in $\Ca\ell(M)$ is the following
equation ($\hb=c=1$):

\begin{equation} \label{be33}
\pa\psi_\Sig \ga_1\ga_2 + m \psi_\Sig \ga_0 = 0 \; .
\end{equation}
The proof is as follows:

Consider the {\it complexification\/} $\Ca\ell_C(M)$ of $\Ca\ell(M)$
called the {\it complex Clifford bundle\/}. Then $\Ca\ell_C(M) = \C \ot
\Ca\ell(M)$ and by the results of section B1 it is trivial to see that
the typical fiber of $\Ca\ell_C(M)$ is $\Ca\ell_{4,1}=\C \ot
\Ca\ell_{1,3}$, the Dirac algebra. Now let $\{\Ga_0, \Ga_1, \Ga_2,
\Ga_3, \Ga_4\} \sub \sec \bwe^1 (M) \sub \sec \Ca\ell_C(M)$ be an
orthonormal basis with
\begin{eqnarray}
&& \Ga_a \Ga_b + \Ga_b \Ga_a = 2 g_{ab} \; , \\
&& g_{ab} = diag(+1,+1,+1,+1,-1) \; . \nb
\end{eqnarray}

Let us identify $\ga_\mu = \Ga_\mu \Ga_4$ and call $I = \Ga_0
\Ga_1 \Ga_2 \Ga_3 \Ga_4$. Since $I^2 = -1$ and $I$ commutes with all
elements of $\Ca\ell_{4,1}$ we identify $I$ with $i=\sqrt{-1}$ and
$\ga_\mu$ with the fundamental set of $\Ca\ell(M)$. Then if $\A \in
\sec \Ca\ell_C(M)$ we have

\begin{equation}
\A = \Phi_s + A^\mu_C \ga_\mu + \frac{1}2 B^{\mu\nu}_C \ga_\mu
\ga_\nu + \frac{1}{3!} \tau^{\mu\nu\rho}_C \ga_\mu \ga_\nu
\ga_\nu+\Phi_p \ga_5 ,
\end{equation}
where $\Phi_s$, $\Phi_p$, $A^\mu_C$, $B^{\mu\nu}_C$,
$\tau^{\mu\nu\rho}_C \in \sec \C \ot \bwe^0(M) \sub \sec \Ca\ell_C(M)$,
{\em i.e.\/}, $\for x \in M$, $\Phi_s(x)$, $\Phi_p(x)$, $A^\mu_C(x)$,
$B^{\mu\nu}_C (x)$, $\tau^{\mu\nu\rho}_C(x)$ are complex numbers.

Now,
\[
f = \frac{1}2 (1+\ga_0) \frac{1}2 (1+i \ga_1 \ga_2)\, ; \quad f^2 = f\,
, 
\] 
is a primitive idempotent field of $\Ca\ell_C(M)$. We can show that
$if = \ga_2 \ga_1 f$. From (\ref{be33}) we can write the following
equation in $\Ca\ell_C(M)$: 
\begin{eqnarray} 
&& \pa \psi_\Sig \ga_1 \ga_2 f + m \psi_\Sig \ga_0 f = 0 \\ 
&& \pa \psi_\Sig i f - m \psi_\Sig f = 0 
\end{eqnarray} 
and we have the following equation for
$\Psi=\psi_{\Sig} f$: 
\begin{equation} \label{be35} 
i \pa \Psi - m \Psi = 0.  
\end{equation}

Using for $\ga_\mu$ the standard matrix representation (denoted here by
$\und\ga_\mu$) we get that the matrix representation of eq.(\ref{be35})
is
\begin{equation}
i \und \ga^\mu \pa_\mu |\Psi\ran - m |\Psi\ran = 0
\end{equation}
where now $|\Phi\ran$ is a usual Dirac spinor field.

We now define a {\it potential} for the Dirac-Hestenes field
$\psi_\Sig$. Since $\psi_\Sig \in \sec \Ca\ell^+(M)$ it is clear that
there exist $A$ and $B \in \sec \bwe^1(M) \sub \sec\Ca\ell(M)$ such
that
\begin{equation}
\psi_\Sig = \pa(A + \ga_5 B),
\end{equation}
since
\begin{eqnarray}
\pa(A+\ga_5B) & = & \pa . A + \pa\we A - \ga_5 \pa . B - \ga_5 \pa
\we B \\
& = & S + B + \ga_5 P ;
\end{eqnarray}
\[
S = \pa.A; \quad B = \pa \we A - \ga_5 \pa \we B; \quad P = - \pa. B 
.
\]
We see that when $m=0$, $\psi_\Sig$ satisfies the {\it Weyl equation}
\begin{equation} \label{be38}
\pa\psi_\Sig = 0 \; .
\end{equation}
Using eq.(\ref{be38}) we see that
\begin{equation}
\pa^2 A = \pa^2 B = 0.
\end{equation}
This last equation allows us to find UPWs solutions for the Weyl
equation once we know UPWs solutions of the scalar wave equation $\Box
\Phi = 0$, $\Phi \in \sec \bwe^0(M) \sub \sec \Ca\ell(M)$. Indeed it is
enough to put $\A=(A+\ga_5 B) = \Phi(1 + \ga_5)v$, where $v$ is a
constant 1-form field. This result has been be used in$^{[48]}$ to
present subluminal and superluminal solutions of the Weyl equation.

We know (see appendix A, section A5) that the Klein-Gordon equation have
superluminal solutions. Let $\Phi_>$ be a superluminal solution of $\Box
\Phi_> + m^2 \Phi_> = 0$. Suppose $\Phi_>$ is a section of
$\Ca\ell_C(M)$. Then in $\Ca\ell_C(M)$ we have the following
factorization:
\begin{equation} \label{be40}
(\Box + m^2) \Phi = (\pa+i m)(\pa-i m) \Phi=0.
\end{equation}
Now
\begin{equation} \label{be41}
\Psi_> = (\pa-i m) \Phi_> f
\end{equation}
is a Dirac spinor field in $\Ca\ell_C(M)$, since
\begin{equation}
(\pa+i m) \Psi_> = 0
\end{equation}
If we use for $\Phi$ in eq.(\ref{be40}) a subluminal or a luminal UPW
solution and then use eq.(\ref{be41}) we see that Dirac equation also
has UPWs solutions with arbitrary speed $0 \leq v < \infty$.

\vspace{1cm}

\cent{\large\bf References}

\begin{description}
\item{1.} R.\ Courant and D.\ Hilbert, {\it Methods of Mathematical
Physics}, vol. II, pp 760, John Wiley and Sons, New York, 1966.

\item{2.} T.\ Waite, ``The Relativistic Helmholtz Theorem and
Solitons,''
Phys. Essays {\bf 8}, 60-70 (1995).

\item{3.} T.\ Waite, A.\ O.Barut and J.\ R.\ Zeni, ``The Purely
Electromagnetic Electron Re-visited,'' in publ. in J.\ Dowling (ed.)
{\it Electron Theory and Quantum Electrodynamics}. Nato Asi Series
Volume, Plenum Press, London (1995).

\item{4.} A.\ M.\ Shaarawi, ``An Electromagnetic Charge-Current Basis for
the de Broglie Double Solution,'' preprint Dep. Eng. Phys. and Math.,
Cairo Univ., Egypt (1995).

\item{5.} J.-Y.\ Lu and J.\ F.\ Greenleaf, ``Limited Diffraction Solutions
to
Maxwell and Schr\"odinger Equations,''
preprint
Biodynamics Res. Unity, Mayo Clinic and Foundation, Rochester (1995);
subm. for publication in J.\ de Physique.

\item{6.} W.\ A.\ Rodrigues Jr and Q.\ A.G.\ de Souza, ``The Clifford Bundle
and the Nature of the Gravitational Field,'' Found. of Phys. {\bf
23}, 1465-1490 (1993).

\item{7.}  Q.\ A.G.\ de Souza  and W.\ A.\ Rodrigues Jr., ``The Dirac
Operator and
the Structure of Riemann-Cartan-Weyl Spaces,'' in P.\ Letelier and
W.\ A.Rodrigues Jr (eds.) {\it Gravitation: The Spacetime Structure},
pp.
179-212. World Sci. Publ. Co., Singapore (1994).

\item{8.} W.\ A.\ Rodrigues Jr., Q.\ A.G.\ de Souza, J.\ Vaz Jr and P.
Lounesto, ``Dirac Hestenes Spinor Fields in Riemann-Cartan Spacetime'',
 Int. J.\ of Theor. Phys. {\bf 35}, 1849-1900 (1996).

\item{9.} D.\ Hestenes and G.\ Sobczyk, {\it Clifford Algebra to
Geometric Calculus}, D.\ Reidel Publ. Co., Dordrecht (1984).

\item{10.} B.\ Jancewicz, ``{\it Multivectors and Clifford Algebras in
Electrodynamics}", World Sci. Publ. Co., Singapore (1988).

\item{11.} H.\ Bateman, {\it Electrical and Optical Motion}, Cambridge
Univ. Press, Cambridge (1915).

\item{12.} L.\ Mackinnon, ``A Non-Dispersive De~Broglie Wave Packet,''
Found. Phys. {\bf 8}, 157--170 (1978).

\item{13.} Ph. Gueret and J.\ P.\ Vigier, ``De Broglie Wave Particle
Duality
in the Stochastic Interpretation of Quantum Mechanics: A Testable
Physical
Assumption,'' Found.\ Phys.\ {\bf 12}, 1057--1083 (1982);
{\bf 38}, 125 (1983).

\item{14.} A.\ O.\ Barut, ``$E=\hbar\om$", Phys. Lett. {\bf A143}, 349-352
(1990).

\item{15.} A.\ O.\ Barut and J.\ Bracken, ``Particle-like Configurations
of the Electromagnetic Field: An Extension of de Broglie's Ideas,''
Found. Phys. {\bf 22}, 1267--1289 (1992).

\item{16.} J.\ Durnin, ``Exact Solutions for Non Diffracting Beams I.
The Scalar Theory,'' J.\ Opt. Soc. Am. {\bf 4}, 651-654 (1987).

\item{17.} J.\ Durnin, J.\ J.\ Miceli Jr.\ and J.\ H.\ Eberly,
``Diffraction-Free Beams,'' Phys. Rev. Lett. {\bf 58}, 1499-1501
(1987).

\item{18.} J.\ Durnin, J.\ J.\ Miceli Jr.\ and J.\ H.\ Eberly, ``Experiments
with Non Diffracting Needle Beams," Opt. Soc. Am., Washington, DC,
available from IEEE Service Center (list. no. 87CH2391-1), Piscataway,
NJ., pp. 208 (1987).

\item{19.} J.\ A.\ Stratton, {\it Electromagnetic Theory}, Mc Graw-Hill
Book Co., New York and London, (1941).

\item{20.} D.\ K.\ Hsu, F.\ J.\ Margeton and D.\ O.\ Thompson, ``Bessel Beam
Ultrasonic Transducers: Fabrication Method and Experimental Results,''
Appl. Phys. Lett {\bf 55}, 2066-2068 (1989).

\item{21.} J.-Y.\ Lu and J.\ F.\ Greenleaf, ``Ultrasonic Nondiffracting
transducer for Medical Imaging,'' IEEE Trans. Ultrason. Ferroelec.
Freq. Contr. {\bf 37}, 438-477 (1990).

\item{22.} J.-Y.\ Lu and J.\ F.\ Greenleaf, ``Pulse-echo Imaging Using a
Nondiffracting Beam Transducer,'' Ultrasound Med. Biol. {\bf 17},
265-281 (1991).

\item{23.} J.-Y.\ Lu and J.\ F.\ Greenleaf, ``Simulation of Imaging
Contrast
of Non Diffracting Beam Transducer,'' J.\ Ultrasound Med. {\bf 10}, 54
 (1991) (Abstract).

\item{24.} J.\ A.\ Campbell and S.\ Soloway. ``Generation of a Non
Diffracting Beam with Frequency Independent Beam Width", J.\ Acoust.
Soc. Am. {\bf 88}, 2467-2477 (1990).

\item{25.} M.\ S.\ Patterson and F.\ S.\ Foster, ``Acoustic Fields of
Conical Radiators", IEEE Trans. Sonics Ultrason. {\bf SU-29}
no. 2, 83-92 (1982).

\item{26.} J.-Y.\ Lu, Z.\ Hehong and J.\ F.\ Greenleaf,  ``Biomedical
Ultrasound Beam Forming", Ultrasound in Med. \& Biol. {\bf 20},
403-428 (1994).

\item{27.} J.\ N.\ Brittingham, ``Focus Waves Modes in Homogeneous
Maxwell's Equations: Transverse Electric Mode", J.\ Appl. Phys. {\bf
54},
1179 (1983).

\item{28.} P.\ A.\ B\'elanger, ``Packet-like Solutions of the Homogeneous
Wave Equation" J.\ Opt. Soc. Am {\bf A1}, 723-724 (1986).

\item{29.} A.\ Sezginer, ``A General Formulation of Focused Wave
Modes", J.\ Opt.  Soc. Am. {\bf A1}, 723-724(1984).

\item{30.} R.\ W.\ Ziolkowski, ``Exact Solutions of the Wave Equation
with Complex Source Locations", J.\ Math. Phys. {\bf 26}, 861-863
(1985).

\item{31.} A.\ M.\ Shaarawi, I.\ M.\ Besieris and R.\ Ziolkowski, ``Localized
Energy Pulse Trains Launched from an Open, Semi-Infinite Circular
Waveguide,'' J.\ Appl. Phys. {\bf 62}, 805 (1988).

\item{32.} I.\ M.\ Besieris, A.\ M Shaarawi and R.\ W.\ Ziolkowski, ``A
Bidirectional Traveling Plane Wave Representation of Exact Solutions
of the Scalar Wave Equation,'' J.\ Math. Phys. {\bf 30}, 1254 (1989).

\item{33.} R.\ W.\ Ziolkowski, ``Localized Transmission of Electromagnetic
Energy", Phys. Rev. {\bf A39}, 2005-2033 (1989).

\item{34.} R.\ W.\ Ziolkowski, D.\ K.\ Lewis, and B.\ D.\ Cook, ``Experimental
Verification of the Localized Wave Transmission Effect", Phys. Rev.
Lett. {\bf 62}, 147 (1989).

\item{35.} A.\ M.\ Shaarawi, I.\ M.\ Besieris and R.\ W.\ Ziolkowski, ``{A}
Novel  Approach to the Synthesis of Non-Dispersive Wave Packets
Solutions to the Klein-Gordon and Dirac Equations", J.\ Math. Phys.
{\bf 31}, 2511-2519 (1996).

\item{36.} P.\ Hillion, ``More on Focus Wave Modes in Maxwell
Equations", J.\ Appl. Phys. {\bf 60}, 2981-2982 (1986).
\item{37.} P.\ Hillion, ``Spinor Focus Wave Modes", J.\ Math. Phys. {\bf
28}, 1743-1748 (1987).

\item{38.} P.\ Hillion, ``Nonhomogeneous Nondispersive Electromagnetic
Waves", Phys. Rev. {\bf A45}, 2622-2627 (1992).

\item{39.} P.\ Hillion, ``Relativistic Theory of Scalar and Vector
Diffraction by Planar Aperture", J.\ Opt. Soc. Am. {\bf A9}, 1794-1880
(1992).

\item{40.} E.\ C.\ de Oliveira, ``On the Solutions of the Homogeneous
Generalized Wave Equation,'' J.\ Math. Phys. {\bf 33}, 3757-3758,
(1992).

\item{41.} W.\ Band, ``Can Information Be Transfered Faster Than
Light? I.\ A Gedanken Device for Generating Electromagnetic Wave
Packets with Superoptic Group Velocity,'' Found Phys. {\bf 18}, 549-562
(1988).

\item{42.} W.\ Band, Can Information Be Transfered Faster than Light?
II.\ The Relativistic Doppler Effect on Electromagnetic Wave Packets
with Suboptic and Superoptic Group Velocities, Found Phys. {\bf 18},
625-638 (1988).

\item{43.} J.-Y.\ Lu and J.\ F.\ Greenleaf, ``Nondiffracting $X$-Waves -
Exact Solutions to Free-Space Scalar Wave Equation  and Their Finite
Aperture Realizations", IEEE Transact. Ultrason. Ferroelec. Freq.
Contr. {\bf 39}, 19-31 (1992).

\item{44.} J.-Y.\ Lu and J.\ F.\ Greenleaf, ``Experimental Verification of
Nondiffracting $X$-Wave", IEEE Trans. Ultrason. Ferroelec. Freq.
Contr. {\bf 39}, 441-446, (1992).

\item{45.} R.\ Donnelly and R.\ Ziolkowski, ``{A} Method for
Constructing Solutions of Homogeneous Partial Differential Equations:
Localized Waves,'' Proc. R.\ Soc. London {\bf A437}, 673-692 (1992).

\item{46.} R.\ Donnelly and R.\ Ziolkowski, ``Designing Localized
Waves", Proc. R.\ Soc. London {\bf A460}, 541-565 (1993).

\item{47.} A.\ O.\ Barut and H.\ C.\ Chandola ``Localized Tachyonic Wavelet
Solutions of the Wave Equation", Phys. Lett. {\bf A180}, 5-8 (1993).

\item{48.} W.\ A.\ Rodrigues Jr and J.\ Vaz Jr, ``Subluminal and
Superluminal Solutions in Vacuum of the Maxwell Equations and the
Massless Dirac Equation", Advances in
Appl. Clifford Algebras {\bf 7} (S), 457-466 (1997). 

\item{49.} J.\ Vaz Jr and W.\ A.\ Rodrigues Jr, ``On the Equivalence of
Maxwell and Dirac Equations, and Quantum Mechanics",  Int. J.\ Theor.
Phys.
{\bf 32}, 945-958 (1993).

\item{50.} J.\ Vaz Jr.\ and W.\ A.\ Rodrigues Jr ``Maxwell and Dirac
Theories as an  Already Unified Theory'', 
Advances in Appl. Clifford Algebras {\bf 7} (S), 369-386 (1997).

\item{51.}  A.\ M.\ Steinberg, P.\ G.\ Kwiat and R.\ Y.\ Chiao,``Measurement
of the Single Photon Tunneling Time", Phys. Rev. Lett. {\bf 71},
708-711 (1993).

\item{52.} A.\ Enders and G.\ Nimtz, ``Photonic Tunneling Experiments,''
Phys. Rev. {\bf B47}, 9605-9609 (1993).

\item{53.} W.\ Heitman and G.\ Nimtz, ``On Causality Proofs of
Superluminal Barrier Traversal of Frequency Band Limited Wave
Packets,''
Phys. Lett. {\bf A196}, 154-158 (1994).

\item{54.} A.\ V.\ Oppenheim and R.\ W.\ Schafer, {\it Digital Signal
Processing}, chap. 5, Englewood Cliffs, Prentice-Hall, Inc. N.\ J.,
(1975).

\item{55.} A.\ Einstein, Sitzungsberichte der Preussischen Akad. D.
Wissenschaften (1919), as translated in H.\ A.\ Lorentz, A.\ Einstein,
H.\ Minkowski and H.\ Weyl, {\it The Principle of Relativity}, p. 142,
Dover,
N.\ Y., 1952.

\item{56.} H.\ Poincar\'e, ``Sur la Dynamique de L'electron".
R.\ C.\ Circ. Mat. Palermo {\bf 21}, 129-175 (1906).

\item{57.} P.\ Ehrenfest, ``Die Translation Deformierbarer Elektron und
der Fl\"achensatz,'' Ann. Phys., Lpz {\bf 23}, 204-205 (1907).

\item{58.} D.\ Reed, ``Archetypal Vortex Topology in Nature", Spec.
Sci. and Tech. {\bf 17}, 205-224 (1994).

\item{59.} M.\ W.\ Evans, ``Classical Relativistic Theory of the
Longitudinal Ghost Fields in Electromagnetism", Found. Phys. {\bf 24},
1671-1688 (1994).

\item{60.} R.\ K.\ Sachs and H.\ Wu, {\it General Relativity for
Mathematicians}, Springer, New
York (1977).

\item{61} W.\ A.\ Rodrigues Jr and M.\ A.F.\ Rosa, ``The Meaning of Time in
Relativity and Einsteins's Later View of the Twin Paradox", Found.
Phys. {\bf 19}, 705-724 (1989).

\item{62.} W.\ A.\ Rodrigues, M.\ E.F.\ Scanavini and L.\ P.\ de Alc\^antara,
``Formal Structures, The Concepts of Covariance, Invariance, Equivalent
Reference Frames and the Principle of Relativity,'' Found. Phys.
Lett. {\bf 3}, 59-79 (1990).

\item{63.} N.\ Bourbaki, {\it Th\'eorie des Ensembles}, chap. 4,
Hermann,
Paris (1957).

\item{64.} H.\ Reichenbach, {\it The Philosophy of Space and Time},
Dover,
New York, 1958.

\item{65.} E.\ Recami, {\it Classical Tachyons and Applications},
Rivista N.\ Cimento {\bf 9}, 1-178 (1986).

\item{66.} W.\ A.\ Rodrigues Jr., Q.\ A.G.\ de Souza and Y.\ Bozhkov, ``The
Mathematical Structure of Newtonian Spacetime: Classical Dynamics and
Gravitation", Found. Phys. {\bf 25}, 871-924 (1995).

\item{67.} W.\ A.\ Rodrigues and J.\ Tiomno, ``On Experiments to Detect
Possible Failures of Relativity Theory", Found. Phys. {\bf 15},
995-961 (1985).

\item{68.} R.\ M.\ Santilli, ``Lie Isotopic Lifting of Special
Relativity for Extended Particles'', Lett. N.\ Cimento {\bf 37},
545-555 (1983).

\item{69.} R.\ M.\ Santilli, ``Nonlinear, nonlocal and noncanonical
isotopies of the Poincar\'e symmetry'', J.\ Moscow Phys. Soc.
{\bf 3}, 255-280 (1993).

\item{70.} R.\ M.\ Santilli, ``Limitations of the Special and
General Relativities and their Isotopic Generalizations'',
Chinese J.\ of Syst. Eng. \& Electr. {\bf 6}, 157-176 (1995).

\item{71.} R.\ M.\ Santilli, {\em in\/} T.\ L.\ Giulli (ed.), {\em New
Frontiers in Hadronic Mechanics\/}, Hadronic Press (1996) (in
publication).

\item{72.} R.\ M.\ Santilli, {\em Elements of Hadronic
Mechanics\/}, vols. I and II (second ed.), Naukora Dumka Publ.,
Ukraine Acad. Sci., Kiev (1995).

\item{73.} R.\ M.\ Santilli, ``Isospecial Relativity with
Applications to Quantum Gravity, Antigravity and Cosmology'',
Balkan Geom. Press, Budapest (in press).

\item{74} G.\ Nimtz, ``New Knowledge of Tunneling from Photonic
Experiments'', to appear in {\it Proc. of the Adriatico Research
Conference: Tunneling and its Implications\/}, 07/30--08/02, 1996, Word
Scientific.

\item{75.} E.\ W.\ Otten, ``Squeezing the Neutrino Mass with New
Instruments",  Nucl. Phys. News {\bf 5}, 11-16 (1995).

\item{76.} E.\ Gianetto {\em et al\/}, ``Are Neutrinos Faster than Light
Particles~?", Phys. Lett. {\bf B178}, 115-118 (1986).

\item{77.} J.\ R.\ Primack, J.\ Holtzman, A.\ Klypin and D.\ O.
Caldwell, ``Cold+hot dark matter cosmology with $m(\nu_\mu)
\simeq m(\nu_\tau) \simeq 2.4 \,\mbox{eV}$'', Phys. Rev. Lett.
{\bf 74}, 2160 (1995).

\item{78.} E.\ T.\ Whittaker, {\it A History of the Theories of Aether and
Electricity\/}, vols. I and II, Humanities Press, NY, 1973.

\item{79.} P.\ M.\ Morse and H.\ Feshbach, {\it Methods of Theoretical
Physics},
Vols. I and II, Mc Graw-Hill Book Co. Inc., New York, (1953).

\item{80.} W.\ A.\ Rodrigues Jr and J.\ E.\ Maiorino, ``A Unified Theory for
Construction of Arbitrary Speeds ($0 \leq v \leq \infty$) Solutions of
the Relativistic Wave Equations'', Random Operators and Stochastic
Equations {\bf 4}, 355-400 (1996).

\item{81.} W.\ A.\ Rodrigues Jr, Q.\ A.G.\ de Souza and J.\ Vaz Jr,
``Spinor Fields and Superfields as Equivalence Classes of Exterior
Algebra
Fields", in R.\ Ablamowicz and P.\ Lounesto (eds.) {\it Clifford
Algebras and Spinor Structures}, pp. 177-198,  Kluwer Acad. Publ.,
Dordrecht (1995).

\item{82.} D.\ Hestenes, {\em Spacetime Algebra}, Gordon and
Breach Sci. Pub., New York (1969).

\item{83.} P.\ Lounesto, ``Clifford Algebras and Hestenes Spinors'',
Found. Phys. {\bf 23}, 1203-1237 (1993).

\item{84.} W.\ A.\ Rodrigues Jr.\ and V.\ L.\ Figueiredo, ``Real
Spin-Clifford Bundles and the Spinor Structure of Spacetime'',
Int. J.\ Theor. Phys. {\bf 29}, 413-424 (1990).

\item{85.} W.\ A.\ Rodrigues Jr.\ and E.\ C.\ de Oliveira, ``Dirac and
Maxwell Equations in the Clifford and Spin-Clifford Bundles'',
Int. J.\ Theor. Phys. {\bf 29}, 397-412 (1990).

\item{86.} I.\ Porteous, {\em Topological Geometry}, van Nostrand,
London (1969).

\end{description}

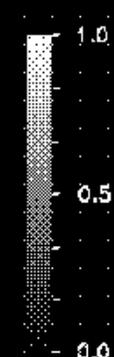
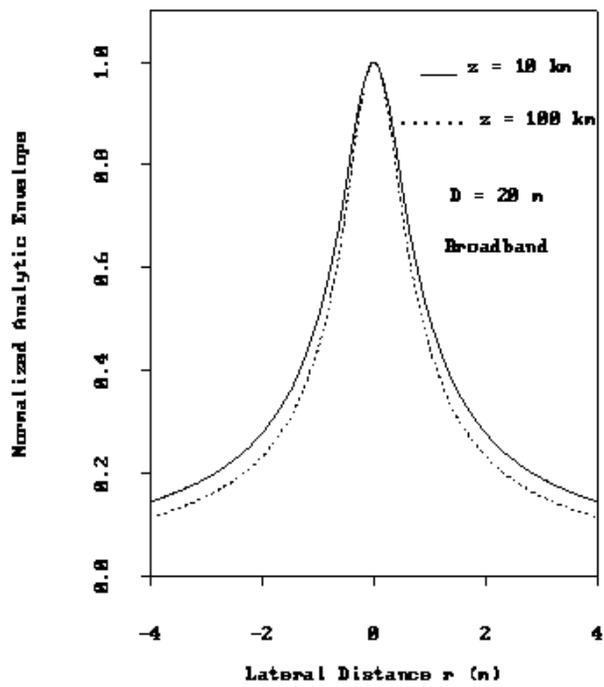
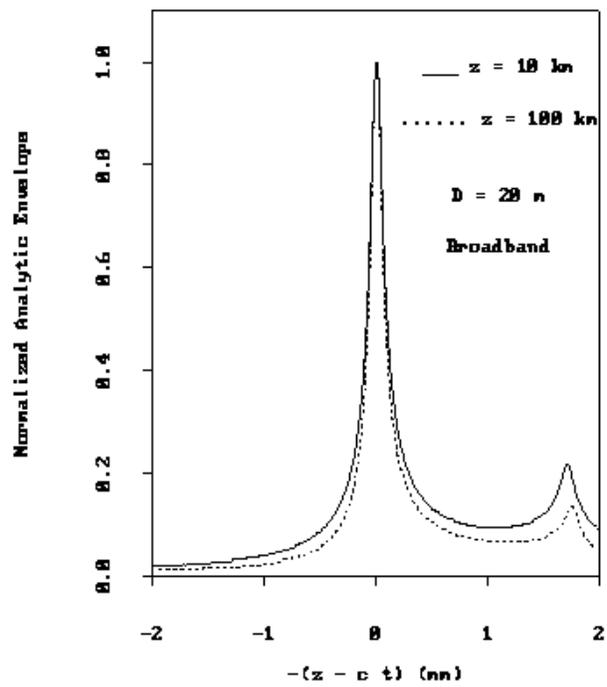
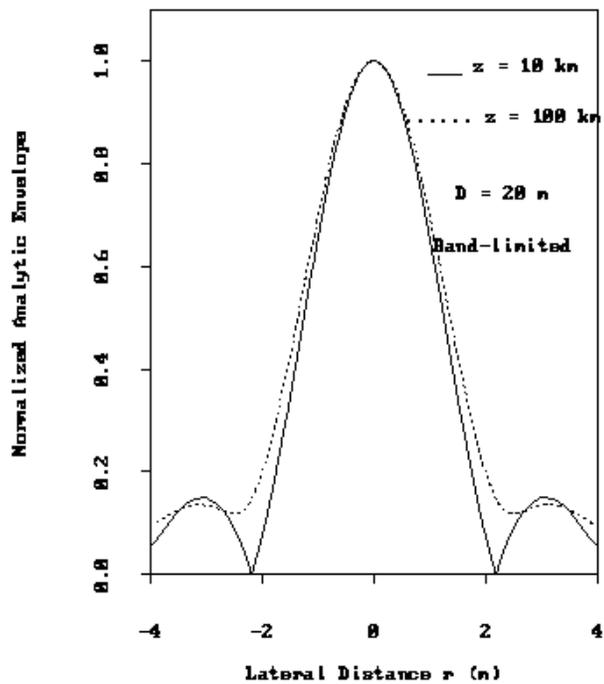
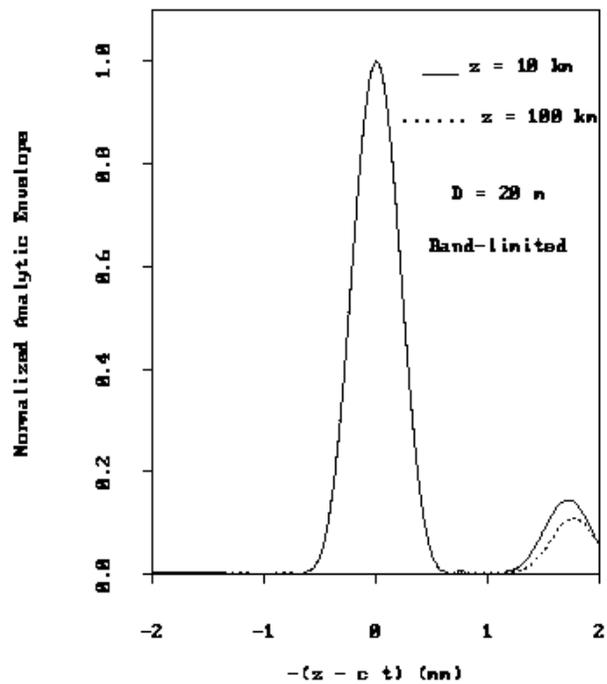
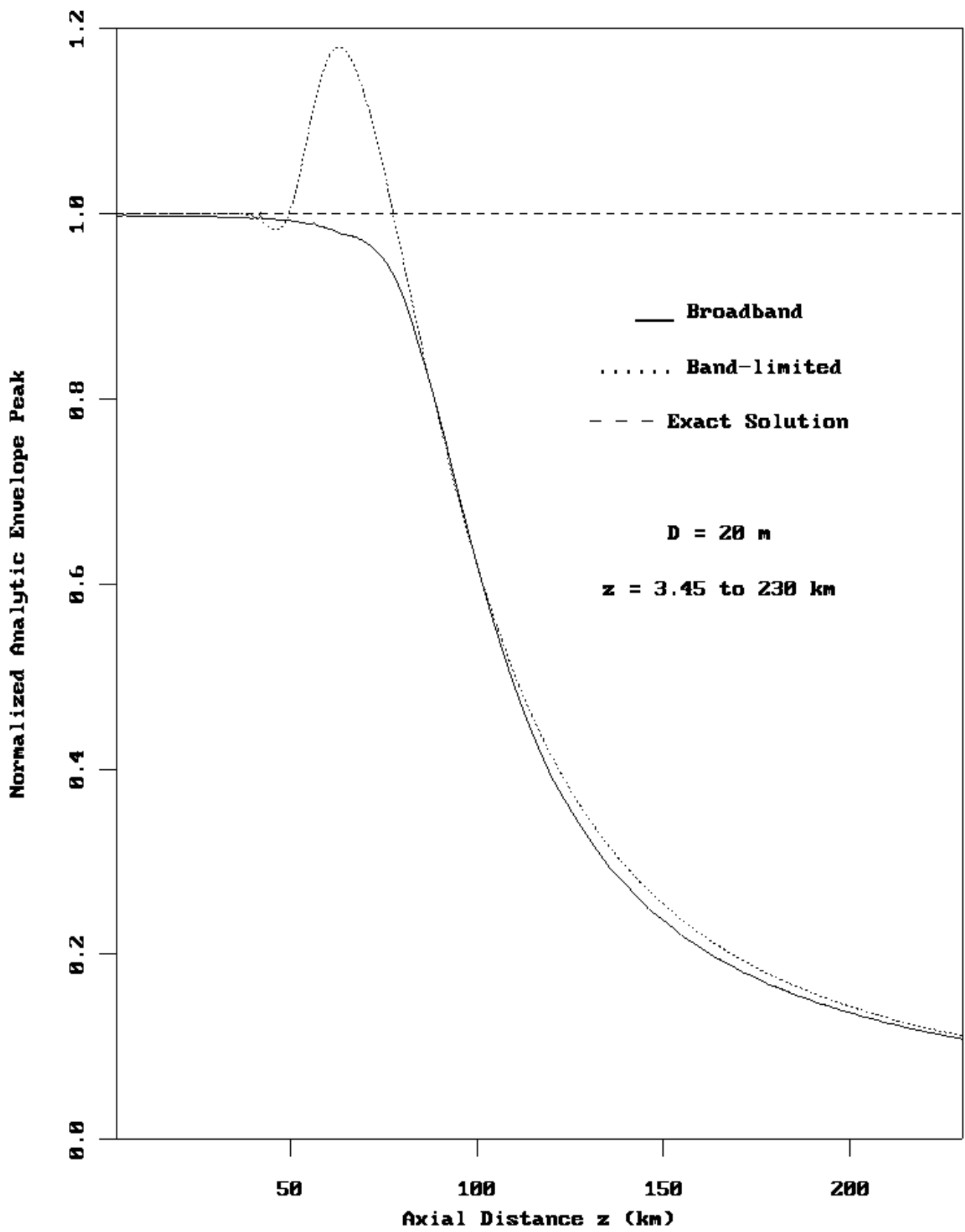
\end{document}